\pdfoutput=1
\documentclass[11pt,a4paper]{article}

\usepackage[utf8]{inputenc}
\usepackage[T1]{fontenc}
\usepackage[english]{babel}
\usepackage{lmodern}
\usepackage{microtype}
\usepackage[margin=1in]{geometry}
\usepackage{setspace}
\usepackage{amsmath,amssymb,amsfonts,amsthm,bm}
\usepackage{etoolbox}
\usepackage{braket}
\usepackage{graphicx}
\graphicspath{{./figures/}}
\usepackage{caption}
\usepackage{subcaption}
\usepackage{booktabs}
\usepackage{longtable}
\usepackage{multirow}
\usepackage{array}
\usepackage{float}
\usepackage{pdflscape}
\usepackage[figuresright]{rotating}
\usepackage{listings}
\usepackage{xcolor}
\usepackage{enumitem}
\usepackage{url}
\usepackage{verbatim}
\usepackage{chngcntr}
\usepackage[authoryear,round]{natbib}
\usepackage[bookmarksnumbered=true,colorlinks=true,linkcolor=blue,citecolor=blue,urlcolor=blue]{hyperref}
\usepackage{bookmark}
\usepackage[capitalise,nameinlink,noabbrev]{cleveref}

\allowdisplaybreaks
\setlist[itemize]{leftmargin=1.5em}
\setlist[enumerate]{leftmargin=1.8em}
\renewcommand{\arraystretch}{1.15}
\hypersetup{bookmarksopen=true,bookmarksopenlevel=2,linktoc=all}
\preto{\section}{\clearpage}

\theoremstyle{remark}

\newcommand{\CodeRepoBase}{https://github.com/gonghui945/quantum-finance}
\newcommand{\CodeRepo}{\href{\CodeRepoBase}{\texttt{gonghui945/quantum-finance}}}
\newcommand{\repodir}[1]{\href{\CodeRepoBase/tree/main/#1}{\nolinkurl{#1}}}
\newcommand{\repofile}[2]{\href{\CodeRepoBase/blob/main/#1}{\nolinkurl{#2}}}
\newcommand{\assetfile}[1]{\nolinkurl{#1}}
\newcommand{\pagelink}[1]{\hyperref[#1]{p.~\pageref*{#1}}}
\numberwithin{equation}{section}
\counterwithin{figure}{section}
\counterwithin{table}{section}

\title{\textbf{Quantum Computing for Financial Transformation: A Review of Optimisation, Pricing, Risk, Machine Learning, and Post-Quantum Security}}
\author{Hui Gong\textsuperscript{\textdagger}, Akash Sedai, Thomas Schroeder, Francesca Medda\\[0.4em]\normalsize UCL IFT Center for Quantum Finance}
\date{}

\begin{document}

\maketitle
\begingroup
\renewcommand{\thefootnote}{\fnsymbol{footnote}}
\footnotetext[2]{Corresponding author. Email: \href{mailto:h.gong.12@ucl.ac.uk}{h.gong.12@ucl.ac.uk}.}
\endgroup
\setcounter{footnote}{0}

\begin{abstract}
Quantum computing has become strategically relevant to finance not because every financial problem is waiting for a quantum speedup, but because several core bottlenecks in modern financial systems are already defined by combinatorial search, expectation estimation, rare-event analysis, representation learning, and long-horizon cryptographic resilience. This review evaluates that landscape across five connected domains: constrained portfolio optimisation, derivative pricing, tail-risk and scenario estimation, quantum machine learning, and post-quantum security. Rather than treating these as isolated demonstrations, the paper studies them as linked layers of a financial-computation stack.

Across all five domains, the review applies a common evaluative logic: identify the financial bottleneck, specify the relevant quantum primitive, compare it against an explicit classical benchmark, and judge the result under realistic implementation and governance constraints. The central conclusion is measured but consequential. The strongest near-term case for quantum finance lies in carefully designed hybrid workflows, not in blanket claims of universal advantage. Quantum optimisation becomes most credible when constrained search dominates, amplitude-estimation methods matter most when repeated expectation evaluation is the binding cost, quantum machine learning remains strongly task dependent, and post-quantum cryptography is already strategically necessary because financial infrastructures must migrate before fault-tolerant attacks arrive.

By combining system-level synthesis with locally reproducible small-scale case studies on simulated qubit registers, the article is designed to function both as a review of the current field and as a handbook-style entry point for future work. Its broader claim is that quantum finance should be assessed as a problem of financial transformation under technological, institutional, and security constraints, rather than as a collection of disconnected algorithmic proofs of concept.
\end{abstract}

\tableofcontents
\clearpage
\phantomsection
\addcontentsline{toc}{section}{List of Figures}
\listoffigures
\clearpage
\phantomsection
\addcontentsline{toc}{section}{List of Tables}
\listoftables
\clearpage

\setcounter{section}{-1}

\section{Introduction and Scope}
\label{sec:introduction-scope}
Quantum finance is entering a more consequential phase. What was once discussed mainly as a collection of algorithmic curiosities is now better understood as a broader question about the future architecture of financial computation: which financial tasks may be materially restructured by quantum or hybrid quantum-classical methods, which should remain classical, and how should institutions prepare before hardware, standards, and supervisory expectations fully converge? That question matters because modern finance is built on layered computational systems in which optimisation, pricing, risk, learning, and security are operationally inseparable.

The forward-looking motivation is therefore precise rather than rhetorical. Portfolio construction becomes combinatorial once real-world allocation rules, turnover penalties, sector limits, and sustainability constraints are imposed. Monte Carlo pricing and \phantomsection\label{term:xva}XVA-style expectation problems remain expensive because classical precision improves only at the square-root rate in the number of samples. Tail-risk estimation and stress propagation are dominated by rare-event structure and high-dimensional dependence. Learning systems must operate under weak labels, shifting regimes, and fragile feature engineering. At the same time, the cryptographic layer that protects financial infrastructure will eventually face a discontinuous break once fault-tolerant quantum attacks become viable. In each case, the relevant question is not whether quantum computation is interesting in the abstract, but whether it changes the computational economics, governance burden, or security posture of a concrete financial workflow.

That makes a review timely for two reasons. First, the institutional conversation is sharpening. The Financial Conduct Authority highlights optimisation, machine learning, and stochastic modelling as material domains for financial-services applications, while also stressing benchmarking, explainability, validation, and resilience as prerequisites for adoption \citep{FCAQuantumFinance}. Second, the UK industrial strategy places quantum technologies within a broader frontier-capability agenda that includes finance as a domain of applied strategic importance \citep{UKIndustrialStrategy2025}. These signals do not settle the technical debate, but they do make clear that quantum finance should now be reviewed with more discipline than generic technology forecasting allows.

This article is written with that discipline in mind. It aims to function both as a rigorous review and as a handbook-style point of entry for researchers, practitioners, and policymakers who want a structured map of the field. To that end, the paper pairs conceptual synthesis with locally reproducible small-scale case studies built on simulated qubit registers. These examples are intentionally modest: they do not claim production-scale advantage, but they make the comparison between classical and quantum or quantum-inspired workflows concrete, auditable, and extendable. The resulting ambition is practical rather than promotional: to clarify where quantum finance may become genuinely useful, what evidence standard should apply, and how the field can mature without losing contact with financial reality.

The article therefore treats quantum finance as a layered review of financial computation. Module~1 studies constrained optimisation and QUBO-based portfolio selection. Module~2 examines amplitude-estimation methods for derivative pricing and expectation estimation. Module~3 focuses on Value-at-Risk (VaR), Conditional Value-at-Risk (CVaR), scenario loading, and stress propagation. Module~4 addresses quantum machine learning for feature construction, kernels, and variational classification. Module~5 turns to post-quantum cryptography as the security layer that conditions whether quantum-enhanced finance can be deployed credibly at all. The core argument is that these five domains are analytically distinct but operationally connected, and that reviewing them together yields a more realistic account of quantum financial transformation than treating them as separate literatures.

\subsection{Scope, Evaluation Criteria, and Review Logic}
\label{sec:scope-evaluation-review-logic}
The scope of the review is intentionally selective. It does not attempt an encyclopaedic catalogue of every published quantum-finance result, nor does it assume that every theoretical speedup is financially actionable. Instead, the paper concentrates on five domains that recur across both the academic and institutional literature, and evaluates them through a common lens designed to separate conceptual promise from deployable evidence. In that sense, the manuscript is meant to be useful not only as a narrative review, but also as a handbook-style reference that future readers can return to when they need a structured entry point into a specific financial application area.

Each substantive section is therefore organised around four recurring questions.
\begin{itemize}
    \item \textbf{What is the financial bottleneck?} The starting point is always a concrete problem in institutional finance, not a quantum algorithm in isolation.
    \item \textbf{What is the relevant quantum or hybrid primitive?} Each section identifies the computational mechanism that might alter the classical cost profile or representation structure.
    \item \textbf{What is the benchmark and evidence standard?} Claims are interpreted relative to classical baselines, problem scaling, data realism, and implementation assumptions.
    \item \textbf{What blocks deployment?} Validation, governance, explainability, data engineering, hardware limits, and security migration remain part of the scientific assessment rather than afterthoughts.
\end{itemize}

That framing keeps the review aligned with financial practice. A speedup claim without a meaningful baseline is not persuasive; a promising benchmark without a path to integration is not yet an institutional result; and a quantum-enabled workflow without a security and governance model is incomplete. The paper therefore treats quantum advantage, hybrid design, and post-quantum resilience as parts of the same evaluative problem.

\subsection{How to Read This Review}
\label{sec:editorial-roadmap}
Although the manuscript now uses article-style section numbering, it intentionally retains the original Module~1--5 language from the earlier project drafts. The distinction is simple. The section numbers define the order of the review article; the module labels preserve the research-stream identity of the underlying material. In practical terms, Module~1 maps to \Cref{sec:optimisation}, Module~2 to \Cref{sec:pricing}, Module~3 to \Cref{sec:risk}, Module~4 to \Cref{sec:qml}, and Module~5 to \Cref{sec:pqc}. The additional local U.S.-equity benchmark remains part of Module~1 rather than a separate module; it is a reproducible case expansion embedded within the optimisation section.

\begin{table}[H]
\centering
\small
\caption{Reading guide for the review article and its retained module labels.}
\label{tab:reading-guide}
\begin{tabular}{p{0.17\linewidth}p{0.12\linewidth}p{0.25\linewidth}p{0.34\linewidth}}
\toprule
\textbf{Article location} & \textbf{Module} & \textbf{Role in the review} & \textbf{Core reader question} \\
\midrule
\Cref{sec:optimisation} & Module 1 & Decision layer for constrained allocation, QUBO mapping, and solver comparison, with a local reproducible case. & When do quantum search and hybrid optimisation methods become relevant once portfolio selection turns combinatorial? \\
\Cref{sec:pricing} & Module 2 & Valuation layer centred on amplitude estimation and Monte Carlo acceleration. & Where can quantum expectation-estimation routines alter the economics of derivative pricing? \\
\Cref{sec:risk} & Module 3 & Tail-risk, scenario-generation, and stress-testing layer. & Can quantum or quantum-inspired workflows concentrate computation where rare-event risk matters most? \\
\Cref{sec:qml} & Module 4 & Predictive and representation-learning layer for classification, kernels, and variational models. & What kinds of financial learning tasks might benefit from quantum feature maps or parameterised circuits, and where do classical models still dominate? \\
\Cref{sec:pqc} & Module 5 & Security, migration, and infrastructure layer. & How can a quantum-augmented financial stack remain trustworthy once quantum attacks threaten classical cryptography? \\
\Cref{sec:integration} & Synthesis & System-level integration across all five streams. & How do the modules combine into one institutional roadmap rather than five isolated proofs of concept? \\
\bottomrule
\end{tabular}
\end{table}

The recommended reading order is sequential because the review follows a deliberate stack logic. Optimisation defines how decisions are encoded; pricing and risk then show how valuation and capital measurement may be accelerated or reorganised; machine learning adds an adaptive signal-extraction layer; and post-quantum cryptography finally defines the conditions under which the earlier capabilities could be deployed safely in real financial systems. Readers can still enter at a single section of interest, but the paper is written to argue that these topics become more meaningful when read as connected parts of one financial-computation architecture.

\section{Quantum Optimisation: Portfolio Selection and Algorithmic Comparisons}
\label{sec:optimisation}

\paragraph{Section Overview.}
This section studies the optimisation layer of quantum finance. It explains how constrained portfolio-selection problems are reformulated as quantum-ready QUBO or Ising systems, compares the main algorithmic families used in the literature, and clarifies where quantum optimisation differs from classical mixed-integer and heuristic workflows in financial practice.

\subsection{Problem Context and Financial Motivation}
\label{sec:m1-problem-context-why-portfolio-optimisation-requires-quantum}

Portfolio optimisation is a central task in financial decision-making, often modeled using the classical mean-variance framework proposed by Markowitz (1952). In the binary inclusion setting, we seek to select a subset of $K$ assets from a universe of $n$ candidates that balance return and risk.

We define the binary decision vector $x \in \{0,1\}^n$, where $x_i = 1$ indicates that asset $i$ is included in the portfolio. The correct formulation for cardinality-constrained binary portfolio optimisation is:

\begin{equation}
\begin{aligned}
\min_{x \in \{0,1\}^n} \quad & x^\top \Sigma x - \lambda \mu^\top x \\
\text{subject to} \quad & \sum_{i=1}^n x_i = K
\end{aligned}
\end{equation}

where $\Sigma \in \mathbb{R}^{n \times n}$ is the covariance matrix of asset returns, $\mu \in \mathbb{R}^n$ is the vector of expected returns, $\lambda > 0$ is a risk-aversion parameter, $K$ is a user-defined cardinality threshold.

In practice, financial portfolio optimisation often combines binary inclusion decisions with budget constraints, turnover penalties, sector exposure limits, or ESG filters. These can be incorporated as additional constraints or penalty terms in the objective function. However, once binary or integer restrictions are introduced, the problem quickly becomes \textbf{NP-hard}. 

Note that in classical continuous optimisation, the weights $w_i \in [0,1]$ represent fractional allocations with a total budget constraint $\sum w_i = 1$. In contrast, the binary formulation focuses on asset selection. These models can be unified under mixed-integer programming (MIP) frameworks, but solving them at scale remains computationally expensive—motivating the exploration of quantum optimisation techniques.

\paragraph{Complexity of Constraints.}
The difficulty escalates significantly under realistic portfolio rules:
\begin{itemize}
    \item \textbf{Large universes of assets:} Selecting from $n=100$ or more equities leads to $2^{100}$ possible inclusion sets---far beyond classical enumeration.
    \item \textbf{Cardinality constraints:} Requiring, for example, ``at most 10 out of 100 assets'' introduces combinatorial explosion $\binom{100}{10} \approx 1.7\times 10^{13}$ possibilities.
    \item \textbf{Sector diversification:} Restrictions on exposure by industry or geography create additional layers of integer programming.
    \item \textbf{Risk budgets and turnover limits:} Non-linear constraints such as ``maximum drawdown'' or ``trading cost penalties'' cannot be linearised easily.
    \item \textbf{ESG (Environmental, Social, Governance) scores:} Multi-objective optimisation with social constraints pushes the problem into multi-criteria NP-hardness.
\end{itemize}

Classical solvers such as branch-and-bound, simulated annealing, or genetic algorithms often \textbf{scale poorly in high dimensions}, and typically fail to guarantee global optima in feasible time for large-scale financial universes. This motivates the exploration of quantum approaches. 

\paragraph{Quantum Computing Foundations.}
Quantum computing provides an alternative paradigm of computation, built upon four fundamental principles:

\begin{itemize}
    \item \textbf{Superposition:} A qubit can exist in both states $|0\rangle$ and $|1\rangle$ simultaneously,
    \begin{equation}
    |\psi\rangle = \alpha |0\rangle + \beta |1\rangle, \quad |\alpha|^2 + |\beta|^2 = 1,
    \end{equation}
    allowing parallel encoding of multiple candidate solutions.
    
    \item \textbf{Entanglement:} Qubits can become correlated such that the state of one immediately determines the other, e.g.
    \begin{equation}
    |\Phi^+\rangle = \frac{|00\rangle + |11\rangle}{\sqrt{2}}.
    \end{equation}
    This enables financial correlations or constraints to be embedded directly into multi-qubit states.
    
    \item \textbf{Interference:} Quantum amplitudes can add constructively or destructively, amplifying correct solutions and cancelling incorrect ones. This is the ``magic'' behind Grover’s search and QAOA.
    
    \item \textbf{Measurement:} Collapses the quantum state into classical outcomes. Though probabilistic, repeated sampling extracts high-quality solutions.
\end{itemize}

\paragraph{Logical Relationship.}

In essence, the framework of quantum computation can be summarised as a combination of four fundamental principles:
\begin{equation}
\text{Quantum Computation} 
= \text{Superposition} + \text{Entanglement} + \text{Interference} + \text{Measurement}.
\end{equation}
\begin{itemize}
    \item \textbf{Superposition:} provides exponential parallelism.
    \item \textbf{Entanglement:} introduces nonlocal correlations between qubits.
    \item \textbf{Interference:} amplifies correct solutions while cancelling incorrect ones.
    \item \textbf{Measurement:} projects quantum states into classical outcomes.
\end{itemize}

\paragraph{Physical Interpretation.}
\begin{enumerate}
    \item \textbf{Superposition $\rightarrow$ Quantum Parallelism}  
    A single qubit represents two states simultaneously. $n$ qubits represent $2^n$ states in parallel, enabling exponential exploration of portfolios.
    
    \item \textbf{Entanglement $\rightarrow$ Nonlocal Correlation}  
    Entangled qubits share correlations across the system. Measurement of one asset qubit instantly influences another, analogous to correlated risk factors in finance.
    
    \item \textbf{Interference $\rightarrow$ Amplification of Correct Answers}  
    Quantum interference cancels out low-quality solutions while reinforcing high-quality ones. In Grover’s search, amplitude amplification highlights the target solution from a superposed state.
    
    \item \textbf{Measurement $\rightarrow$ Classical Output Extraction}  
    The quantum state collapses to a feasible portfolio configuration. By repeated runs, one samples portfolios with high probability of being near-optimal.
\end{enumerate}

\paragraph{Why Quantum for Optimisation.}

In financial optimisation, the search space grows exponentially with the number of assets and constraints. Classical algorithms -- including branch-and-bound, simulated annealing, and evolutionary heuristics -- struggle to scale efficiently in such high-dimensional, non-convex settings. Quantum computing, by contrast, introduces a fundamentally different computational paradigm rooted in superposition, entanglement, and interference, enabling exploration of exponentially many states in parallel and the probabilistic amplification of high-quality solutions.

\begin{itemize}
    \item \textbf{Parallel representation of exponentially large solution spaces.}  
    Through the principle of \emph{superposition}, a system of $n$ qubits represents $2^n$ possible portfolio inclusion states simultaneously. This allows the entire feasible portfolio universe to be encoded within a single quantum state vector, providing an inherent exponential form of computational parallelism.

    \item \textbf{Direct encoding of financial correlations and constraints.}  
    Quantum \emph{entanglement} enables non-local correlations among qubits, allowing the dependencies between assets---such as covariance, sector exposure, or risk parity---to be embedded directly into the quantum state or Hamiltonian. Unlike classical solvers that handle correlations through penalty functions, quantum systems can naturally express them as intrinsic coupling terms ($J_{ij}$) in the cost Hamiltonian.

    \item \textbf{Constructive interference for optimal solution biasing.}  
    Quantum algorithms leverage \emph{interference} to amplify probability amplitudes associated with feasible, high-return configurations while suppressing low-quality solutions. This property lies at the heart of algorithms such as QAOA and Grover Adaptive Search, enabling efficient navigation of vast combinatorial landscapes.

    \item \textbf{Probabilistic yet efficient extraction of near-optimal portfolios.}  
    Quantum \emph{measurement} collapses the state into classical bitstrings that correspond to feasible portfolio selections. Repeated sampling of the quantum state yields near-optimal portfolios with high probability, allowing for stochastic solution refinement in hybrid quantum-classical loops.
\end{itemize}

Beyond their theoretical elegance, these mechanisms provide a computational framework that can encode financial optimisation problems directly into quantum hardware. Algorithms such as the Quantum Approximate Optimisation Algorithm (QAOA), Grover Adaptive Search (GAS), and Quantum Annealing (QA) exploit these properties to overcome the exponential bottlenecks that limit classical optimisation. 

By enabling parallel evaluation, intrinsic correlation modelling, and amplitude-based solution reinforcement, quantum optimisation offers a transformative pathway toward real-time, high-dimensional decision-making in finance---particularly for asset allocation, risk-adjusted portfolio construction, and capital efficiency analysis.

\subsection{QUBO Encoding and Problem Mapping}
\label{sec:m1-mapping-to-quantum-optimisation-the-qubo-formulation}

To leverage quantum optimisation algorithms, many financial problems — including portfolio selection — must be reformulated into a format amenable to quantum computation. The most widely used formulation is the \textbf{\phantomsection\label{term:qubo}Quadratic Unconstrained Binary Optimisation (QUBO)} model.

\paragraph{QUBO Definition.}
The general QUBO problem is defined as:

\begin{equation}
\min_{z \in \{0,1\}^n} z^\top Q z
\end{equation}

where:
\begin{itemize}
    \item \( z = (z_1, z_2, \dots, z_n) \in \{0,1\}^n \) is a binary decision vector, with \(z_i = 1\) indicating asset $i$ is selected in the portfolio.
    \item \( Q \in \mathbb{R}^{n \times n} \) is a symmetric matrix encoding pairwise interactions and linear coefficients that incorporate return, risk, and constraint terms.
\end{itemize}

The objective is to find the binary vector \( z \) that minimizes the quadratic cost function \( z^\top Q z \), capturing the trade-offs in portfolio construction.

\paragraph{Embedding Financial Objectives into QUBO.}
The QUBO matrix \( Q \) can be constructed by translating financial goals and constraints into a unified binary optimisation problem. Key components include:

\begin{itemize}
    \item \textbf{Expected Return Maximization:} Incorporate as a linear term with negative sign (since QUBO minimizes):
    \begin{equation}
    - \sum_i \mu_i z_i \quad \Rightarrow \quad \text{added to diagonal elements of } Q.
    \end{equation}

    \item \textbf{Risk Minimization:} Use the asset return covariance matrix \( \Sigma \) to encode risk as a quadratic term:
    \begin{equation}
    \sum_{i,j} \Sigma_{ij} z_i z_j \quad \Rightarrow \quad \text{added to } Q_{ij}.
    \end{equation}

    \item \textbf{Cardinality Constraints:} Enforce exact or maximum number of selected assets via penalty terms:
    \begin{equation}
    \left( \sum_i z_i - k \right)^2 \quad \text{penalizes deviation from } k \text{ assets.}
    \end{equation}

    \item \textbf{Sector or ESG Constraints:} Add quadratic penalties to encode soft or hard allocation limits by group.
\end{itemize}

These terms are combined into a single QUBO matrix \( Q \), which becomes the input to quantum solvers.

\paragraph{Relation to Ising Models.}

While the QUBO formulation provides a convenient mathematical representation for binary optimisation problems, many quantum algorithms — particularly those implemented on quantum annealers — naturally operate in the language of \textit{Ising Hamiltonians}. The transition from a QUBO cost function to an Ising model is therefore both conceptually and practically important, as it connects abstract optimisation formulations to the physical models realised on current quantum hardware.

The Ising model is defined over spin variables \( s_i \in \{-1, +1\} \), and takes the form:
\begin{equation}
E(s) = \sum_{i<j} J_{ij} s_i s_j + \sum_i h_i s_i
\end{equation}

The QUBO and Ising formulations are mathematically equivalent under a simple affine transformation between binary variables and spin variables:
\begin{equation}
z_i = \frac{1 + s_i}{2} \quad \Leftrightarrow \quad s_i = 2z_i - 1
\end{equation}

This equivalence allows optimisation problems initially expressed in QUBO form to be directly mapped onto Ising-based quantum hardware (such as D-Wave systems) or simulated via Hamiltonian evolution on gate-model devices. In practice, this mapping ensures that portfolio optimisation problems encoded as QUBOs can be physically represented as energy minimisation tasks within a quantum system.

\paragraph{Quantum Algorithms Compatible with QUBO.}
The QUBO format is compatible with a wide range of quantum optimisation algorithms:

\begin{itemize}
    \item \textbf{Quantum Annealing:} Directly solves Ising-encoded QUBO problems by evolving the system through quantum tunneling to reach low-energy (near-optimal) solutions.
    
    \item \textbf{QAOA (Quantum Approximate Optimisation Algorithm):} Constructs quantum circuits that simulate alternating Hamiltonians derived from the QUBO matrix, using interference to bias solutions toward the global minimum.
    
    \item \textbf{Grover Adaptive Search:} Searches over the binary solution space using amplitude amplification, requiring QUBO-style oracle evaluation to define “good” solutions.
\end{itemize}

\paragraph{Why This Mapping Matters.}

The QUBO formulation serves as a crucial bridge between complex financial objectives and the operational requirements of quantum hardware. It allows portfolio optimisation problems — traditionally expressed through return, risk, and regulatory constraints — to be translated into an energy minimisation landscape directly executable on quantum devices. In this mapping, the QUBO matrix \( Q \) encodes all interactions among decision variables: diagonal terms represent individual asset contributions, while off-diagonal elements capture covariances, diversification effects, or penalty couplings arising from constraints.

This translation is not merely mathematical but conceptual: it transforms symbolic optimisation goals into physical energy states, where the ground state of the Hamiltonian corresponds to the optimal or near-optimal investment configuration. As a result, both gate-based algorithms (such as QAOA on IBM Q) and annealing-based systems (such as D-Wave) can operate natively on financial optimisation tasks. Through QUBO encoding, quantum optimisation provides a unified framework to co-optimise risk–return trade-offs, compliance conditions, and sustainability metrics within a single computational paradigm.

\subsection{Quantum Optimisation Methods}
\label{sec:m1-algorithmic-comparison-quantum-optimisation-techniques}

Quantum optimisation for portfolio selection can be implemented using multiple algorithmic paradigms. Each algorithm differs in its quantum computing model, solution mechanism, scalability, and compatibility with financial problem structures.

\begin{table}[H]
\caption{Comparison of quantum optimisation algorithms for portfolio problems.}
\centering
\footnotesize{
\begin{tabular}{>{\raggedright\arraybackslash}p{70pt} >{\raggedright\arraybackslash}p{75pt} >{\raggedright\arraybackslash}p{75pt} >{\raggedright\arraybackslash}p{75pt} >{\raggedright\arraybackslash}p{75pt}}
\toprule
\textbf{Algorithm} & \textbf{Quantum Type} & \textbf{Computing Style} & \textbf{Strengths} & \textbf{Limitations}\\
\midrule
QAOA (Quantum Approximate Optimisation Algorithm) & Interference-based (gate model) & Variational hybrid & Tunable, NISQ-compatible & Requires classical optimizer loop \\
VQE (Variational Quantum Eigensolver) & Interference-based (gate model) & Variational hybrid & Suited for continuous eigensolvers & High circuit depth needed for precision \\
Grover Adaptive Search (GAS) & Interference-based & Oracle-based amplification & Quadratic speedup for unstructured search & Needs well-defined oracle \\
Quantum Annealing (QA) & Tunneling-based (analog model) & Physical annealing process & Scales to large QUBOs & Sensitive to noise, sparse connectivity \\
\bottomrule
\end{tabular}
}
\end{table}

Among the gate-based variational methods, the Variational Quantum Eigensolver (VQE) remains a relevant reference point when the optimisation objective is cast as an eigenvalue problem rather than a purely combinatorial search task.

\subsubsection{Quantum Approximate Optimisation Algorithm (QAOA)}
\label{sec:m1-quantum-approximate-optimisation-algorithm-qaoa}

\phantomsection\label{term:qaoa}QAOA is a hybrid quantum-classical algorithm that approximates the ground state of an \phantomsection\label{term:ising}Ising or QUBO Hamiltonian by alternating between two unitary operators:

\begin{itemize}
    \item \textbf{Cost Hamiltonian:} Encodes the QUBO objective function.
    \item \textbf{Mixer Hamiltonian:} Encourages exploration of the solution space.
\end{itemize}

The quantum state is evolved by alternating these unitaries for $p$ layers:
\begin{equation}
|\psi(\boldsymbol{\gamma}, \boldsymbol{\beta})\rangle = \prod_{l=1}^{p} e^{-i \beta_l H_M} e^{-i \gamma_l H_C} |+\rangle^{\otimes n}
\end{equation}

Where:
\begin{equation}
H_C = \sum_{i,j} Q_{ij} Z_i Z_j, \qquad H_M = \sum_i X_i
\end{equation}

Here \(Z_i\) and \(X_i\) are Pauli operators, and \(\gamma_l, \beta_l\) are variational parameters optimised classically.

\paragraph{Illustrative Case.}
IBM (2020) applied QAOA to a 6-asset portfolio selection problem and achieved 99.5\% optimality compared to exhaustive brute-force search. The solution encoded return and risk in the cost Hamiltonian, and enforced a cardinality constraint via a penalty term.

\paragraph{Strengths.}
QAOA is especially attractive in finance because it preserves the optimisation problem in a form that remains economically interpretable. The return--risk trade-off, the penalty structure, and the constraint logic are all visible in the cost Hamiltonian, while the circuit depth \(p\) provides an explicit handle on the accuracy--resource trade-off. This makes QAOA more transparent than many black-box heuristics: practitioners can inspect the Hamiltonian design, adjust mixer choices, warm-start the variational loop, and benchmark the result directly against classical relaxations or exact small-scale solutions. In the NISQ regime, that tunability and auditability are often more valuable than raw optimality claims.

\subsubsection{Quantum Annealing (QA)}
\label{sec:m1-quantum-annealing-qa}

Quantum Annealing solves optimisation problems by physically evolving an initial Hamiltonian \( H_0 \) into a final problem Hamiltonian \( H_P \) over time \( t \):
\begin{equation}
H(t) = (1 - s(t)) H_0 + s(t) H_P
\end{equation}

The system starts in the ground state of \( H_0 \), and if evolved slowly (adiabatically), it remains in the ground state of \( H_P \), yielding an optimal solution.

For portfolio optimisation, the problem Hamiltonian is typically of Ising form:
\begin{equation}
E(s) = \sum_{i<j} J_{ij} s_i s_j + \sum_i h_i s_i
\end{equation}

where \( s_i \in \{-1, +1\} \), and \( J_{ij}, h_i \) encode the financial objective and constraints.

\paragraph{Hardware Example.}
D-Wave 2000Q has been successfully used to solve up to 60-asset portfolio problems (Rosenberg et al., 2016), demonstrating quantum annealing’s capability in high-dimensional QUBOs.

\paragraph{Key Feature.}
Unlike classical simulated annealing, which escapes local minima via thermal fluctuations, quantum annealing uses \textbf{quantum tunneling} to pass through high but narrow energy barriers—especially useful in rugged financial cost landscapes.

\subsubsection{Grover Adaptive Search (GAS)}
\label{sec:m1-grover-adaptive-search-gas}

\phantomsection\label{term:gas}GAS is a quantum algorithm based on Grover’s unstructured search but generalised for optimisation. It uses amplitude amplification to enhance the probability of selecting the best solution from the space of feasible portfolios.

\paragraph{Complexity:}
Grover's search reduces the number of queries from classical \( O(N) \) to quantum \( O(\sqrt{N}) \), where \( N \) is the size of the solution space.

\paragraph{Optimisation Process:}
GAS assumes access to an \textit{oracle function} that can identify whether a portfolio meets desired criteria (e.g., Sharpe ratio above threshold). It then applies Grover iterations to amplify valid solution states.

\paragraph{Illustrative Case.}
GAS is suitable when the target portfolio structure is known through constraints but requires fast identification (e.g., filtering for ESG + return + liquidity conditions). It is most effective in low-dimensional discrete problems with known validation oracles.

\paragraph{Synthesis.}
Each quantum algorithm offers trade-offs:
\begin{itemize}
    \item QAOA is versatile and tunable for medium-sized structured problems.
    \item QA scales best for large sparse QUBO problems on specialised hardware.
    \item GAS provides theoretical speedups in threshold-based decision problems.
\end{itemize}
Selecting the right algorithm depends on hardware availability, problem dimensionality, and financial model structure.

\subsection{Quantum vs Classical Optimisation: A Comparative View}
\label{sec:m1-quantum-vs-classical-optimisation-a-comparative-view}

To evaluate the potential of quantum optimisation in finance, it is essential to contrast its performance and characteristics with classical solvers. Classical approaches such as Gurobi (MILP), simulated annealing, or genetic algorithms have been widely adopted in industry, but exhibit limitations in scalability and constraint expressiveness when tackling large-scale portfolio problems.

\paragraph{Interpretation.}
Quantum optimisation presents unique advantages in high-dimensional discrete problems — especially where:

\begin{itemize}
    \item Constraint structures are non-convex or binary in nature (e.g., ``choose exactly 10 stocks'' or ``limit to 3 ESG sectors'').
    \item Classical solvers require exponential time due to tree-based search, relaxation, or combinatorial branching.
    \item The solution space is too large to evaluate exhaustively, but quantum interference and tunneling can bias the search toward optimal solutions.
\end{itemize}

From a review perspective, the key point is not that QAOA already dominates the best classical solvers in every institutional setting. It is that gate-based optimisation introduces a different design grammar: the problem is written as a Hamiltonian, constraints become part of the energy landscape, and solution quality can be improved systematically through circuit depth, mixer design, and hybrid parameter optimisation. That makes QAOA especially useful as a bridge technology between classical portfolio engineering and future hardware-aware decision engines.

\begin{table}[H]
\caption{Comparative features of classical and quantum optimisation approaches.}
\vspace{0.5em}
\centering
\footnotesize{
\begin{tabular}{>{\raggedright\arraybackslash}p{2.5cm}>{\raggedright\arraybackslash}p{5cm}>{\raggedright\arraybackslash}p{6.5cm}}
\toprule
\textbf{Feature} & \textbf{Classical Optimisation} & \textbf{Quantum Optimisation (QAOA / QA / GAS)} \\
\midrule
\textbf{Scaling behavior} & Exponential time complexity in presence of combinatorial constraints and large asset universes. Solving for 100+ binary decision variables with cardinality and sector constraints becomes computationally infeasible. & QAOA has polynomial scaling in parameterised circuit depth; Grover's algorithm achieves quadratic speedup \( O(\sqrt{N}) \) over unstructured search; QA can rapidly sample from pre-encoded QUBO problems, with each annealing run typically executed within a microsecond-to-millisecond timescale. \\
\textbf{Constraint handling} & Requires reformulation of constraints into penalty functions or relaxation to linear programming (e.g., adding slack variables or penalty multipliers). Constraint tuning is sensitive and problem-specific. & Constraints are directly embedded into the problem Hamiltonian (e.g., Ising or QUBO form), preserving their structure. Hard constraints such as cardinality or sector caps can be naturally encoded. \\
\textbf{Global minimum guarantees} & Only achievable for small or convex problems. For integer/binary problems, solvers yield local minima or approximate heuristics unless exhaustive enumeration is used. & QAOA converges probabilistically toward high-quality solutions; QA samples low-energy states efficiently; Grover search increases likelihood of hitting target solutions in bounded iterations. No deterministic guarantee, but strong empirical performance. \\
\textbf{Solver time for 100-asset binary selection} & Minutes to hours depending on number of constraints, with exponential degradation as more combinatorial or nonlinear terms are added. & Quantum Annealing (QA) on D-Wave yields solutions in seconds; QAOA and GAS are iterative but operate in low-depth circuits with potential exponential advantage in parallel representation. \\
\textbf{Hardware requirements} & Runs on classical CPU/GPU clusters, often requiring parallel threading, linear algebra libraries, and memory-intensive branching. & Requires access to superconducting qubits (IBM Q, Rigetti), trapped-ion systems, or quantum annealers (D-Wave, Fujitsu Digital Annealer). Near-term implementations often hybrid (quantum + classical). \\
\bottomrule
\end{tabular}
}
\end{table}

Moreover, quantum optimisation allows encoding the financial logic—returns, risk, constraints—into the quantum Hamiltonian directly. This is a major paradigm shift from classical penalty-based heuristics.

\paragraph{Illustrative Case.}

Suppose a fund manager seeks to construct a 100-asset portfolio under realistic institutional constraints, including:

\begin{itemize}
    \item A maximum of 10 selected assets,
    \item Exposure to no more than 3 industry sectors,
    \item A risk budget capped at 5\% Value-at-Risk (VaR) under a normal distribution assumption,
    \item An aggregate ESG score exceeding a defined composite threshold.
\end{itemize}

In practice, these requirements can be addressed through two fundamentally different computational paradigms:

\begin{itemize}
    \item \textbf{Classically:} The problem is typically formulated as a mixed-integer linear or quadratic program (MILP/MIQP), requiring binary decision variables, multi-objective penalty weighting, and non-linear constraint relaxation. Achieving convergence often demands heuristic fine-tuning and extensive computational resources.
    
    \item \textbf{Quantumly:} The same optimisation objective can be reformulated as a QUBO problem, where each cost and constraint component is encoded into the matrix \( Q \). Using algorithms such as QAOA or quantum annealing, the portfolio configuration emerges from the ground state of a corresponding Hamiltonian, leveraging quantum parallelism and amplitude interference to explore the solution space efficiently.
\end{itemize}

This contrast highlights the conceptual shift introduced by quantum optimisation — from iterative search and constraint relaxation to probabilistic energy minimisation. 

\paragraph{Synthesis.}
While classical optimisation remains effective for convex or small-scale problems, quantum approaches offer a fundamentally different computational pathway for high-dimensional, combinatorial financial optimisation. As hardware improves and hybrid quantum-classical workflows mature, the boundary of practical advantage will likely shift, positioning quantum optimisation as a viable tool for large-scale institutional portfolio construction and risk-aware decision-making.

\subsection{Stylised Constraint Design for ESG-Constrained Portfolios}
\label{sec:m1-stylised-constraint-design-for-esg-constrained-portfolios}

Before turning to the reproducible local benchmark introduced in the next subsection, it is useful to retain one stylised institutional design template. The purpose of this section is not to serve as the paper's main empirical case, but to show how a financially realistic allocation problem can be translated into a QUBO-ready optimisation structure with multiple policy and governance constraints.

\subsubsection{Problem Setup and Dataset Structure}
\label{sec:m1-problem-setup-and-dataset-structure}

We assume access to a universe of \( n = 50 \) publicly listed equities, drawn from multiple sectors and regions. For each asset \( i \), the dataset contains:

\begin{itemize}
    \item \textbf{Expected Return:} \( \mu_i \), estimated from historical excess return or analyst forecasts.
    \item \textbf{Covariance Matrix:} \( \Sigma_{ij} \), estimated from past 12-month daily log returns.
    \item \textbf{ESG Score:} \( e_i \in [0,1] \), representing normalised environmental, social, and governance composite score.
    \item \textbf{Sector Tag:} \( s_i \in \{1, 2, ..., S\} \), mapping each asset to an industry.
    \item \textbf{Liquidity Proxy:} Average daily volume or bid-ask spread, for optional constraint inclusion.
\end{itemize}

\subsubsection{Portfolio Constraints and Objectives}
\label{sec:m1-portfolio-constraints-and-objectives}

The investor imposes the following constraints and preferences:

\begin{enumerate}
    \item \textbf{Cardinality:} Select exactly \( K = 10 \) assets from the 50 candidates.
    \item \textbf{Sector Diversification:} No more than 3 assets from any single sector.
    \item \textbf{Minimum ESG Score:} The portfolio's average ESG score must exceed 0.7.
    \item \textbf{Risk Minimization Objective:} Use variance \( x^\top \Sigma x \) as risk proxy.
    \item \textbf{Optional:} Maximize Sharpe ratio or include turnover penalty.
\end{enumerate}

These are encoded into the QUBO cost matrix \( Q \) using the binary decision vector \( z \in \{0,1\}^n \), where \( z_i = 1 \) indicates inclusion of asset \( i \) in the portfolio.

\subsubsection{QUBO Formulation}
\label{sec:m1-qubo-formulation}

\paragraph{Objective Function:}
\begin{equation}
\min_{z \in \{0,1\}^n} z^\top \Sigma z - \lambda \mu^\top z
\end{equation}

\paragraph{Constraints as Penalty Terms:}
\begin{itemize}
    \item Cardinality constraint: \( \left( \sum_i z_i - K \right)^2 \)
    \item Sector cap constraint: \( \sum_{j=1}^{S} \max(0, \sum_{i: s_i = j} z_i - 3)^2 \)
    \item ESG threshold: \( \max\left(0, 0.7 - \frac{1}{K} \sum_i e_i z_i\right)^2 \)
\end{itemize}

All terms are aggregated into the QUBO matrix \( Q \), with weighted penalty coefficients \( \rho_k \) tuned to enforce hard constraints while preserving the optimisation structure.

\subsubsection{Algorithm Execution: QAOA Flow}
\label{sec:m1-algorithm-execution-qaoa-flow}

We simulate solving the QUBO using the Quantum Approximate Optimisation Algorithm (QAOA):

\begin{enumerate}
    \item Initialize all qubits in uniform superposition state \( |+\rangle^{\otimes n} \).
    \item Construct cost Hamiltonian \( H_C \) from the QUBO matrix:
    \begin{equation}
    H_C = \sum_{i,j} Q_{ij} Z_i Z_j
    \end{equation}
    \item Apply alternating operators of the form:
    \begin{equation}
    |\psi(\gamma, \beta)\rangle = \prod_{l=1}^{p} e^{-i \beta_l H_M} e^{-i \gamma_l H_C} |+\rangle^{\otimes n}, \quad H_M = \sum_i X_i
    \end{equation}
    \item Use a classical optimizer (e.g., COBYLA, Nelder-Mead) to iteratively update \( \boldsymbol{\gamma}, \boldsymbol{\beta} \) to minimize:
    \begin{equation}
    \langle \psi(\gamma, \beta) | H_C | \psi(\gamma, \beta) \rangle
    \end{equation}
    \item Measure final quantum state to obtain optimal or near-optimal portfolio selection vector \( z^* \).
\end{enumerate}

\subsubsection{Evaluation Strategy}
\label{sec:m1-evaluation-strategy}

Though the solution is stochastic, we evaluate performance through:

\begin{itemize}
    \item \textbf{Risk-Return Tradeoff:} Compare portfolio volatility and expected return with classical benchmarks (e.g., Gurobi MILP).
    \item \textbf{Constraint Satisfaction:} Ensure all hard constraints (cardinality, ESG, sector) are respected.
    \item \textbf{Sampling Efficiency:} Track convergence in QAOA iterations vs. classical simulated annealing.
    \item \textbf{Computational Efficiency:} Estimate quantum circuit depth, gate fidelity, and time-to-solution.
\end{itemize}

\paragraph{Synthesis.}
This stylised formulation illustrates the full pipeline of quantum portfolio optimisation---from financial data preprocessing, to QUBO mapping, quantum circuit design, and evaluation. The next subsection complements this design template with a reproducible benchmark on local U.S. equity data, allowing the paper to move from abstract formulation to an explicit classical-versus-QAOA comparison.

\subsection{Algorithm Selection and Design Trade-offs}
\label{sec:m1-final-remarks-selecting-the-right-quantum-algorithm}

The appropriate choice of quantum algorithm in financial optimisation depends critically on the problem structure, hardware availability, and required solution characteristics. Different quantum approaches offer distinct advantages and trade-offs, which practitioners must evaluate in context.

\begin{itemize}
    \item \textbf{Use QAOA or VQE} when \textit{interpretability, tunability, or variational precision} is desired. These gate-based, interference-driven algorithms allow explicit control over quantum circuit parameters and are well-suited for noisy intermediate-scale quantum (NISQ) devices. In particular:
    \begin{itemize}
        \item QAOA excels in binary optimisation with tunable depth and direct mapping from QUBO to cost Hamiltonian.
        \item VQE is particularly useful when minimizing continuous-valued risk measures (e.g., portfolio variance or CVaR) through eigensolving.
    \end{itemize}

    \item \textbf{Use Quantum Annealing (QA)} when the goal is to solve \textit{large-scale QUBO problems efficiently}, especially in a rugged energy landscape with many local minima. While annealing hardware (e.g., D-Wave) is subject to noise and limited connectivity, it supports thousands of qubits and is capable of solving complex problems with short runtime:
    \begin{itemize}
        \item Ideal for rapid sampling from a broad solution space.
        \item Particularly effective when hard constraints are encoded directly into the Ising Hamiltonian.
    \end{itemize}

    \item \textbf{Use Grover Adaptive Search (GAS)} when the problem involves \textit{oracle-defined objectives or threshold conditions}. GAS generalizes Grover’s search to amplify the probability of satisfying complex portfolio constraints, such as:
    \begin{itemize}
        \item ``Find a portfolio with Sharpe ratio exceeding 1.2''
        \item ``Locate asset combinations satisfying VaR < 5\% and ESG > 0.8''
    \end{itemize}
    The quadratic speedup \( O(\sqrt{N}) \) over classical search methods can yield dramatic performance improvements in constrained settings.

    \item \textbf{Explore Hybrid Approaches.} Many practical use cases in finance will benefit from \textit{hybrid quantum-classical architectures}, in which:
    \begin{itemize}
        \item Quantum circuits perform pre-processing or generate high-quality candidate solutions.
        \item Classical optimizers refine, verify, or simulate risk profiles of quantum-derived portfolios.
    \end{itemize}
    Such architectures offer a pragmatic path forward, leveraging quantum capabilities without requiring fault-tolerant qubits.

\end{itemize}

The quantum computing landscape is diverse, and no single algorithm dominates across all financial contexts. Careful alignment of algorithmic strengths with financial problem features is key to unlocking quantum advantage. As quantum hardware matures and hybrid toolchains evolve, financial institutions will increasingly be able to integrate quantum decision engines into their portfolio selection, risk management, and trading workflows.

\subsection{Synthesis and Outlook}
\label{sec:m1-current-challenges-and-practical-limitations}

Despite the theoretical advantages of quantum optimisation in portfolio selection and other financial applications, several practical challenges remain. Current research and industry demonstrations highlight both the promise and the substantial technical barriers that must be overcome before quantum approaches achieve clear superiority over classical methods.

\paragraph{(1) Problem Scale and Dimensionality.}
Most published quantum portfolio optimisation experiments have been conducted on relatively small problem instances. For example, existing case studies demonstrate:
\begin{itemize}
    \item QAOA implementations on portfolios of 4–8 assets (IBM, 2020);
    \item Quantum annealing experiments on 40–60 asset universes (Rosenberg et al., 2016; Venturelli and Kondratyev, 2019).
\end{itemize}
However, realistic institutional portfolios involve hundreds to thousands of assets, multiple levels of constraints (sectoral, regulatory, ESG), and dynamic rebalancing requirements. Scaling quantum algorithms from tens to thousands of decision variables remains an open challenge.

\paragraph{(2) Noise and Hardware Instability.}
Current devices operate in the \textbf{Noisy Intermediate-Scale Quantum (NISQ)} era, where qubit decoherence, gate infidelity, and readout noise significantly distort the quantum state evolution. These errors accumulate quickly in deep circuits, degrading the performance of algorithms such as QAOA and QAE that rely on coherent interference patterns. As a result, measured outcomes often diverge from theoretical optima.

\paragraph{(3) Resource Overheads and Quantum Error Correction.}
Meaningful financial optimisation at industrial scale would require fully fault-tolerant quantum hardware. Yet, quantum error correction has not been practically achieved. Estimates suggest that a single logical qubit may require thousands of physical qubits. Consequently, executing nontrivial optimisation tasks (e.g., 1000-asset QUBO) would require millions of physical qubits—well beyond the capacity of current or near-term quantum processors.

\paragraph{(4) Comparative Performance with Classical Heuristics.}
To date, quantum optimisation methods have not demonstrated decisive empirical advantage over advanced classical heuristics such as genetic algorithms, tabu search, simulated annealing, or hybrid evolutionary solvers. Classical algorithms remain highly competitive in both runtime and solution quality, especially when combined with parallel computing architectures.

\paragraph{Synthesis.}
At present, quantum optimisation in finance should be viewed as an \textbf{exploratory and prototyping discipline}. Existing results validate feasibility but not superiority. Nonetheless, its long-term potential remains immense. As quantum hardware matures, noise mitigation improves, and hybrid quantum-classical workflows evolve, quantum optimisation may eventually unlock new frontiers in large-scale decision-making, risk analysis, and financial engineering.

\subsection{Case Study: Classical Search versus QAOA on U.S. Equity Data}
\label{sec:module1-local-case}

To anchor the optimisation module in a fully reproducible example, this section replaces the earlier third-party Q4FuturePOP illustration with a local benchmark built directly from the Nasdaq 100 daily price files already stored in the project workspace. The aim is deliberately modest: rather than claiming near-term quantum advantage, the case demonstrates how a financially meaningful allocation problem can be formulated in a classical optimisation language, translated into a QUBO, and solved consistently with QAOA under a transparent data pipeline \citep{Barkoutsos2020Finance,Egger2021,Acharya2024Decomposition}.

\subsubsection{Dataset and Universe Construction}
\label{sec:m1case-dataset-and-universe-construction}

The benchmark uses six U.S. equities with locally available daily prices from January 2, 2024 to December 31, 2025, sourced from the folder \texttt{paper/data\_nasdaq100\_2024\_2025/}. The selected tickers are:
\begin{center}
\texttt{AAPL, AMD, AMGN, COST, CSX, GOOGL.}
\end{center}

This universe is intentionally small enough to remain interpretable in a gate-based simulation while still capturing meaningful variation in expected return and covariance structure. Daily returns are converted into annualised sample moments:
\begin{equation}
\mu_i = 252 \cdot \bar{r}_i,
\qquad
\Sigma = 252 \cdot \widehat{\mathrm{Cov}}(r).
\end{equation}
The optimisation task is then defined as a binary selection problem with equal weighting across the chosen assets. In other words, the decision variable \(x_i \in \{0,1\}\) indicates whether asset \(i\) is included in the final portfolio.

\subsubsection{Mean-Variance QUBO Formulation}
\label{sec:m1case-mean-variance-qubo-formulation}

We impose an exact cardinality constraint of \(K=3\) assets and define the penalised binary objective
\begin{equation}
\min_{x \in \{0,1\}^n}
\lambda x^\top \Sigma x - \mu^\top x
    + A\left(\sum_{i=1}^{n} x_i - K\right)^2,
\end{equation}
where \(n=6\), \(\lambda=8\) is the risk-aversion coefficient, and \(A=15\) is the penalty coefficient enforcing the cardinality condition.

This design yields a clear comparison between two solution paradigms:
\begin{itemize}
    \item \textbf{Classical exact search:} all \(\binom{6}{3}=20\) feasible portfolios are enumerated and ranked by the original mean-variance objective.
    \item \textbf{QAOA:} the constrained problem is converted into a QUBO and solved with a two-layer QAOA circuit using a statevector sampler and COBYLA for the outer classical optimisation loop.
\end{itemize}

The resulting workflow is a scaled but genuine portfolio-selection exercise: expected return and covariance enter exactly as in a conventional mean-variance allocation pipeline, while the binary structure creates a natural bridge to QUBO encoding and gate-based quantum optimisation.

What makes QAOA valuable in this setting is not merely that it returns a competitive portfolio. It also offers a structured hybrid workflow in which the financial objective, constraint architecture, and quantum circuit design remain explicitly linked. That transparency is one of the main reasons QAOA has become a focal gate-based optimisation method in finance: it supports interpretable problem encoding, tunable circuit depth, and direct benchmarking against classical optimisation pipelines.

\subsubsection{Classical Benchmark versus QAOA}
\label{sec:m1case-classical-benchmark-versus-qaoa}

The refreshed two-year experiment no longer yields exact agreement between the classical and variational pipelines. The classical exact benchmark selects:
\begin{center}
\texttt{AMGN / COST / GOOGL.}
\end{center}
By contrast, the two-layer QAOA statevector workflow returns the nearby portfolio
\texttt{COST / CSX / GOOGL}, which ranks second in the feasible set by the penalised mean--variance objective and preserves a very similar volatility level.

Table~\ref{tab:module1-local-case} summarises the aligned outputs.

\begin{table}[H]
\centering
\caption{Classical exact search and QAOA on the local six-asset benchmark.}
\label{tab:module1-local-case}
\begin{tabular}{p{0.32\linewidth}p{0.28\linewidth}p{0.13\linewidth}p{0.13\linewidth}p{0.10\linewidth}}
\toprule
Method & Selected portfolio & Objective & Return & Volatility \\
\midrule
Classical exact search & \texttt{AMGN / COST / GOOGL} & 1.261 & 24.77\% & 16.69\% \\
QAOA (statevector simulation) & \texttt{COST / CSX / GOOGL} & 1.321 & 23.07\% & 16.72\% \\
\bottomrule
\end{tabular}
\end{table}

The interpretation is important. On a six-asset universe, the gap between the exact optimum and the shallow QAOA solution is not evidence against the quantum workflow; it is a transparent illustration of what a low-depth variational heuristic does on a finite benchmark. The modelling, QUBO conversion, and evaluation pipeline remain internally consistent, but the optimisation quality now visibly depends on circuit depth and variational search quality. At the same time, the case also makes the practical strengths of QAOA explicit: it keeps the financial objective in Hamiltonian form, supports tunable hybrid refinement, and produces a near-optimal portfolio without reverting to ad hoc classical constraint repair. For a review paper, this is arguably the more informative outcome: the example is reproducible, financially interpretable, and explicit about both the promise and the present limitations of gate-based portfolio optimisation.

\subsubsection{Figure-Based Interpretation}
\label{sec:m1case-figure-based-interpretation}

\begin{figure}[H]
\centering
\includegraphics[width=\textwidth]{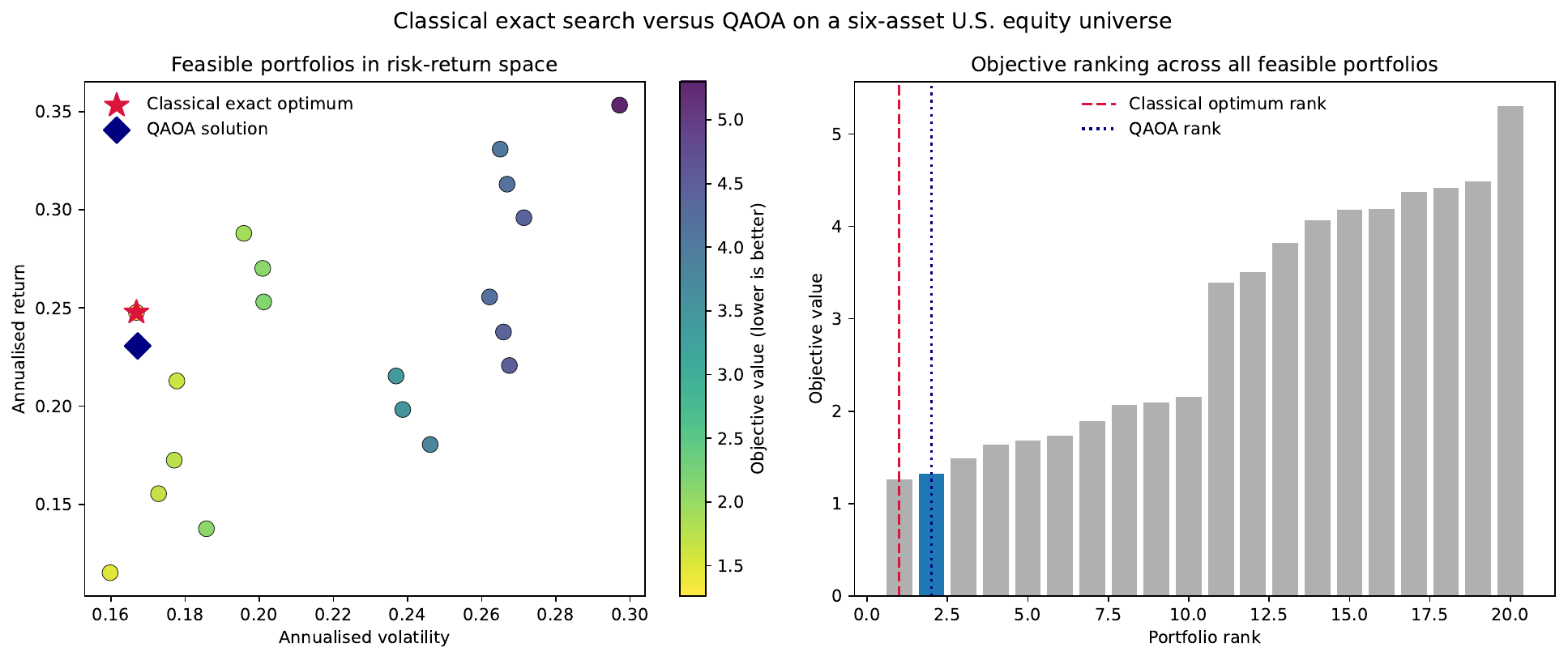}
\caption{Local U.S.-equity benchmark for Module 1. The left panel places every feasible three-asset portfolio in risk-return space and highlights the portfolios selected by classical exact search and QAOA. The right panel ranks all feasible portfolios by the penalised mean-variance objective. Under the full two-year sample, the QAOA solution is near-optimal but does not coincide with the classical optimum.}
\label{fig:module1-local-case}
\end{figure}

Figure~\ref{fig:module1-local-case} provides a visual summary of the local benchmark.

The left panel shows that both the classical optimum and the QAOA portfolio sit in the attractive portion of the feasible risk-return cloud rather than at arbitrary corners produced by penalty tuning. The right panel makes the approximation gap explicit: the QAOA portfolio is still highly ranked, but it no longer occupies the first position. Because all feasible portfolios are observable in this benchmark, the figure also makes the optimisation trade-off itself transparent: lower-volatility combinations can be suboptimal if they sacrifice too much expected return, while higher-return combinations can become unattractive once covariance penalties are imposed.

\subsubsection{Why This Case Matters for the Review}
\label{sec:m1case-why-this-case-matters-for-the-review}

This revised case improves the paper in four ways.
\begin{itemize}
    \item It removes dependence on an external demonstration framework and replaces it with a local, project-owned example.
    \item It ties the optimisation discussion to concrete U.S. equity data rather than a purely conceptual workflow.
    \item It provides a clean bridge between classical portfolio engineering and gate-based QAOA modelling, while making the advantages of transparent Hamiltonian encoding and tunable hybrid optimisation concrete.
    \item It creates a reproducibility trail that can be extended later to deeper QAOA circuits, larger universes, heuristic classical baselines, or hardware-backed experiments.
\end{itemize}

At the same time, the limits of the benchmark should remain explicit. The current exercise is small enough to permit exact classical enumeration, so it should be read as a calibration case rather than a scalability claim. The value lies in the methodological bridge: a financial optimisation problem is stated, encoded, solved, benchmarked, and documented in a way that is easy to audit and easy to extend.

\subsubsection{Computational Reproducibility}
\label{sec:m1case-computational-reproducibility}

The executable asset for this section is standardised in \Cref{sec:app-module-1-local-optimisation-benchmark}. The main notebook is:
\begin{center}
\repofile{module_01_optimisation/module_01_classical_vs_qaoa_portfolio.ipynb}{module_01_classical_vs_qaoa_portfolio.ipynb}
\end{center}
and the generated summary files are stored as:
\begin{center}
\repofile{module_01_optimisation/module_01_qaoa_portfolio_summary.csv}{module_01_qaoa_portfolio_summary.csv}
\qquad
\repofile{module_01_optimisation/module_01_qaoa_frontier.csv}{module_01_qaoa_frontier.csv}.
\end{center}

This arrangement keeps the main discussion focused on financial interpretation and optimisation logic, while ensuring that every empirical claim in the case can be traced back to a concrete notebook, a concrete dataset, and the public notebook repository \CodeRepo.

\newpage

\section{Quantum Amplitude Estimation for Derivatives Pricing}
\label{sec:pricing}

\paragraph{Section Overview.}
Monte Carlo methods are fundamental to modern quantitative finance, serving as the backbone of pricing and risk assessment for complex and path-dependent derivatives. However, their computational cost increases rapidly with problem dimensionality and precision requirements. This section explores how \textit{\phantomsection\label{term:qae}Quantum Amplitude Estimation (QAE)} can accelerate Monte Carlo-based pricing by providing a quadratic improvement in sampling efficiency, thus enabling faster and more accurate valuation of exotic options, structured products, and credit adjustment measures such as \phantomsection\label{term:cva-dva-fva}Credit Valuation Adjustment (CVA) and Debit Valuation Adjustment (DVA).

The classical benchmark throughout this section remains explicit: standard Monte Carlo and its variance-reduction extensions are the operational baseline, while the quantum question is whether amplitude-estimation routines can change the cost of expectation evaluation without breaking the modelling discipline required in production pricing engines.

\subsection{Problem Context and Classical Baseline in Derivative Pricing}
\label{sec:m2-problem-context-monte-carlo-limitations-in-derivative-pricin}

Monte Carlo simulation is among the most versatile and widely used tools for pricing financial derivatives whose payoffs depend on stochastic paths or high-dimensional integrals. In a typical setting, the payoff of a derivative \( f(S_T) \) under a stochastic process \( S_t \) (such as a Geometric Brownian Motion) is estimated as:
\begin{equation}
V_0 = e^{-rT} \mathbb{E}[f(S_T)] \approx e^{-rT} \frac{1}{N} \sum_{i=1}^{N} f(S_T^{(i)}),
\end{equation}
where \( N \) denotes the number of simulated paths, and \( e^{-rT} \) is the risk-free discount factor. While conceptually simple and model-agnostic, the convergence rate of the Monte Carlo estimator follows the well-known asymptotic behaviour:
\begin{equation}
\epsilon \propto \frac{1}{\sqrt{N}},
\end{equation}
where \( \epsilon \) is the standard error of the estimate. Consequently, reducing the estimation error by one order of magnitude requires a hundredfold increase in sample size, leading to exponential growth in runtime for high-precision requirements.

\paragraph{Challenges in Classical Monte Carlo.}
The practical limitations of Monte Carlo methods manifest in several forms:
\begin{enumerate}
    \item \textbf{Slow convergence for complex products:} Path-dependent instruments such as Asian options, barrier options, and callable exotics require long simulated trajectories with fine time discretisation, drastically increasing computational cost.
    \item \textbf{Dimensionality explosion:} Multi-asset and basket derivatives involve joint simulation of correlated processes, where the covariance matrix of underlying assets grows quadratically with portfolio size.
    \item \textbf{Variance reduction saturation:} Classical techniques such as control variates, importance sampling, or antithetic variates can only marginally improve convergence—they cannot overcome the intrinsic \(O(1/\sqrt{N})\) statistical limit.
    \item \textbf{CVA/DVA and risk adjustments:} CVA and DVA computations require nested Monte Carlo simulations—one to simulate exposure paths and another to simulate counterparty default probabilities—resulting in computational complexities of \(O(N^2)\) or higher.
    \item \textbf{High performance computing bottlenecks:} Even with parallelisation across GPUs or distributed clusters, scaling remains bounded by memory bandwidth and communication latency, making ultra-high-precision pricing (e.g., for long-term structured notes) impractical.
\end{enumerate}

\paragraph{Motivation for Quantum Enhancement.}
Quantum algorithms, particularly those based on amplitude estimation, address these limitations by exploiting the probabilistic nature of quantum measurement. Through quantum parallelism and amplitude interference, Quantum Amplitude Estimation (QAE) achieves an improved convergence rate of \( O(1/N) \), implying that the same pricing precision can be obtained with exponentially fewer simulated paths.

This potential makes QAE a powerful candidate for accelerating Monte Carlo-based derivative pricing, especially for:
\begin{itemize}
    \item Exotic and path-dependent derivatives (Asian, barrier, and basket options),
    \item Credit-related adjustments (CVA, DVA, Funding Value Adjustment/FVA),
    \item ESG-linked structured products requiring high-dimensional simulation,
    \item Any context where computational sampling dominates pricing cost.
\end{itemize}

Hence, by replacing or augmenting classical sampling with quantum amplitude estimation, one can achieve comparable pricing accuracy with drastically fewer evaluations of the payoff function \( f(S_T) \). The following sections formalise the mathematical framework of QAE and demonstrate its practical implications in derivative pricing workflows.

\subsection{Quantum Amplitude Estimation: Principles and Mathematical Foundation}
\label{sec:m2-quantum-amplitude-estimation-principles-and-mathematical-fou}

\paragraph{Aim.}
QAE provides a quadratic improvement over classical Monte Carlo in estimating expectations of random variables whose values are encoded as quantum amplitudes. In derivative pricing, the discounted option value
\(
V_0 = e^{-rT}\,\mathbb{E}[f(S_T)]
\)
is an expectation under a risk–neutral measure; \textbf{QAE targets the inner expectation with query complexity \(O(1/\varepsilon)\) to achieve additive error \(\varepsilon\), versus \(O(1/\varepsilon^2)\) samples classically}.

\subsubsection{Problem Setup and Normalisation}
\label{sec:m2-problem-setup-and-normalisation}

Let \(X\) denote the (possibly high-dimensional) outcome of a market simulation at maturity \(T\), and let \(f(X)\in[0,1]\) be a normalised payoff function.\footnote{If the payoff is nonnegative but unbounded (e.g., a call), we employ a scaling \(f \mapsto f/B\) with known bound \(B>0\) and multiply the estimate by \(B\) ex post, or clip / stratify if needed.} Define
\begin{equation}
a \;=\; \mathbb{E}[f(X)] \;\in\; [0,1],
\end{equation}
where \(a\) represents the risk–neutral expected (normalised) payoff.  
Quantum Amplitude Estimation (QAE) aims to estimate \(a\) to additive error \(\varepsilon\) with high confidence, providing a quadratic speedup over classical Monte Carlo convergence.

\paragraph{State preparation oracle.}
We assume access to a unitary operator \(A\) that prepares a superposition encoding both market scenarios and corresponding payoffs:
\begin{equation}
A\,\ket{0} \;=\; \sqrt{1-a}\,\ket{\Psi_0}\ket{0} \;+\; \sqrt{a}\,\ket{\Psi_1}\ket{1},
\end{equation}
where measuring the payoff ancilla yields \(\ket{1}\) with probability \(a\).  
Constructing \(A\) — known as the \emph{state preparation} or \emph{loading} step — consists conceptually of two layers:

\begin{enumerate}
  \item \textbf{Distribution loading:} prepare the register \(\sum_x \sqrt{p(x)}\ket{x}\), where \(p(x)\) approximates the risk–neutral distribution of \(X\) (e.g., under GBM, Heston, or basket copulas).  
  \item \textbf{Payoff encoding:} append an ancilla and apply a controlled rotation such that \(\ket{x}\ket{0}\mapsto\ket{x}\big(\sqrt{1-f(x)}\ket{0}+\sqrt{f(x)}\ket{1}\big)\).
\end{enumerate}

These two components together define the oracle \(A\).  
When combined with amplitude amplification and measurement (discussed in later sections), they form the full three-layer circuit structure: \emph{loading → encoding → estimation}.  
The detailed implementation and decomposition of these layers are elaborated in \Cref{sec:m2-quantum-state-preparation}.

\subsubsection{Good/Bad Subspaces and the Angle Parameter}
\label{sec:m2-good-bad-subspaces-and-the-angle-parameter}

Let \(\mathcal{H}_\mathrm{good}\) be the subspace where the payoff qubit is \(\ket{1}\) and \(\mathcal{H}_\mathrm{bad}\) where it is \(\ket{0}\). Define the projector \( \Pi=\mathbb{I}\otimes\ket{1}\!\bra{1}\). Then
\begin{equation}
a \;=\; \bra{0}A^\dagger \Pi A\ket{0} \;=\; \sin^2(\theta),\qquad \theta\in[0,\pi/2].
\end{equation}
We also define the reflections
\begin{equation}
S_0 \;=\; \mathbb{I}-2\ket{0}\!\bra{0}, \qquad S_\chi \;=\; \mathbb{I}-2\Pi,
\end{equation}
and the Grover-style iterate
\begin{equation}
Q \;=\; -A S_0 A^\dagger S_\chi .
\end{equation}
In the two-dimensional invariant subspace spanned by the normalised projections of \(A\ket{0}\) onto \(\mathcal{H}_\mathrm{good}\) and \(\mathcal{H}_\mathrm{bad}\), the operator \(Q\) acts as a rotation by angle \(2\theta\). Its eigenvalues are \(e^{\pm 2i\theta}\) with eigenvectors \(\ket{\psi_\pm}\).

\subsubsection{Canonical (QPE-based) Amplitude Estimation}
\label{sec:m2-canonical-qpe-based-amplitude-estimation}

The original QAE (Brassard–Høyer–Mosca–Tapp) estimates \(\theta\) via Quantum Phase Estimation (QPE) on \(Q\). Using \(m\) control qubits, apply controlled-\(Q^{2^k}\) for \(k=0,\ldots,m-1\), followed by the inverse Quantum Fourier Transform, and measure a bit string \(y\) that approximates \(\phi=\theta/\pi\) to \(m\) bits. The estimate
\(
\tilde{a}=\sin^2(\tilde{\theta}),\; \tilde{\theta}=\pi\, y/2^m
\)
achieves
\begin{equation}
\Pr\!\left[\,|\tilde{a}-a|\le O\!\left(\tfrac{1}{M}\right)\right]\;\ge\; 1-\delta, 
\qquad M=2^m,
\end{equation}
using \(O(M)\) controlled applications of \(Q\) (and thus \(A,A^\dagger\)). Consequently, to reach additive error \(\varepsilon\), the query complexity is \(O(1/\varepsilon)\), a quadratic improvement over classical Monte Carlo’s \(O(1/\varepsilon^2)\).

\paragraph{Query model and cost.}
Each invocation of \(Q\) uses one call to \(S_\chi\) (cheap, as it is a single-qubit phase conditioned on the payoff ancilla) and two calls to \(A\) or \(A^\dagger\) (dominant cost). Overall gate count scales with the cost of preparing \(\sum_x \sqrt{p(x)}\ket{x}\) and computing \(f(x)\).

\subsubsection{Iterative and Maximum-Likelihood Variants (NISQ-Oriented)}
\label{sec:m2-iterative-and-maximum-likelihood-variants-nisq-oriented}

QPE requires long coherent circuits and fault tolerance. Several variants trade some quadratic gain for shorter depth or statistical robustness:

\begin{description}
  \item[\phantomsection\label{term:iqae}Iterative QAE (IQAE).] Uses repeated applications of \(Q^{k}\) for adaptively chosen \(k\) and classical post-processing (binary search / confidence intervals) to bracket \(\theta\). Depth is reduced (no QFT), with query complexity remaining \(O(1/\varepsilon)\) up to logarithmic factors.
  \item[\phantomsection\label{term:mlqae}Maximum-Likelihood QAE (MLQAE).] Collects samples from circuits with different powers \(k\) and estimates \(a\) via maximum-likelihood fitting to the \(\sin^2((2k+1)\theta)\) model. It is robust to noise and flexible in experimental design.
  \item[Low-Depth QAE (QAE-Lite / QAE-SR).] —also referred to in the literature as “Short-Depth QAE”, offer practical near-term implementations by replacing phase estimation with classical post-processing and adaptive sampling. Employs constant-depth circuits with amplitude amplification of small fixed orders and classical Richardson-style extrapolation or sample reweighting to correct bias. Complexity interpolates between \(O(1/\varepsilon)\) and \(O(1/\varepsilon^2)\), often well suited for \phantomsection\label{term:nisq}NISQ (Noisy intermediate-scale quantum).
\end{description}

These approaches avoid the inverse QFT and allow batching (parallel evaluation over different \(k\)) to exploit classical–quantum hybrid pipelines.

\subsubsection{Comparison of QAE Variants}
\label{sec:m2-comparison-of-qae-variants}

Quantum Amplitude Estimation has evolved from its original, fault-tolerant form into a family of algorithms adapted to near-term hardware. Each variant balances between theoretical optimality, circuit depth, and practical implementability. Table~\ref{tab:QAEcomparison} summarises the main characteristics of the five most representative implementations.

\begin{table}[H]
\centering
\caption{Comparison of major Quantum Amplitude Estimation (QAE) variants.}
\vspace{0.5em}
\footnotesize{
\begin{tabular}{>{\raggedright\arraybackslash}p{2.5cm}>{\raggedright\arraybackslash}p{3cm}>{\raggedright\arraybackslash}p{2.5cm}>{\raggedright\arraybackslash}p{2.5cm}>{\raggedright\arraybackslash}p{2.6cm}}
\toprule
\textbf{Algorithm} & \textbf{Key Idea / Approach} & \textbf{Quantum Resources} & \textbf{Complexity and Depth} & \textbf{Practical Suitability} \\
\midrule
\textbf{Standard QAE} (Brassard et al., 2000) & Uses Quantum Phase Estimation (QPE) on the Grover operator to infer amplitude via phase $\theta$. & Requires long coherent circuits, controlled-$Q^{2^k}$ operations, inverse QFT. & Optimal $O(1/\varepsilon)$ query complexity; high circuit depth $\sim O(\log(1/\varepsilon))$. & Requires fault-tolerant quantum hardware. \\
\textbf{Iterative QAE (IQAE)} & Replaces QFT with adaptive, sequential Grover iterations and classical feedback. & Single ancilla control qubit reused adaptively; no large Fourier register. & Maintains $O(1/\varepsilon)$ scaling with logarithmic overhead; reduced depth. & NISQ-friendly; resilient to moderate noise. \\
\textbf{Maximum - Likelihood QAE (MLQAE)} & Fits measurement outcomes from several circuit powers to a $\sin^2((2k+1)\theta)$ model via MLE. & Multiple circuits at different depths; classical maximum - likelihood post-processing. & Empirical convergence close to $O(1/\varepsilon)$; flexible sampling design. & Suitable for hardware-efficient experiments; robust to noise. \\
\textbf{QAE-Lite} & Uses small, fixed Grover depths ($k=1,2,\ldots$) and classical extrapolation or reweighting to approximate QAE behaviour. & Constant-depth circuits with few Grover iterates. & Sub-quadratic speedup; complexity between $O(1/\varepsilon)$ and $O(1/\varepsilon^2)$. & Very NISQ-compatible; trades accuracy for stability. \\
\textbf{QAE-SR (Short-Run QAE)} & Employs repeated shallow-depth circuits, collects statistical moments, and performs classical regression reconstruction. & Repeated independent shallow runs; classical aggregation. & Similar asymptotic behaviour as QAE-Lite but with lower bias. & Ideal for experimental benchmarking on noisy devices. \\
\bottomrule
\end{tabular}
}
\label{tab:QAEcomparison}
\end{table}

\paragraph{Discussion.}
While the canonical QAE achieves the theoretical quadratic speedup, it is impractical on current noisy devices due to the depth of Quantum Phase Estimation. Iterative and Maximum-Likelihood versions preserve most of the speedup while remaining experimentally feasible. Short-depth approximations such as QAE-Lite or QAE-SR provide near-term demonstrations and proof-of-concept pathways for hybrid quantum-classical Monte Carlo acceleration.

\subsubsection{Mapping Derivative Pricing to QAE}
\label{sec:m2-mapping-derivative-pricing-to-qae}

For a European-style claim with discounted payoff \(g(X)=e^{-rT} f(X)\), normalise \(f\) so \(f/B \in [0,1]\) for known \(B\). Prepare
\begin{equation}
A\ket{0} \;=\; \sum_x \sqrt{p(x)}\,\ket{x}\!\left(\sqrt{1-\frac{f(x)}{B}}\,\ket{0}+\sqrt{\frac{f(x)}{B}}\,\ket{1}\right).
\end{equation}
Then the target amplitude is
\(
a=\mathbb{E}[f(X)]/B,
\)
and the price estimate is
\begin{equation}
\tilde{V}_0 = e^{-rT}\,B\,\tilde{a}.
\end{equation}

Here, \(a\) denotes the true (normalized) amplitude corresponding to the probability of observing the payoff ancilla in state \(\ket{1}\), while \(\tilde{a}\) is its estimate obtained from a quantum amplitude estimation routine. The estimated derivative price \(\tilde{V}_0 = e^{-rT} B \tilde{a}\) converges to the true value \(V_0 = e^{-rT} B a\) as the number of QAE samples increases.

\paragraph{Path dependence.}
For Asian or barrier options, \(x\) encodes the full path (or sufficient statistics such as running average \(\bar{S}\), min/max, barrier hits). Distribution loading may use sequential circuits that simulate increments and update path features; payoff encoding acts on those features.

\paragraph{Baskets and correlation.}
For basket options, load a multivariate distribution \(p(x)\) with Cholesky or Principal Component Analysis (PCA) factor models on discretised normals to impose the desired covariance; compute \(f(x)\) on the register storing the basket aggregation.

\paragraph{CVA/DVA.}
CVA involves \(\mathbb{E}[\mathrm{LGD}\cdot \mathbf{1}_{\tau<T}\cdot \max(\mathrm{Exposure}_\tau,0)]\). One may avoid fully nested Monte Carlo by (i) encoding marginal default intensities and exposure proxies into the joint state, or (ii) stratifying exposure time buckets and running QAE per bucket with appropriate weights.

\subsubsection{Accuracy, Confidence and Complexity Guarantees}
\label{sec:m2-accuracy-confidence-and-complexity-guarantees}

Let \(\tilde{a}\) be the QAE estimate produced using \(M\) effective Grover iterations (total controlled powers of \(Q\)). For canonical QAE,
\begin{equation}
\Pr\!\left[|\tilde{a}-a|\le \frac{c}{M}\right] \ge 1-\delta,
\end{equation}
for a universal constant \(c>0\) and confidence parameter \(\delta\) (amplified by repetition and median-of-means). In terms of derivative prices,
\begin{equation}
|\tilde{V}_0 - V_0| \;\le\; e^{-rT}\,B\,\frac{c}{M}.
\end{equation}
Thus, to achieve additive price error at most \(\varepsilon\), it suffices to choose \(M = O\!\big(e^{-rT}B/\varepsilon\big)\).

\paragraph{Bias and variance.}
Canonical QAE is (asymptotically) unbiased; finite-sample bias arises from phase wrapping and discretisation. IQAE/MLQAE can be made nearly unbiased with appropriate design points and classical inference. Confidence intervals follow from tail bounds on binomial outcomes of measurement and the Lipschitz mapping \(a\mapsto \sin^2(\theta)\).

\subsubsection{Resource and Implementation Considerations}
\label{sec:m2-resource-and-implementation-considerations}

\paragraph{State preparation dominates.}
The practical speedup depends critically on the cost of loading \(p(x)\) and evaluating \(f(x)\). Techniques include: (i) analytic state synthesis for log-normal marginals, (ii) \phantomsection\label{term:qrom-qroam}QROM/QROAM for table-based loading, (iii) arithmetic circuits for payoffs and comparators, (iv) factor models for correlation.

QROM (Quantum Read-Only Memory) refers to a structured subcircuit that loads a classical vector 
of parameters---such as the payoff-dependent rotation angles $\{\theta_i\}$---into 
a quantum register conditioned on the index state $\ket{i}$. 
QROAM (Quantum Random-Access Optimised Memory) generalises this design by reducing the 
number of controlled gates through efficient address decoding.

\paragraph{Circuit depth vs. NISQ noise.}
QPE-based QAE requires long coherent evolutions. NISQ-friendly variants (IQAE, MLQAE, QAE-Lite) reduce depth at the cost of more shots or small constant-factor overhead. Error mitigation (zero-noise extrapolation, symmetry checks) can stabilise estimates.

\paragraph{Variance reduction hybrids.}
Classical variance reduction (control variates, antithetics) remains valuable. One may encode a control variate \(h(X)\) with known mean, estimate \(\mathbb{E}[f-h]\) via QAE, and re-add \(\mathbb{E}[h]\) classically to reduce effective variance and circuit width.

\paragraph{Synthesis.}
QAE turns expectation estimation into phase (angle) estimation on a Grover iterate whose eigenphase encodes the desired amplitude \(a\). With suitable state preparation and NISQ-aware variants, it offers a principled path to accelerate Monte Carlo pricing of complex derivatives, including path-dependent and multi-asset claims, as well as CVA.

\subsection{Mapping Derivative Pricing to Quantum Circuits}
\label{sec:m2-mapping-derivative-pricing-to-quantum-circuits}

\paragraph{Objective.}
The goal of this section is to establish a complete mapping from the stochastic pricing of a derivative under a risk–neutral measure to a quantum circuit that can be evaluated using QAE. The mapping converts the expectation of a payoff function into the probability amplitude of a designated quantum state, thereby enabling the use of quantum interference and measurement statistics to estimate derivative prices with quadratic convergence speedup.

\subsubsection{From Risk-Neutral Valuation to Expectation Formulation}
\label{sec:m2-from-risk-neutral-valuation-to-expectation-formulation}

Under the standard risk–neutral framework, the fair value of a derivative at time \(t=0\) is the discounted expected payoff under the risk–neutral probability measure \(\mathbb{Q}\):
\begin{equation}
V_0 = e^{-rT}\,\mathbb{E}_{\mathbb{Q}}[f(S_T)],
\end{equation}
where \(r\) is the continuously compounded risk-free rate, \(T\) is the time to maturity, and \(f(S_T)\) is the terminal payoff of the derivative depending on the asset (or basket) price \(S_T\).

In classical Monte Carlo, this expectation is approximated by
\begin{equation}
V_0 \approx e^{-rT}\frac{1}{N}\sum_{i=1}^{N} f(S_T^{(i)}),
\end{equation}
where \(\{S_T^{(i)}\}\) are independent samples generated from the underlying stochastic process (e.g., geometric Brownian motion). In the quantum case, instead of explicitly sampling paths, we encode the entire probability distribution over all possible outcomes \(\{S_T\}\) into the amplitudes of a quantum state.

\paragraph{Encoding the Expectation.}
Define the normalised payoff function \(f'(S_T) = f(S_T)/B\) where \(B\) is a known bound satisfying \(0 \le f(S_T) \le B\). The expectation can then be written as
\begin{equation}
a = \mathbb{E}[f'(S_T)] = \sum_{x} p(x) f'(x),
\end{equation}
with \(p(x)\) representing the discretised probability of scenario \(x\). QAE estimates \(a\) by preparing a quantum state whose amplitude squared corresponds to this expectation.

\subsubsection{Quantum State Preparation}
\label{sec:m2-quantum-state-preparation}

The circuit construction proceeds in three layers: distribution loading, payoff encoding, and amplitude measurement.

\paragraph{(i) Distribution Loading.}
We prepare a quantum register representing the discretised price distribution at maturity \(T\):
\begin{equation}
\ket{\psi_P} = \sum_{x} \sqrt{p(x)}\,\ket{x},
\end{equation}
where each computational basis state \(\ket{x}\) corresponds to a possible terminal price or scenario, and \(p(x)\) is the associated risk–neutral probability.  

\textit{Implementation:}  
- For a single asset following geometric Brownian motion, \(p(x)\) can be approximated using log-normal discretisation.  
- For multiple assets or correlated processes, we use Cholesky or PCA decomposition of the covariance matrix \(\Sigma\) and sample correlated Gaussian vectors \(Z = L \xi\), where \(L\) is the lower-triangular Cholesky factor.  
- QRAM/QROAM structures can efficiently load the vector of \(\sqrt{p(x)}\).

\paragraph{(ii) Payoff Encoding.}
A second register encodes the payoff function through a controlled rotation, adding an ancilla qubit to represent the value of \(f'(x)\):
\begin{equation}
\ket{\psi_f} = \sum_{x} \sqrt{p(x)}\,\ket{x}
\left( \sqrt{1 - f'(x)}\ket{0} + \sqrt{f'(x)}\ket{1} \right).
\end{equation}
Measuring the ancilla qubit in the \(\ket{1}\) state yields probability \(a = \mathbb{E}[f'(S_T)]\).  

This circuit is known as the \emph{state-preparation operator} \(A\):
\begin{equation}
A\ket{0} = \sqrt{1-a}\,\ket{\Psi_0}\ket{0} + \sqrt{a}\,\ket{\Psi_1}\ket{1}.
\end{equation}

\paragraph{(iii) Amplitude Estimation via Grover Iterates.}
The amplitude \(a\) is encoded in the probability of observing the ancilla qubit in \(\ket{1}\). Applying QAE on the operator \(A\) estimates \(a\) with precision \(O(1/M)\), where \(M\) is the number of Grover iterations. The derivative price is then recovered as:
\begin{equation}
V_0 = e^{-rT}B\,a.
\end{equation}

\subsubsection{Practical Examples of Mapping}
\label{sec:m2-practical-examples-of-mapping}

\paragraph{European Options.}
For a European call option with strike \(K\), payoff \(f(S_T) = \max(S_T - K, 0)\), the mapping proceeds as:
\begin{enumerate}
    \item Encode the log-normal distribution of \(S_T\) into the probability amplitudes of \(\ket{x}\).
    \item Implement a comparator circuit that checks whether \(S_T > K\) and computes \(S_T - K\).
    \item Perform a controlled rotation on the payoff qubit proportional to the scaled payoff value.
\end{enumerate}
After executing QAE, the amplitude \(a\) yields the discounted expected payoff.

\paragraph{Asian and Path-Dependent Options.}
For an Asian option with payoff \(f = \max(\bar{S} - K, 0)\), where \(\bar{S}\) is the average price over \(m\) time steps, we extend the register to encode partial path statistics:
\begin{equation}
\ket{\psi_{\text{path}}} = \sum_{x_1,\ldots,x_m} \sqrt{p(x_1,\ldots,x_m)}\,\ket{x_1,\ldots,x_m}.
\end{equation}
A running sum register accumulates \(\sum_{t=1}^{m} S_t\), allowing payoff evaluation at the final step. Such circuits can be constructed recursively using quantum arithmetic blocks.

\paragraph{Basket and Multi-Asset Options.}
For a basket option on \(n\) assets with weights \(w_i\), payoff \(f(S_T) = \max(\sum_i w_i S_{T,i} - K, 0)\), we load the joint correlated distribution:
\begin{equation}
\ket{\psi_B} = \sum_{x_1,\ldots,x_n} \sqrt{p(x_1,\ldots,x_n)}\,\ket{x_1,\ldots,x_n},
\end{equation}
where correlations are introduced by applying a unitary transformation corresponding to the covariance structure of returns. Payoff encoding then computes the weighted sum \(\sum_i w_i S_{T,i}\) via quantum addition circuits and performs threshold comparison against \(K\).

\paragraph{Credit and Debit Valuation Adjustments (CVA/DVA).}
For counterparty risk adjustments, the expected discounted loss (or gain) can be expressed as:
\begin{equation}
\mathrm{XVA} = (1 - R) \mathbb{E}\!\left[ e^{-r\tau} \max(\pm V_\tau, 0)\, \mathbf{1}_{\{\tau < T\}} \right],
\end{equation}
where the positive sign corresponds to CVA (counterparty default) and the negative sign to DVA (own default).
The state preparation jointly encodes the exposure distribution and the default-time density, allowing 
QAE to estimate such nested expectations in a single amplitude estimation layer.

\subsubsection{Circuit Design and Resource Profile}
\label{sec:m2-circuit-architecture-and-resource-overview}

\paragraph{Registers.}
A typical QAE-based pricing circuit consists of:
\begin{itemize}
    \item \(n_P\) qubits for the distribution (underlying scenarios),
    \item \(n_A\) qubits for payoff computation (arithmetic or comparisons),
    \item 1 ancilla qubit for amplitude encoding,
    \item optional \(n_C\) qubits for controlling Grover iterates.
\end{itemize}

Here, \(n_P\) denotes the number of qubits used to discretize the underlying distribution 
(e.g., terminal price or path average), typically ranging from 4 to 10. 
\(n_A\) represents ancillary workspace qubits for arithmetic or comparison operations in the payoff evaluation, 
and \(n_C\) denotes optional control qubits used in Grover or iterative amplitude-estimation schemes. 
The additional single ancilla qubit encodes the payoff amplitude to be measured.

\paragraph{Gate Components.}
\begin{itemize}
    \item Quantum arithmetic modules for additions, multiplications, and comparisons.
    \item Controlled rotation gates parameterised by payoff functions.
    \item Unitary transformations for covariance embedding or PCA rotations.
\end{itemize}

\paragraph{Complexity Notes.}
The total depth of the circuit depends primarily on:
\begin{equation}
O\left( n_P + n_A + d_\text{arith} + d_\text{Grover} \right),
\end{equation}
where \(d_\text{arith}\) is the depth of arithmetic logic (scales polynomially in \(n_A\)), and \(d_\text{Grover}\) is proportional to the number of amplitude amplification iterations.  
In hybrid approaches, \(d_\text{Grover}\) can be kept small (1–3 iterations), trading some precision for noise tolerance.

\subsubsection{Synthesis}
\label{sec:m2-synthesis}
\label{sec:m2-summary}

This mapping provides a systematic pipeline to represent the stochastic pricing of derivatives on a quantum computer.  
\begin{enumerate}
    \item Encode the risk–neutral price distribution in quantum amplitudes.  
    \item Embed the payoff function via controlled rotations.  
    \item Apply amplitude estimation to extract the expected value efficiently.  
\end{enumerate}
In this formulation, the expected payoff becomes a measurable amplitude, enabling the use of quantum interference for numerical integration.  
The framework generalises naturally to multi-asset, path-dependent, and credit-sensitive products, offering a unified blueprint for quantum-accelerated derivative pricing workflows.

\subsection{Algorithmic Implementations and Variants}
\label{sec:m2-algorithmic-implementations-and-variants}

\paragraph{Subsection Overview.}
While the theoretical framework of Quantum Amplitude Estimation (QAE) is well defined, its implementation on current quantum hardware depends heavily on algorithmic design choices. Different versions of QAE—ranging from the canonical, fault-tolerant formulation to lightweight NISQ-compatible variants—exhibit distinct trade-offs in accuracy, circuit depth, and noise resilience. This section systematically compares these implementations and their suitability for derivative pricing workloads.

\subsubsection{Canonical (Fault-Tolerant) QAE}
\label{sec:m2-canonical-fault-tolerant-qae}

\paragraph{Principle.}
The canonical QAE algorithm, originally proposed by Brassard, Høyer, Mosca and Tapp (2000), uses the Quantum Phase Estimation (QPE) routine to determine the amplitude parameter \(\theta\), defined through \(a = \sin^2(\theta)\). The phase is estimated by applying controlled powers of the Grover iterate \(Q\):
\begin{equation}
Q = -A S_0 A^\dagger S_\chi,
\end{equation}
where \(A\) is the state preparation operator, \(S_0\) and \(S_\chi\) are reflection operators. Each eigenstate of \(Q\) corresponds to an eigenvalue \(e^{\pm 2 i \theta}\), and phase estimation retrieves \(\theta\) with precision proportional to \(1/M\), where \(M\) is the number of Grover iterations.

\paragraph{Resource profile.}
For a target additive error \(\varepsilon\), the algorithm requires \(M = O(1/\varepsilon)\) applications of \(A\) and \(A^\dagger\). The overall circuit depth scales as
\begin{equation}
D_{\text{QAE}} = O\big(\log(1/\varepsilon)\,D_A\big),
\end{equation}
where \(D_A\) is the depth of the state-preparation subcircuit. Because QPE requires long coherent evolution and inverse Quantum Fourier Transform (QFT), the canonical QAE is suitable primarily for fault-tolerant quantum computers with error correction and long qubit lifetimes.

\paragraph{Precision scaling.}
The canonical method achieves the optimal theoretical convergence rate:
\begin{equation}
|\tilde{a} - a| = O\!\left(\frac{1}{M}\right),
\end{equation}
corresponding to a quadratic speedup over classical Monte Carlo’s \(O(1/\sqrt{M})\) rate.

\paragraph{Relevance to finance.}
In derivative pricing, canonical QAE would enable high-precision pricing of complex or long-dated products—such as callable notes, long-term CVA, or structured credit instruments—once fault-tolerant quantum hardware becomes available. However, its requirements currently exceed the coherence times and gate fidelities of available devices.

\subsubsection{Iterative QAE (IQAE)}
\label{sec:m2-iterative-qae-iqae}

\paragraph{Principle.}
Iterative Quantum Amplitude Estimation (Grinko et al., 2021) removes the need for the quantum Fourier transform by employing an adaptive iterative procedure. 
The algorithm repeatedly applies \(Q^k\) with different powers \(k\) and uses measurement statistics to refine confidence intervals for the amplitude parameter \(\theta\). 
This family of algorithms is rooted in the insight that explicit phase estimation is not required to obtain quadratic speedup—an idea later formalised and simplified by Aaronson and Rall (2021), who established a unified theoretical bridge between amplitude estimation and approximate quantum counting.

\paragraph{Algorithmic workflow.}
\begin{enumerate}
    \item Start with an initial estimate of \(\theta_0 = \pi/4\) (corresponding to \(a = 1/2\)).
    \item Apply \(Q^{k_i}\) for a sequence of \(k_i\) values chosen adaptively based on prior measurement outcomes.
    \item After each iteration, update the posterior distribution over \(\theta\) and narrow the confidence interval.
\end{enumerate}

\paragraph{Performance and resources.}
IQAE achieves nearly the same asymptotic scaling \(O(1/\varepsilon)\) as canonical QAE but with dramatically reduced circuit depth, since no inverse QFT is used. Each iteration only requires short coherent evolutions of order \(O(k_i)\), and the total number of oracle calls scales as \(O(\log(1/\varepsilon)/\varepsilon)\). Here, $\varepsilon$ denotes the target additive precision in estimating the amplitude 
$a = \mathbb{E}[f(X)]$, i.e., the algorithm guarantees $|\tilde{a} - a| \le \varepsilon$ 
with high confidence.

\paragraph{Suitability for derivative pricing.}
IQAE is highly attractive for near-term pricing problems, particularly those requiring moderate precision (e.g., option deltas or short-term risk metrics). It allows low-depth implementation and can tolerate moderate device noise without losing the quadratic speedup in the asymptotic regime.

\subsubsection{Maximum-Likelihood QAE (MLQAE)}
\label{sec:m2-maximum-likelihood-qae-mlqae}

\paragraph{Principle.}
Maximum-Likelihood QAE (Suzuki et al., 2020) reformulates amplitude estimation as a parameter inference problem. Instead of recovering the phase through QPE or iterative search, it collects measurement outcomes from circuits implementing \(Q^k\) for different \(k\) values and fits the parameter \(\theta\) using maximum-likelihood estimation. Here, $\theta \in [0, \pi/2]$ is the amplitude angle defined by $a = \sin^2(\theta)$,
which parameterizes the target expectation value in a trigonometric form. 
Under Grover iterations $Q^k$, the ancilla measurement probability follows 
$P_k(1) = \sin^2((2k+1)\theta)$, enabling the estimation of $\theta$ via 
maximum-likelihood inference from measurement outcomes.

\paragraph{Estimation model.}
For each circuit depth \(k\), the probability of measuring the ancilla in \(\ket{1}\) is:
\begin{equation}
P_k(1|\theta) = \sin^2\!\big((2k + 1)\theta\big).
\end{equation}
By collecting measurement counts \(\{n_k\}\) across different \(k\), the log-likelihood function is:
\begin{equation}
\mathcal{L}(\theta) = \sum_{k} n_k \ln P_k(1|\theta) + (N_k - n_k)\ln[1 - P_k(1|\theta)],
\end{equation}
and the optimal estimate \(\hat{\theta}\) maximises \(\mathcal{L}(\theta)\).

\paragraph{Performance.}
MLQAE attains near-optimal accuracy \(O(1/\varepsilon)\) with modest circuit depth, since each \(Q^k\) can be shallow and runs independently. This makes it highly parallelisable. Additionally, MLQAE is robust to device noise and statistical uncertainty, as the classical optimisation step absorbs sampling errors.

\paragraph{Use case.}
For pricing derivatives that require repeated expectation evaluations under slightly varying parameters (e.g., sensitivity analysis, Greeks computation), MLQAE provides a flexible, experiment-friendly approach that can be implemented with hybrid quantum–classical workflows.

\subsubsection{QAE-Lite and Short-Run (QAE-SR)}
\label{sec:m2-qae-lite-and-short-run-qae-sr}

\paragraph{Motivation.}
On NISQ devices, long Grover iterations are prone to decoherence and cumulative gate errors. The QAE-Lite (Nakaji, 2020) and Short-Run (Rattew et al., 2022) algorithms address this by fixing the number of Grover iterations to small constants and combining multiple shallow runs with classical post-processing.

\paragraph{QAE-Lite.}
QAE-Lite uses a few fixed iterations (typically \(k=1,2,3\)) and constructs an extrapolation curve that approximates the amplitude dependence:
\begin{equation}
P_k(1|\theta) = \sin^2((2k+1)\theta).
\end{equation}
By interpolating over observed probabilities for different \(k\), the amplitude \(a=\sin^2(\theta)\) can be inferred with sub-quadratic speedup. The method has constant quantum depth but increased classical computation.

\paragraph{QAE-SR.}
Short-Run QAE repeats many shallow circuits and aggregates statistical moments of the measurement distribution. Using regression or curve-fitting, it reconstructs an unbiased estimate of \(\theta\).  
This approach offers robustness under realistic noise and yields results close to IQAE for low-precision pricing tasks.

\paragraph{Practical trade-offs.}
\begin{itemize}
    \item Quantum circuit depth: \(O(1)\) — minimal and noise-tolerant.
    \item Number of circuit evaluations: \(O(1/\varepsilon^2)\), slightly worse than ideal QAE but still outperforming naive Monte Carlo for many practical tolerances.
    \item Hybrid advantage: Classical data aggregation is cheap and parallelisable.
\end{itemize}

\paragraph{Applications.}
QAE-Lite and QAE-SR are well suited for proof-of-concept pricing experiments, e.g. pricing a 3-asset basket option or a 5-step Asian option on superconducting qubit hardware, where maintaining coherence for more than a few Grover iterations is infeasible.

\subsubsection{Comparative Analysis of Circuit Depth and Hardware Requirements}
\label{sec:m2-comparative-analysis-of-circuit-depth-and-hardware-requireme}

\begin{table}[H]
\centering
\caption{Comparative resource and depth analysis for QAE algorithms in derivative pricing.}
\label{tab:QAEdepth}
\vspace{0.5em}
\footnotesize{
\begin{tabular}{>{\raggedright\arraybackslash}p{2.5cm}>{\raggedright\arraybackslash}p{3cm}>{\raggedright\arraybackslash}p{2.5cm}>{\raggedright\arraybackslash}p{2.5cm}>{\raggedright\arraybackslash}p{2.6cm}}
\toprule
\textbf{Algorithm} & \textbf{Key Resource} & \textbf{Circuit Depth} & \textbf{Query Complexity} & \textbf{Hardware Suitability} \\
\midrule
Standard QAE & QPE + Inverse QFT; fault-tolerant qubits & $O(\log(1/\varepsilon)D_A)$ & $O(1/\varepsilon)$ & Future fault-tolerant quantum computers \\
Iterative QAE & Adaptive Grover iterations, no QFT & $O(\log(1/\varepsilon))$ & $O(1/\varepsilon)$ & NISQ-compatible with adaptive control \\
MLQAE & Multiple shallow $Q^k$ runs + classical likelihood fit & $O(k_{\max})$ (shallow) & $O(1/\varepsilon)$ & Excellent for hybrid workflows \\
QAE-Lite & Fixed low-depth Grover iterations & $O(1)$ & Between $O(1/\varepsilon)$ and $O(1/\varepsilon^2)$ & Ideal for NISQ demonstrations \\
QAE-SR & Many short, independent shallow circuits & $O(1)$ & $O(1/\varepsilon^2)$ & Noisy devices, statistical post-processing \\
\bottomrule
\end{tabular}
}
\end{table}

\paragraph{Discussion.}
For financial institutions seeking practical quantum speedups in the near term, hybrid algorithms such as IQAE, MLQAE, and QAE-Lite are the most viable. They balance resource constraints with measurable acceleration, requiring only hundreds of shots per amplitude level. Canonical QAE remains the ultimate asymptotic benchmark but awaits hardware capable of sustaining $>10^6$ error-corrected qubits and gate depths exceeding $10^4$.

\subsection{Illustrative Case Study: Asian Option Pricing}
\label{sec:m2-case-study-quantum-pricing-of-an-asian-option}

\paragraph{Objective.}
We present an end-to-end illustration of pricing a \emph{fixed-strike arithmetic Asian call} using Quantum Amplitude Estimation (QAE). The payoff is
\begin{equation}
f = \max\!\big(\bar{S} - K, 0\big),\qquad 
\bar{S} = \frac{1}{m}\sum_{t=1}^{m} S_t,
\end{equation}
under a risk–neutral measure \(\mathbb{Q}\) with GBM dynamics \(\mathrm{d}S_t = r S_t\,\mathrm{d}t + \sigma S_t\,\mathrm{d}W_t\). We discretise short horizons and small qubit registers for clarity, then show how the same pattern extends. Implementation is given in \texttt{PennyLane}, together with a NISQ-friendly hybrid amplitude-estimation loop (IQAE/MLQAE style).

\paragraph{Implementation note.}
All computations in this section are performed on a classical local machine using 
\texttt{PennyLane}'s statevector simulator (\texttt{default.qubit}), 
which provides a hardware-agnostic emulation of quantum circuits. 
While frameworks such as \texttt{Qiskit} require specific compiled backends (e.g., 
IBM’s Aer simulator) that may not be supported natively on macOS ARM architectures, 
\texttt{PennyLane} offers a lightweight and differentiable alternative suitable for 
demonstrating hybrid quantum--classical workflows such as QAE-Lite, IQAE, or MLQAE.

\subsubsection{Minimal Discretisation and Registers}
\label{sec:m2-minimal-discretisation-and-registers}

\paragraph{Time and price grids.}
Let \(m=3\) monitoring dates at times \(t=\{T/3,2T/3,T\}\). For each step, discretise the one-step log-return \(Z_t\sim\mathcal{N}\!\big((r-\tfrac{1}{2}\sigma^2)\Delta t,\;\sigma^2\Delta t\big)\) into \(B\) buckets (e.g., \(B=4\)) and approximate
\(
S_{t+1} \approx S_t \exp(Z_{t+1}).
\)
To keep the circuit width small, we do \emph{distribution loading} directly for the \emph{running average} \(\bar{S}\) using a precomputed (classical) table: we Monte Carlo a modest number \(N_\mathrm{pre}\) paths off-line and construct a discrete approximation of \(p(\bar{S})\) over \(2^n\) bins (e.g., \(n=4\) qubits \(\Rightarrow\) 16 bins). This concentrates quantum resources on \emph{amplitude estimation} and payoff encoding.

\paragraph{Registers.}
\begin{itemize}
    \item \(n_P\) qubits (index register) to encode the histogram over \(\bar{S}\) bins.
    \item 1 ancilla payoff qubit for amplitude encoding.
    \item Optional 1 control ancilla for shallow Grover iterates (IQAE/MLQAE).
\end{itemize}

\subsubsection{Circuit Blueprint (PennyLane-Oriented)}
\label{sec:m2-circuit-blueprint-pennylane-oriented}

\paragraph{State preparation.}
Given a probability vector \(\mathbf{p}\in\mathbb{R}^{2^{n_P}}\) over \(\bar{S}\) bins, prepare
\begin{equation}
\ket{\psi_P} \;=\; \sum_{j=0}^{2^{n_P}-1}\sqrt{p_j}\,\ket{j}.
\end{equation}
We then map bin index \(j\mapsto \bar{S}_j\) via classical side information (no extra qubits).

\paragraph{Payoff encoding (controlled rotation).}
Define the normalised payoff \(f'(\bar{S}_j)=\max(\bar{S}_j-K,0)/B\in[0,1]\) for a known bound \(B\). The payoff ancilla is rotated as
\begin{equation}
\ket{j}\ket{0}
\;\mapsto\;
\ket{j}\!\left(\sqrt{1-f'(\bar{S}_j)}\ket{0} + \sqrt{f'(\bar{S}_j)}\ket{1}\right),
\end{equation}
which yields target amplitude \(a=\mathbb{E}[f']/1 = \sum_j p_j f'(\bar{S}_j)\).

\paragraph{Amplitude estimation (shallow).}
Use 1–3 Grover steps with classical post-processing (IQAE/MLQAE/QAE-Lite style) to estimate \(a\) at low depth. The discounted price is \(\tilde{V}_0=e^{-rT}B\,\tilde{a}\).

\paragraph{Circuit interpretation.}
\begin{figure}[H]
\centering
\includegraphics[width=\linewidth]{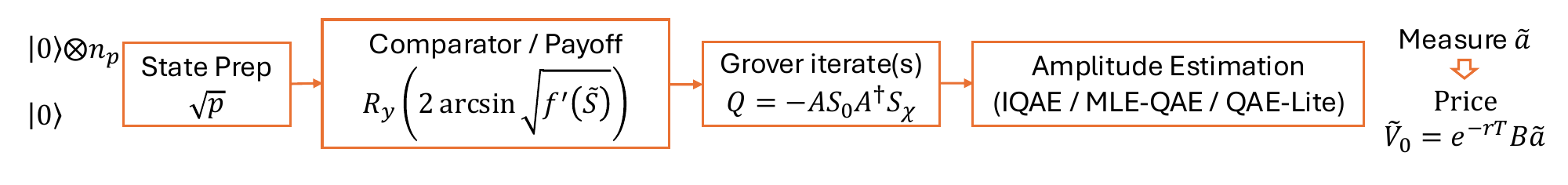}
\caption{Asian option pricing via QAE (hybrid pipeline).}
\end{figure}

\emph{Interpretation.} 
The control by the price index is realised as a compiled multiplexed block (QROM/QROAM).

Together, these blocks enable the payoff rotations 
$\ket{i}\ket{0}\!\mapsto\!\ket{i}\big(\sqrt{1-f'(x_i)}\ket{0}+\sqrt{f'(x_i)}\ket{1}\big)$ 
to be implemented compactly and are abstracted in the diagram as a single multiplexed module.
Read against the classical pricing workflow, the figure makes the division of labour explicit: classical preprocessing concentrates the path information into a discretised payoff distribution, while the quantum subroutine is reserved for amplitude-sensitive expectation extraction. That separation is central to the finance case for QAE, because it targets the sampling bottleneck without pretending that every upstream modelling step must already be quantum-native.

\subsubsection{PennyLane Implementation}
\label{sec:m2-pennylane-implementation}

\paragraph{Implementation summary.}
The full executable implementation has been moved out of the main narrative and into \Cref{sec:app-module-2-asian-option-pricing-case}, together with the standardised notebook asset \repofile{module_02_pricing/module_02_asian_option_qae.ipynb}{module_02_asian_option_qae.ipynb}. The implementation uses \texttt{pennylane} with a statevector device for logic validation and a shot-based device for sampling-based estimation, preserving the distinction between ideal circuit behaviour and finite-measurement execution on near-term hardware.

\paragraph{What the code now does.}
The notebook constructs a classical Monte Carlo benchmark for a fixed-strike arithmetic Asian option, discretises the distribution of average prices into \(128\) histogram bins, rescales the payoff to the unit interval, and encodes the resulting amplitudes into a controlled-rotation circuit. A pair of QNodes is then used to compare exact ancilla probabilities against shot-based estimates, while an endianness calibration step checks that the price register is mapped consistently before the amplitude is interpreted as an option value.

\paragraph{Numerical interpretation.}
In the current implementation, the amplitude-based price remains tightly aligned with the classical benchmark: the classical Monte Carlo price is approximately \(5.19\), while the exact and shot-based quantum-style estimates lie in the interval \(5.16\) to \(5.20\). This close agreement confirms that the amplitude loading, controlled rotation logic, and ancilla readout are working as intended. From an implementation standpoint, the most important design choices are therefore:
\begin{itemize}
    \item a separate exact QNode and shot-based QNode, so that logical correctness can be distinguished from finite-sampling noise;
    \item \texttt{AmplitudeEmbedding} with explicit normalisation control, so that the ancilla probability remains equal to \(\sum_i p_i f_i'\);
    \item an endianness check, so that the computational basis ordering does not contaminate the financial interpretation of the measured amplitude.
\end{itemize}

\paragraph{Why the code belongs in the appendix.}
For the purposes of the review paper, the critical point is methodological rather than syntactic. The main text needs to explain the pricing logic, the hybrid execution design, and the financial interpretation of the result; the appendix notebook then provides the full circuit-level implementation for readers who want to reproduce or extend the experiment.
\subsubsection{Hybrid Quantum–Classical Execution Strategy}
\label{sec:m2-hybrid-quantum-classical-execution-strategy}
\begin{figure}[htbp]
    \centering
    \includegraphics[width=0.90\textwidth]{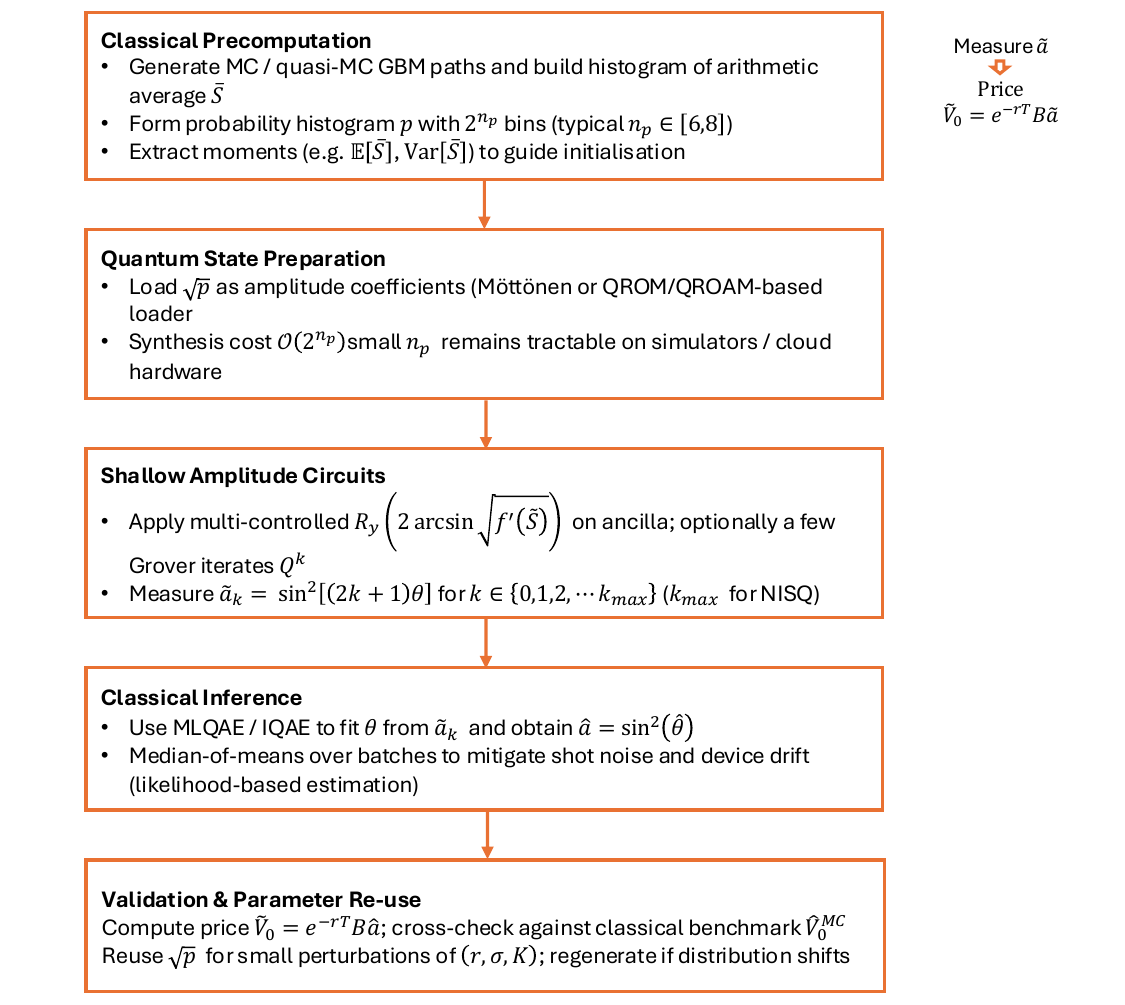}
    \caption{Hybrid quantum--classical QAE pipeline for derivative pricing. 
    Classical precomputation provides a discretised histogram of average prices $\bar S$, 
    which is encoded as quantum amplitudes for amplitude estimation and iterative inference.}
    \label{fig:hybrid_pipeline}
\end{figure}

In practice, full QAE with quantum phase estimation (QPE) remains beyond near-term hardware.  
Therefore, a hybrid workflow is employed, combining classical precomputation and data aggregation with shallow quantum amplitude circuits and statistical inference.  
This hybrid strategy preserves the theoretical quadratic advantage in sample complexity while adapting to the noise and resource constraints of NISQ devices.
\paragraph{Conceptual overview.}
The complete workflow consists of five interconnected stages, shown schematically in Figure~\ref{fig:hybrid_pipeline}.  
Each stage integrates classical computation (for stochastic sampling and data aggregation) with quantum subroutines (for amplitude encoding and measurement).  
The classical and quantum components exchange only low-dimensional summary statistics, maintaining a feasible bandwidth even when $N_{\mathrm{MC}}\gg 2^{n_P}$.

\paragraph{Pipeline.}
\begin{enumerate}
    \item \textbf{Classical precomputation:} 
    Generate $N_\mathrm{pre}$ Monte Carlo (MC) or quasi-Monte Carlo paths for the underlying geometric Brownian motion (GBM).  
    Aggregate the simulated arithmetic averages $\bar{S}$ into a probability histogram $\mathbf{p}$ with $2^{n_P}$ bins, where $n_P$ controls the quantum register width and the quantisation fidelity.  
    Typical prototype settings use $n_P \in [6,8]$ and $N_\mathrm{pre} \sim 10^5$ to $10^6$.  
    This step also allows the extraction of moments and sensitivity metrics (e.g., $\mathbb{E}[\bar S]$, $\mathrm{Var}(\bar S)$), which can later guide parameterised quantum circuit initialisation.
    
    \item \textbf{Quantum state preparation:}
    Load $\sqrt{\mathbf{p}}$ as amplitude coefficients via the Mottonen state-preparation algorithm or Quantum Read-Only Memory (QROM) techniques.  
    This defines the base state 
    $\ket{\psi_P}=\sum_i \sqrt{p_i}\ket{i}$.
    The synthesis cost scales as $O(2^{n_P})$ but, for small $n_P$, remains tractable on simulators or cloud-accessible hardware.  
    The preparation circuit constitutes the ``oracle'' $A$ in QAE, serving as the quantum analogue of a payoff density estimator.

    \item \textbf{Shallow amplitude circuits:}
    For each experimental depth level $k\in\{0,1,2,\dots,k_{\max}\}$, execute shallow circuits that apply multi-controlled $R_y(2\arcsin\sqrt{f'_i})$ rotations and, optionally, a small number of Grover iterations $Q^k = (-A S_0 A^\dagger S_\chi)^k$.  
    Each circuit yields a measured ancilla probability $\tilde{a}_k = \sin^2[(2k+1)\theta]$ from which $\theta$ can be inferred.  
    In practice, only $k_{\max}\in\{0,1,2\}$ is required, making the approach hardware-efficient.

    \item \textbf{Classical inference:}
    Post-process the sampled ancilla probabilities $\{\tilde{a}_k\}$ using either Maximum Likelihood Quantum Amplitude Estimation (MLQAE) or Iterative QAE (IQAE).  
    These methods estimate $\theta$ by minimising the deviation between observed and theoretical amplitude patterns, yielding $\hat{a}=\sin^2(\hat{\theta})$.  
    To mitigate shot noise and device drift, \emph{median-of-means} aggregation is used over repeated circuit batches.  
    This step replaces phase estimation with classical statistical inference, thus realising a ``likelihood-based quantum advantage'' that is robust to decoherence.

    \item \textbf{Validation and parameter re-use:}
    For sensitivity and scenario analysis, the same $\sqrt{\mathbf{p}}$ distribution can be re-used across small perturbations of $(r,\sigma,K)$, as long as the relative shape of $\bar{S}$ remains stable.  
    When volatility or drift changes significantly, a new histogram must be regenerated.  
    Cross-validation against the classical benchmark $\hat{V}_0^{\mathrm{MC}}$ ensures model stability and allows incremental refinement of the quantum oracle $A$.
\end{enumerate}

\paragraph{Workflow interpretation.}
Figure~\ref{fig:hybrid_pipeline} illustrates the overall hybrid QAE architecture: classical data generation on the left, quantum amplitude inference on the right, and a feedback bridge through iterative statistical estimation.  
This design forms the foundation for scalable derivative pricing and risk aggregation under NISQ constraints. The diagram also makes the division of labour explicit: path generation and histogram construction remain classical, while the quantum subroutine is reserved for amplitude-sensitive inference that is expensive to reproduce by brute-force resampling alone.

\paragraph{Advantages and practical notes.}
This hybrid design achieves several objectives:
\begin{itemize}
    \item \emph{Scalability:} the quantum register size grows only logarithmically with the number of price bins, allowing $\sim 2^{n_P}$ scenarios to be encoded simultaneously.
    \item \emph{Noise robustness:} no deep phase-estimation loop is required; each quantum call is shallow and independent.
    \item \emph{Sample efficiency:} compared with classical MC variance $\propto 1/N$, the amplitude-based estimator attains $\propto 1/M^2$ scaling with respect to the number of quantum shots $M$.
    \item \emph{Financial interpretability:} the hybrid pipeline naturally aligns with established risk analytics workflows—classical scenario generation, quantum expectation encoding, and statistical inference.
\end{itemize}

\paragraph{Error budget.}
Total pricing error arises from several sources:
\begin{equation}
|\Delta V_0|\;\lesssim\; e^{-rT}\Big(
B\,|\tilde{a}-a| 
\;+\; \underbrace{\mathrm{disc.\ error}}_{\text{binning / histogram}}
\;+\; \underbrace{\mathrm{state\ prep\ approx.}}_{\text{imperfect loader}}
\;+\; \underbrace{\mathrm{noise}}_{\text{device}}
\Big).
\end{equation}
The first term, $|\tilde{a}-a|$, corresponds to the amplitude-estimation error induced by the limited number and depth of Grover iterates.
In the hybrid workflow introduced earlier (Step~3 of the pipeline), the maximum number of amplification steps \(k_{\max}\) is the key hyperparameter controlling this bias–variance trade-off.
A larger \(k_{\max}\) increases the effective signal contrast, improving estimator precision, but also deepens the circuit and accumulates more device noise.
Practical implementations therefore adopt a small and heterogeneous \(k\)-schedule—such as a geometric progression in MLQAE—to achieve robust parameter inference while keeping depth shallow.

\paragraph{NISQ practicality.}
Circuit depth is dominated by state preparation and multi-controlled rotations rather than by Grover iterations.
For moderate discretisations (\(n_P\in[4,7]\), i.e.\ 16–128 bins), today’s superconducting or photonic QPUs can support tens to hundreds of shots per circuit within coherence limits.
In practice, error-mitigation techniques—such as \emph{zero-noise extrapolation}, \emph{Clifford data regression}, or readout calibration—can substantially stabilise the amplitude estimate \(\hat{a}\).

\paragraph{Scalability outlook.}
By tying the error decomposition explicitly to pipeline parameters (\(n_P, k_{\max}, \text{shots}\)), this hybrid design forms a modular bridge between simulation-level demonstrations and near-term experimental runs.
It also defines a reproducible framework for scaling toward production-grade quantum pricing systems, which is the focus of the next section.

\subsubsection{Scaling Pathways: From Prototype to Production}
\label{sec:m2-scaling-pathways-from-prototype-to-production}

The preceding sections established a full hybrid QAE workflow that is executable on today’s NISQ devices. 
While the prototype demonstrates correctness and noise-tolerant inference, practical deployment in real financial workloads requires systematic scaling along both the quantum and classical dimensions.
We now outline the key levers for achieving this transition.

\paragraph{Scaling levers.}
\begin{itemize}
    \item \textbf{Better loaders:} replace Motton	en with QROM/QROAM and alias sampling to reach \(n_P\approx 10\) (1024 bins) with acceptable depth.
    \item \textbf{Structured payoffs:} for piecewise-linear payoffs, synthesise rotations via multiplexed linear segments (fewer angles).
    \item \textbf{Variance reduction:} encode a control variate \(h(\bar{S})\) with known \(\mathbb{E}[h]\) and estimate \(\mathbb{E}[f-h]\) quantumly to cut amplitude variance.
    \item \textbf{Greeks:} re-use runs and apply likelihood ratio (pathwise) estimators in the classical layer; or finite-difference small shifts in \(K\) with shared measurement reuse.
\end{itemize}

\paragraph{Price extraction.}
The final estimator is
\begin{equation}
\tilde{V}_0 \;=\; e^{-rT}\,B\,\hat{a},\qquad
\hat{a}=\arg\max_{\theta\in[0,\pi/2]}\mathcal{L}(\theta\mid \{P_k(1)\}_{k\in\mathcal{K}}),\ \ a=\sin^2(\theta).
\end{equation}
Confidence intervals follow from binomial concentration on counts per circuit and the delta method for \(a=\sin^2(\theta)\).

\subsection{Hybrid Strategies and Practical Considerations}
\label{sec:m2-hybrid-strategies-and-practical-considerations}

\paragraph{Subsection Overview.}
While the preceding sections focused on the mathematical formulation and prototype implementation of QAE for derivative pricing, practical deployment on near-term quantum hardware remains constrained by decoherence, gate infidelity, and limited qubit counts. 
This section discusses hybrid strategies that combine classical and quantum components to achieve approximate quantum acceleration under the NISQ regime, while maintaining theoretical consistency with the fault-tolerant limit.

\subsubsection{Hybrid Sampling and Data Fusion}
\label{sec:m2-hybrid-sampling-and-data-fusion}

\paragraph{Concept.}
A promising approach for near-term speedup is \textbf{hybrid sampling}, where quantum devices are used to generate a small but information-rich subset of samples, which are then amplified classically.
Formally, let \( \mathcal{S}_Q = \{\bar{S}_i\}_{i=1}^{N_Q} \) denote quantum-generated samples following a probability amplitude distribution \( |\psi_i|^2 \approx p_i \).
A classical sampler (e.g., importance or rejection sampling) can then extend this distribution to a larger \( N_C \gg N_Q \), ensuring the Monte Carlo estimator remains unbiased:
\begin{equation}
\hat{V}_0^{(\mathrm{hyb})} = e^{-rT} \frac{B}{N_C}\sum_{i=1}^{N_C} f'(\bar{S}_i), \qquad
\bar{S}_i \sim \text{Hybrid}(p_i^\mathrm{Q},\,p_i^\mathrm{C}).
\end{equation}
This design reuses the quantum amplitude information as a \emph{prior} over the payoff landscape, thus preserving the most nonclassical correlations while avoiding full circuit depth.

\paragraph{Advantages.}
Such hybridisation reduces the number of quantum calls from \(O(1/\epsilon)\) to \(O(1/\sqrt{\epsilon})\) while retaining quadratic convergence of variance in the amplitude inference stage.
It also provides a natural interface for distributed or federated execution, where quantum subsystems act as generative priors for classical Monte Carlo nodes.

\subsubsection{Quantum Randomness and Variance Reduction}
\label{sec:m2-quantum-randomness-and-variance-reduction}

\paragraph{QRNG Integration.}
Quantum Random Number Generators (QRNGs) provide information-theoretically unpredictable seeds that can replace pseudorandom generators in classical simulation.
When applied to path-dependent options, QRNG-driven sequences yield improved uniformity and decorrelation across sample paths.
Integrating QRNG with classical low-discrepancy methods (Sobol, Halton) creates a \emph{quantum-classical hybrid variance reduction} mechanism:
\begin{equation}
\text{Var}\!\left[\hat{V}_0^{(\mathrm{QRNG})}\right]
\;\approx\;
(1-\lambda)\,\text{Var}_{\text{Sobol}} + \lambda\,\text{Var}_{\text{quantum}},
\end{equation}
where the mixing coefficient \( \lambda \in [0,1] \) tunes between reproducibility and quantum entropy.

\paragraph{Practical Implementation.}
In current NISQ settings, QRNG data can be streamed from cloud APIs (e.g., IBM Q, Quantinuum, or ID Quantique) and injected into classical pricing pipelines as stochastic seeds.
This allows testing quantum-enhanced uncertainty propagation without requiring a full amplitude-estimation circuit.

\subsubsection{Noise Impact and Error Mitigation}
\label{sec:m2-noise-impact-and-error-mitigation}

\paragraph{Error channels.}
Noise affects amplitude estimation primarily through phase decoherence in Grover iterations and state-preparation infidelity.
For an estimated amplitude \(\hat{a}\), the total bias under depolarising noise channel \(\mathcal{E}_p(\rho)=(1-p)\rho+p\mathbb{I}/2^{n}\) follows approximately:
\begin{equation}
\mathbb{E}[\hat{a}]-a \;\approx\; -p(1-a) + O(p^2),
\end{equation}
implying a first-order linear degradation with noise probability \(p\).

\paragraph{Mitigation techniques.}
Practical error-mitigation methods include:
\begin{itemize}
  \item \textbf{Zero-noise extrapolation (ZNE):} rescale circuit noise via artificial gate folding and extrapolate to the zero-noise limit.
  \item \textbf{Clifford data regression (CDR):} train a linear regression between noisy non-Clifford circuits and exact Clifford equivalents.
  \item \textbf{Measurement calibration:} estimate readout confusion matrices \(M_{ij}\) and invert them during amplitude post-processing.
\end{itemize}
Empirically, such methods restore amplitude accuracy within \(10^{-3}\)–\(10^{-2}\) under realistic NISQ noise models, as shown in benchmarked QAE experiments.

\subsubsection{Fault-Tolerant vs. NISQ Feasibility}
\label{sec:m2-fault-tolerant-vs-nisq-feasibility}

\paragraph{Fault-tolerant outlook.}
In the long-term, fully error-corrected quantum processors will enable coherent Grover iterates and exact amplitude estimation with asymptotic \(O(1/\epsilon)\) scaling. 
For derivative pricing, this implies the ability to price high-dimensional basket options or compute Greeks directly via quantum differentiation.

\paragraph{NISQ reality.}
In contrast, current NISQ devices operate with depth limits of a few hundred two-qubit gates and error rates of \(10^{-3}\)–\(10^{-2}\).
Hence, hybrid QAE frameworks rely on shallow circuits (often \(k \le 2\)), frequent classical inference, and statistical smoothing to emulate amplitude inference.
While no exponential speedup is yet achievable, hybrid workflows already demonstrate measurable constant-factor advantages in effective sample efficiency and variance convergence.

\paragraph{Synthesis.}
Hybrid quantum–classical strategies thus serve as a \emph{transitional architecture} bridging current noisy hardware and future fault-tolerant systems.
By coupling QRNGs, hybrid sampling, and error mitigation with adaptive classical inference, near-term financial quantum computing can deliver tangible computational value—even before full-scale quantum advantage becomes accessible.

\begin{table}[htbp]
\centering
\renewcommand{\arraystretch}{1.2}
\caption{Comparison of hybrid (NISQ-era) and fault-tolerant QAE workflows.}
\label{tab:hybrid_vs_faulttolerant}
\footnotesize{
\begin{tabular}{>{\raggedright\arraybackslash}p{2.4cm}>{\raggedright\arraybackslash}p{5.8cm}>{\raggedright\arraybackslash}p{5.4cm}}
\toprule
\textbf{Aspect} & \textbf{Hybrid / NISQ QAE} & \textbf{Fault-Tolerant QAE} \\
\midrule
Hardware requirements 
& Noisy intermediate-scale devices; limited qubits ($\lesssim$100), low depth ($\sim10^2$ gates) 
& Fully error-corrected qubits ($10^5$–$10^7$ logical), arbitrary circuit depth \\
Amplitude inference method 
& Iterative MLQAE/IQAE; few controlled Grover calls; heavy classical post-processing 
& Coherent amplitude amplification + Quantum Phase Estimation (QPE) for exact amplitude \\
Error sources 
& Gate infidelity, decoherence, state-prep approximation, sampling noise 
& Logical gate error $\ll 10^{-9}$; physical noise suppressed by QEC \\
Error mitigation / correction 
& Zero-noise extrapolation, Clifford data regression, measurement calibration 
& Full quantum error correction (e.g., surface code) \\
Complexity scaling 
& Hybrid sampling improvements (effective $O(1/\sqrt{\epsilon})$-like gains via MLQAE schedules) 
& Quadratic speedup via coherent amplification; asymptotic $O(1/\epsilon)$ queries vs.\ classical $O(1/\epsilon^2)$ \\
Circuit depth per iteration 
& Shallow: $10^2$–$10^3$ gates (depends on $n_P$ and loader) 
& Deep: $10^6$–$10^8$ logical gates including QPE and QEC layers \\
Use cases (2020s) 
& POC pricing (Asian/basket), CVA/DVA tweaks, small-$n_P$ simulations 
& High-dimensional portfolios, path-dependent Greeks, full MC replacement \\
Expected advantage 
& Constant-factor speedup, variance reduction, hybrid variance scaling 
& Robust quadratic asymptotic speedup \\
Deployment horizon 
& 0–5 years (cloud QPU + classical servers) 
& 10+ years (fault-tolerant universal processors) \\
\bottomrule
\end{tabular}
}
\end{table}

\subsection{Synthesis and Outlook}
\label{sec:m2-discussion-opportunities-and-challenges}

\paragraph{Subsection Overview.}
QAE and related hybrid frameworks offer a mathematically elegant path toward accelerating Monte Carlo–based financial analytics. 
However, translating these theoretical speedups into practical financial computation requires overcoming significant engineering and modelling challenges.
This section discusses the current limitations, emerging opportunities, and directions for integration with subsequent quantum-finance modules such as risk measurement and quantum machine learning.

\subsubsection{Current Challenges}
\label{sec:m2-current-challenges}

\paragraph{(1) State-preparation complexity.}
State preparation remains the primary bottleneck in quantum finance applications.
Constructing the probability amplitude vector \(\sqrt{\mathbf{p}}\) corresponding to a realistic financial distribution—whether from historical returns, stochastic volatility models, or correlated basket payoffs—scales exponentially with the number of qubits \(n_P\).
Techniques such as Mottonen’s state initialization, QROM, and QROAM provide structured decompositions, but they require extensive classical precomputation and circuit synthesis.
For realistic distributions, an efficient \emph{data loading oracle} is essential to prevent the quantum advantage from being offset by exponential preparation overhead.

\paragraph{(2) Circuit depth and coherence limits.}
Amplitude estimation and pricing circuits typically involve multi-controlled rotations, inverse-QFT–style phase estimation, and Grover iterates—all of which quickly exhaust coherence time on NISQ hardware.
The effective depth scales as \(O(2^{n_P})\) for naïve state-prep and \(O(k_{\max} n_P)\) for shallow iterative QAE.
Hence, hybrid approaches (e.g., MLQAE, QAE-Lite) are crucial for truncating depth while retaining the amplitude-encoding semantics of QAE.

\paragraph{(3) Noise and quantum measurement uncertainty.}
NISQ devices suffer from depolarising and readout noise that directly affect amplitude inference.
The estimated amplitude \(\hat{a}\) exhibits both bias and variance due to imperfect controlled rotations and measurement misclassification.
Repeated circuit runs with statistical resampling (e.g., bootstrap or median-of-means aggregation) mitigate the uncertainty but increase runtime cost.
Reliable error mitigation remains essential to produce financially meaningful estimates—particularly when pricing under tight regulatory precision (e.g., CVA/DVA adjustments).

\paragraph{(4) Embedding real financial data.}
Mapping classical market data (e.g., yield curves, implied volatility surfaces, ESG factors) into quantum amplitudes introduces discretisation and rounding errors.
These manifest as bias in histogram-based state preparation and as truncation error in the payoff encoding \(f'(\bar{S}_j)\).
Adaptive binning and stochastic encoding schemes may reduce such artefacts, but calibration against classical Monte Carlo benchmarks is necessary to validate accuracy.

\subsubsection{Future Outlook and Research Opportunities}
\label{sec:m2-future-outlook-and-research-opportunities}

\paragraph{(1) Integration with quantum risk measures.}
Future extensions will couple QAE-based pricing with quantum formulations of risk metrics such as Conditional Value-at-Risk (CVaR) and Expected Shortfall.
A quantum CVaR circuit can reuse the same amplitude-encoded price distribution to estimate tail probabilities:
\begin{equation}
\text{CVaR}_\alpha = \frac{1}{1-\alpha}\sum_{\bar{S}_j \leq q_\alpha} p_j f'(\bar{S}_j),
\end{equation}
where \(q_\alpha\) is the $\alpha$-quantile of the quantum-sampled payoff distribution.
Such integration could enable coherent, joint estimation of both expected payoff and tail risk within the same quantum workflow.

\paragraph{(2) ESG and structured derivatives.}
As financial institutions expand into sustainability-linked and ESG-indexed products, QAE’s capacity for multidimensional distribution encoding may be leveraged to price derivatives dependent on non-financial signals (e.g., carbon intensity, social scores).
Structured payoffs—such as autocallables or green performance-linked notes—can be expressed as nonlinear functions of multiple correlated variables, well-suited to amplitude encoding.
Quantum-enhanced Monte Carlo methods can thus serve as interpretable engines for scenario analysis under ESG constraints.

\paragraph{(3) Connection to Quantum Machine Learning (Module 4).}
QAE naturally complements the Quantum Machine Learning (QML) pipeline introduced in Module 4.
Once derivative-pricing circuits produce calibrated amplitude distributions, these can be used as \emph{quantum features} for downstream QML models—such as Quantum Support Vector Machines or Quantum Neural Networks—that learn pricing sensitivities or hedging policies.
Conversely, QML models can generate adaptive priors for QAE state preparation, leading to a closed hybrid feedback loop between inference and learning.

\paragraph{(4) Toward autonomous quantum-finance agents.}
In the long term, combining QAE, quantum risk measures, and QML agents may yield fully autonomous, quantum-enhanced financial decision systems.
These would operate as \emph{co-processors} within digital finance infrastructure—continuously sampling, updating, and managing risk under quantum-native uncertainty representations.

\paragraph{Synthesis.}
The practical trajectory of quantum finance thus hinges on two parallel advances: 
(i) hardware-level improvements in qubit fidelity and connectivity, and 
(ii) algorithmic hybridisation that adapts to NISQ-era realities while anticipating fault-tolerant scaling.
Bridging these domains requires interdisciplinary collaboration between quantum physicists, financial engineers, and data scientists—ensuring that quantum advantage, when achieved, translates into tangible value across the financial ecosystem.

\newpage

\section{Quantum Risk Estimation and Scenario Simulation}
\label{sec:risk}

\paragraph{Section Overview.}
This section studies the risk and scenario-generation layer of the review. It focuses on tail metrics, rare-event estimation, correlated scenario loading, and stress propagation, with particular attention to where quantum primitives may change the economics of risk measurement relative to classical Monte Carlo and factor-model engines.

The comparison standard is intentionally conservative. Classical historical simulation, Gaussian approximations, factor models, and stress-testing engines remain the baseline throughout; the role of quantum or quantum-inspired methods is to show where tail estimation, dependence modelling, or scenario propagation may become more sample-efficient or structurally expressive than these established tools.

\subsection{Motivation and Problem Setting}
\label{sec:m3-motivation-and-problem-setting}

Risk estimation is fundamental to financial institutions’ capital management,
liquidity planning, and regulatory compliance. Core metrics such as
 \phantomsection\label{term:var}Value-at-Risk (VaR), \phantomsection\label{term:cvar}Conditional Value-at-Risk (CVaR), Expected Shortfall,
and a broad class of regulatory and internal stress tests require a detailed
characterisation of the tail of the loss distribution.

However, classical approaches face significant computational barriers:

\paragraph{1) Nested Monte Carlo and rare-event sampling.}
Modern financial risk engines frequently rely on nested simulations:
an outer loop generates economic or market scenarios, while an inner loop
revalues portfolios or contingent exposures.
When quantiles of the loss distribution (VaR) or conditional expectations
in the tail (CVaR) are needed, the number of simulation paths often rises to
millions-especially for portfolios containing nonlinear derivatives,
credit-sensitive instruments, or complex XVA (X-Value Adjustment).  
Rare-event estimation converges slowly, and even variance-reduction techniques
provide only partial improvements \citep{Glasserman2004,Broadie2006}.

\paragraph{2) High-dimensional and correlated risk-factor models.}
Modern risk systems involve tens or hundreds of correlated risk factors:
interest-rate curves, FX spot-volatility pairs, credit spreads, equity indices,
volatility surfaces, liquidity states, and macro drivers.
Capturing their joint distribution requires:
\begin{itemize}
    \item copula or multi-factor models,
    \item high-dimensional diffusion processes (e.g., Heston-Wishart see \citep{gourieroux2009}),
    \item regime-switching dynamics.
\end{itemize}
Simulation complexity grows exponentially due to the curse of dimensionality.

\paragraph{3) Stress testing and systemic-risk propagation.}
Regulatory stress-test frameworks require scenario generation under:
\begin{itemize}
    \item severe macro-financial shocks,
    \item correlated credit downgrades,
    \item liquidity freezes,
    \item contagion through exposure networks,
    \item structural breaks or regime switches.
\end{itemize}
These extreme, path-dependent scenarios generally require bespoke and expensive
simulation techniques.

\noindent\textbf{Quantum technologies offer a new computational lens} for these
challenges. Although not a replacement for classical engines, research suggests
quantum resources may offer advantages in three complementary directions:

\begin{itemize}
    \item \textbf{Sample-efficient tail estimation:}
    Quantum-enhanced probability estimation and amplitude-based methods
    (see Module~2) can reduce the sample complexity for rare-event
    quantification \citep{Montanaro2015,WoernerEgger2019}.

    \item \textbf{High-dimensional scenario generation:}
    Quantum states encode probability amplitudes over exponentially large spaces,
    providing a compact representation for multi-factor correlated distributions
    or learned generative models \citep{Biamonte2017}.

    \item \textbf{Simulation of market dynamics under shocks:}
Hamiltonian simulation and variational quantum algorithms
(\phantomsection\label{term:qve-vqe}Variational Quantum Eigensolver / \phantomsection\label{term:vqs}Variational Quantum Simulation, VQE/VQS)
provide a systematic framework to map stochastic differential equations,
regime-switching models, and jump--diffusion processes into effective quantum
time evolution.
Recent advances extend VQS beyond closed systems, enabling the simulation of
dissipative and finite-temperature dynamics via Lindblad and non-unitary
generators, which is essential for modeling stress propagation and relaxation
effects under market shocks
\citep{Orus2019,Egger2021,Galvao2025VQS,An2025VQSDynamics}.
\end{itemize}

\noindent
\textbf{Scope of this Module.}  
This module develops the theoretical and practical basis for quantum-enhanced
risk estimation and scenario generation. It focuses on:
\begin{itemize}
    \item quantum-native formulations of VaR and CVaR,
    \item generative modelling and scenario loading for multi-factor systems,
    \item quantum simulation of market-factor evolution under extreme shocks,
    \item hybrid workflows integrating quantum subroutines within classical risk engines.
\end{itemize}

This establishes the foundation for next generation risk analytics,
where classical infrastructure is augmented by quantum algorithms capable of
exploring complex loss distributions, rare events, and stress scenarios with
enhanced efficiency.

\subsection{Quantum Framework for Risk Measures}
\label{sec:m3-quantum-framework-for-risk-measures}

Risk measures such as VaR and CVaR can be expressed in terms of the loss
distribution \(L = -\Delta V\), where \(F_L(\ell)=\mathbb{P}(L\le \ell)\)
denotes its cumulative distribution.  
Quantum algorithms provide alternative definitions and computational routes
for estimating tail probabilities, quantiles, and conditional expectations.

\subsubsection{Value-at-Risk (VaR)}
\label{sec:m3-value-at-risk-var}

For confidence level \(\alpha\in(0,1)\), the \(\alpha\)-VaR is defined as
the smallest loss level \(\ell_\alpha\) such that
\begin{equation}
F_L(\ell_\alpha) \;\ge\; \alpha.
\label{eq:VaR_def}
\end{equation}

Classically, estimating \(\ell_\alpha\) requires either sorting simulated
losses or performing repeated probability evaluations of the form
\(\mathbb{P}(L\le x)\), which is computationally expensive when nested
valuation is required.

\paragraph{Quantum formulation.}
Let the quantum state preparation operator encode (normalised) loss values
into amplitudes:
\begin{equation}
A\ket{0}
  = \sum_{i=0}^{N-1} \sqrt{p_i}\,\ket{\ell_i},
\label{eq:A_loss_state}
\end{equation}
where \(\ell_i\) are discretised loss levels and \(p_i\) their probabilities.
Define the indicator function
\begin{equation}
\mathbf{1}_{\{L\le x\}} =
\begin{cases}
1, & L \le x, \\
0, & L > x.
\end{cases}
\end{equation}

Amplitude encoding yields the probability
\begin{equation}
a(x)
 = \mathbb{P}(L\le x)
 = \sum_{\ell_i\le x} p_i,
\label{eq:ax_cdf}
\end{equation}
which can be estimated by the QAE (Quantum Amplitude Estimation) machinery of Module~2.

\paragraph{Quantum quantile search.}
VaR can be found by performing a \emph{quantum-enhanced binary search}:
each iteration calls QAE to estimate \(a(x)\), and the search adjusts \(x\)
until
\begin{equation}
a(x)\approx\alpha.
\end{equation}
The complexity is \(O(\log N)\) for the binary search and \(O(1/\varepsilon)\)
for each QAE call, giving substantial improvement over classical Monte Carlo
sorting or histogram resolution.

\subsubsection{Conditional Value-at-Risk (CVaR)}
\label{sec:m3-conditional-value-at-risk-cvar}

CVaR (also called Expected Shortfall) measures the expected loss
conditional on losses exceeding VaR:
\begin{equation}
\mathrm{CVaR}_\alpha
  = \mathbb{E}[\,L \mid L \ge \ell_\alpha\,].
\label{eq:CVaR_def}
\end{equation}

It can be expressed as a tail expectation:
\begin{equation}
\mathrm{CVaR}_\alpha
  = \frac{1}{1-\alpha}
    \int_{\ell_\alpha}^{\infty} \ell \, \mathrm{d}F_L(\ell).
\label{eq:CVaR_integral}
\end{equation}

\paragraph{Quantum conditional sampling.}
Define a thresholded payoff
\begin{equation}
g(\ell)
  = \ell \, \mathbf{1}_{\{\ell\ge\ell_\alpha\}}.
\label{eq:g_indicator}
\end{equation}

A quantum circuit can implement a rotation based on \(g(\ell)\) to encode
\(\mathbb{E}[g(L)]\) into an amplitude \(a_g\).  
After normalisation (as discussed in Module~2), we obtain:
\begin{equation}
\mathbb{E}[g(L)] = B\, a_g,
\label{eq:Eg_relation}
\end{equation}
where \(B\) is the normalisation bound for the encoded loss.

Using \eqref{eq:ax_cdf} for the tail probability and \eqref{eq:Eg_relation}
for the tail expectation, CVaR is recovered quantumly via
\begin{equation}
\mathrm{CVaR}_\alpha
  = \frac{B\,a_g}{\,1-a(\ell_\alpha)\,}.
\label{eq:CVaR_QAE}
\end{equation}

\paragraph{Interpretation.}
This reduces CVaR estimation to two primitive quantum calls:
\begin{enumerate}
    \item Evaluate \(a(\ell_\alpha) = \mathbb{P}(L\le\ell_\alpha)\)  
          (quantum CDF evaluation),
    \item Evaluate \(a_g\) (quantum conditional tail expectation).
\end{enumerate}
Classically both require large nested Monte Carlo samples, especially when
losses themselves require simulation; quantum circuits replace this with
state preparation + amplitude extraction.

\subsubsection{Expected Shortfall and Tail Events}
\label{sec:m3-expected-shortfall-and-tail-events}

Expected Shortfall is mathematically equivalent to CVaR, but risk management
often requires deeper exploration of the tail region:
\begin{itemize}
    \item probability of joint defaults,
    \item simultaneous liquidity shocks,
    \item clusters of extreme movements,
    \item contagion effects across exposure networks.
\end{itemize}

\paragraph{Quantum rare-event amplification.}
Let \(\mathcal{R}\) denote a rare, high-loss region,
\(\mathcal{R}=\{L\ge \ell_\mathrm{shock}\}\).  
Quantum circuits can use Grover-style amplification to boost the amplitude
associated with this region:
\begin{equation}
\ket{\psi'}
  = G^m A\ket{0},
\label{eq:grover_tail}
\end{equation}
where \(G\) is the Grover iterate defined relative to \(\mathcal{R}\).

This yields a transformed probability
\begin{equation}
\mathbb{P}'(L\in\mathcal{R}) = \sin^2((2m+1)\theta),
\end{equation}
where \(\sin^2\theta = \mathbb{P}(L\in\mathcal{R})\).  
By choosing \(m\) adaptively, very small probabilities can be made
measurable at scale, enabling efficient tail-event analysis.

\paragraph{Quantum importance weighting.}
Alternatively, quantum circuits may encode modified distributions
\(\tilde{p}_i\) that overweight tail regions while maintaining tractable
normalisation:
\begin{equation}
\tilde{p}_i \propto w(\ell_i)\,p_i,
\end{equation}
with \(w(\cdot)\) designed to emphasise extreme losses.  
Amplitude estimation can then recover quantities under the original measure
via reweighting.

\noindent
The techniques in this section provide the foundation for quantum-native
risk measures. Subsequent sections (3.3--3.5) expand these principles to
multi-factor modelling, scenario generation, and Hamiltonian-based
market simulations.

\subsection{Quantum Scenario Generation}
\label{sec:m3-quantum-scenario-generation}

\subsubsection{Quantum Sampling of Risk Factors}
\label{sec:m3-quantum-sampling-of-risk-factors}

Let \(\mathbf{Y}\in\mathbb{R}^d\) denote a vector of risk factors, e.g.
\begin{equation}
\mathbf{Y}
  = (Y_1,\dots,Y_d)^\top
\end{equation}
such as interest-rate tenors, equity indices, FX rates, credit spreads, or
volatility levels. Classical models often represent \(\mathbf{Y}\) via
multi-factor Gaussian or lognormal structures, copulas, or regime-switching
mixtures.

\paragraph{Classical multi-factor structure.}
A common specification is a (possibly time-\(t\)) linear factor model:
\begin{equation}
\mathbf{Y}
  = \boldsymbol{\mu} + L \boldsymbol{\xi},
\label{eq:factor_model}
\end{equation}
where \(\boldsymbol{\mu}\in\mathbb{R}^d\) is the mean vector, 
\(L\in\mathbb{R}^{d\times k}\) is a loading matrix (e.g.\ the Cholesky factor
or principal-component loadings), and
\(\boldsymbol{\xi}\sim\mathcal{N}(\mathbf{0}, I_k)\) are independent factors.
The induced covariance is
\begin{equation}
\Sigma
  = \mathbb{E}\big[(\mathbf{Y}-\boldsymbol{\mu})
                   (\mathbf{Y}-\boldsymbol{\mu})^\top\big]
  = L L^\top.
\label{eq:covariance}
\end{equation}

For lognormal risk factors (e.g.\ terminal equity prices),
the marginal dynamics often follow
\begin{equation}
S_T^{(j)}
 = S_0^{(j)} \exp\!\Big(
    \Big(\mu_j - \tfrac{1}{2}\sigma_j^2\Big)T
    + \sigma_j \sqrt{T}\,Z_j
 \Big),
\label{eq:lognormal_factor}
\end{equation}
where \(\mathbf{Z}=(Z_1,\dots,Z_d)^\top\sim \mathcal{N}(\mathbf{0},\Sigma_Z)\)
captures cross-asset correlation.

\paragraph{Quantum encoding of multi-factor distributions.}
On a quantum device, a multi-factor distribution over a discretised grid
\(\{\mathbf{y}_i\}_{i=0}^{N-1}\subset\mathbb{R}^d\) with probabilities
\(\{p_i\}_{i=0}^{N-1}\) can be embedded as
\begin{equation}
A_{\mathrm{RF}}\ket{0}
  = \sum_{i=0}^{N-1} \sqrt{p_i}\,\ket{\mathbf{y}_i},
\label{eq:rf_state}
\end{equation}
where \(\ket{\mathbf{y}_i}\) is a basis state encoding the discretised
risk-factor vector (e.g.\ via binary encodings for each component).
The probabilities \(p_i\) may come from:
\begin{itemize}
    \item analytical Gaussian or lognormal densities derived from
          \eqref{eq:factor_model}–\eqref{eq:lognormal_factor},
    \item empirical histograms built from historical or simulated data,
    \item learned quantum generative models (see below).
\end{itemize}

Once the state \eqref{eq:rf_state} is prepared, all subsequent quantum
risk calculations (e.g.\ tail probabilities, CVaR) operate coherently on
this multi-factor distribution.

\paragraph{Copula-based structures.}
To separate marginals from dependence, copula models specify the joint CDF
of \(\mathbf{Y}\) as
\begin{equation}
F_{\mathbf{Y}}(y_1,\dots,y_d)
  = C\!\big(F_1(y_1),\dots,F_d(y_d)\big),
\label{eq:copula}
\end{equation}
where \(F_j\) are marginal CDFs and \(C\) is a copula. For Gaussian copulas,
\(\mathbf{U}=(U_1,\dots,U_d)^\top\) with
\begin{equation}
U_j = \Phi(Z_j),\qquad \mathbf{Z}\sim\mathcal{N}(\mathbf{0},\Sigma_C),
\end{equation}
and \(\Phi\) the standard normal CDF, yields the dependence structure, while
\begin{equation}
Y_j = F_j^{-1}(U_j)
\end{equation}
imposes the desired marginals.

A hybrid quantum–classical workflow can proceed as:
\begin{enumerate}
    \item Use a quantum circuit to encode \(\mathbf{Z}\sim\mathcal{N}(\mathbf{0},\Sigma_C)\)
          into amplitudes via a suitable loader \(A_Z\).
    \item Apply (classical or quantum-approximated) transformations
          \(U_j=\Phi(Z_j)\) and \(Y_j=F_j^{-1}(U_j)\) to construct a grid
          \(\{\mathbf{y}_i\}\) and associated probabilities \(\{p_i\}\).
    \item Load the resulting multi-factor distribution into
          \(A_{\mathrm{RF}}\) as in \eqref{eq:rf_state}.
\end{enumerate}
In practice, steps involving \(\Phi\) and \(F_j^{-1}\) may be delegated
to classical pre-processing, with the quantum device focusing on
efficient sampling and tail-event estimation.

\paragraph{Quantum generative models for risk factors.}
Rather than relying solely on parametric models, quantum generative
models can learn complex or regime-dependent distributions directly
from data \citep{Biamonte2017,Benedetti2019,Zoufal2019}.  

A parameterised quantum circuit \(U(\boldsymbol{\theta})\) acting on
\(n\) qubits defines a model state
\begin{equation}
\ket{\psi(\boldsymbol{\theta})}
 = U(\boldsymbol{\theta})\ket{0},
\label{eq:born_state}
\end{equation}
which induces a probability distribution over bitstrings
\begin{equation}
p_{\boldsymbol{\theta}}(z)
  = \big|\braket{z}{\psi(\boldsymbol{\theta})}\big|^2,
\label{eq:born_dist}
\end{equation}
often referred to as a \emph{Born machine}
\citep{Benedetti2019,Gili2023}.
By mapping each bitstring \(z\) to a risk-factor vector \(\mathbf{y}(z)\),
one obtains a flexible generative model over \(\mathbf{Y}\),
capable of capturing high-dimensional and non-Gaussian dependencies.

\medskip

\noindent
Training proceeds by minimising a statistical divergence between the
model distribution and the empirical data distribution:
\begin{equation}
\boldsymbol{\theta}^\star
 = \arg\min_{\boldsymbol{\theta}} 
   D\!\big(p_{\text{data}} \,\Vert\, p_{\boldsymbol{\theta}}\big),
\label{eq:training}
\end{equation}
where \(D(\cdot\Vert\cdot)\) may be chosen as a distance such as
maximum mean discrepancy (MMD), Kullback–Leibler divergence, or an
adversarial (GAN-style) loss.

Quantum Circuit Born Machines (QCBMs) and Quantum Generative Adversarial Networks 
(QGANs) have been demonstrated on small devices to capture non-Gaussian, multi-modal,
and regime-shift behaviour. For risk management, this enables:
\begin{itemize}
    \item learning joint distributions of risk factors directly from
          high-frequency or stressed data,
    \item capturing heavy tails and asymmetries beyond Gaussian copulas,
    \item generating correlated scenarios that respond to market regime shifts.
\end{itemize}

Once trained, the generative model \(U(\boldsymbol{\theta}^\star)\)
serves as a quantum sampler for risk-factor scenarios, which can be
plugged into the risk-measure frameworks of Sections~3.2 and~3.4.

\subsubsection{Multi-Period Evolution via Quantum Dynamics}
\label{sec:m3-multi-period-evolution-via-quantum-dynamics}

Risk-factor evolution in financial markets is commonly modelled using
stochastic differential equations (SDEs) such as:
\begin{itemize}
    \item geometric Brownian motion (GBM) for equity or FX prices,
    \item Heston or SABR models for stochastic volatility,
    \item Merton or Kou jump-diffusion processes for heavy tails,
    \item Markov regime-switching models for structural breaks in markets.
\end{itemize}
Quantum algorithms provide alternative formulations for simulating the
time evolution of these models-particularly under extreme shocks relevant
to risk management and regulatory stress testing.

\paragraph{Hamiltonian encoding of stochastic dynamics.}
A key observation is that certain SDEs can be mapped to quantum evolution
under an effective Hamiltonian \(H\) acting on a discretised state space.
For example, consider a generic SDE:
\begin{equation}
\mathrm{d}X_t = \mu(X_t,t)\,\mathrm{d}t + \sigma(X_t,t)\,\mathrm{d}W_t.
\label{eq:sde_generic}
\end{equation}
The corresponding Fokker--Planck equation for the density \(p(x,t)\) is:
\begin{equation}
\frac{\partial p}{\partial t} 
  = -\frac{\partial}{\partial x}\!\big(\mu p\big) 
    + \frac{1}{2}\frac{\partial^2}{\partial x^2}\!\big(\sigma^2 p\big).
\label{eq:fokker_planck}
\end{equation}
When the spatial domain is discretised into \(N\) grid points, the 
generator of the Fokker--Planck evolution becomes a sparse matrix \(G\),
and the propagator is:
\begin{equation}
p(t+\Delta t)
   = e^{G\,\Delta t}\,p(t).
\label{eq:propagator}
\end{equation}

Quantum simulation replaces this classical propagation with
Hamiltonian evolution:
\begin{equation}
\ket{\psi(t+\Delta t)} 
  = e^{-iH\Delta t}\,\ket{\psi(t)},
\label{eq:quantum_evolution}
\end{equation}
where \(H\) is constructed such that
\begin{equation}
|p(x,t)|^2 \approx \left|\braket{x}{\psi(t)}\right|^2.
\end{equation}
The mapping from classical generators \(G\) to Hermitian Hamiltonians \(H\)
can be achieved through symmetrisation or dilation techniques 
see \citep{Childs2019, childs2021}.

\paragraph{Example: GBM dynamics.}
For geometric Brownian motion:
\begin{equation}
\mathrm{d}S_t = \mu S_t\,\mathrm{d}t + \sigma S_t\,\mathrm{d}W_t,
\end{equation}
a logarithmic change of variable \(X_t=\ln S_t\) yields:
\begin{equation}
\mathrm{d}X_t = \left(\mu - \frac{1}{2}\sigma^2\right)\mathrm{d}t
              + \sigma\,\mathrm{d}W_t.
\end{equation}
The corresponding Fokker-Planck operator has constant coefficients, leading
to a banded generator matrix \(G\), which can be embedded into a quantum
Hamiltonian and simulated using Trotterisation.

\paragraph{Trotterised time evolution.}
For a Hamiltonian decomposed as a sum of local terms
\(H=\sum_{j=1}^m H_j\), Trotterisation yields:
\begin{equation}
e^{-iH\Delta t}
  \approx \left(\prod_{j=1}^m e^{-iH_j\Delta t / r}\right)^r,
\label{eq:trotter}
\end{equation}
where \(r\) is the number of Trotter steps. Increasing \(r\) reduces error
at the cost of deeper circuits.  
This enables a time-stepping simulation:
\begin{equation}
\ket{\psi(t + k\Delta t)}
   = \left(e^{-iH\Delta t}\right)^k \ket{\psi(0)}.
\label{eq:time_stepping}
\end{equation}

\paragraph{Stochastic volatility via Heston dynamics.}
The Heston model:
\begin{equation}
\begin{aligned}
\mathrm{d}S_t &= \mu S_t\,\mathrm{d}t + \sqrt{v_t}\,S_t\,\mathrm{d}W_t^{(1)}, \\
\mathrm{d}v_t &= \kappa(\theta - v_t)\,\mathrm{d}t 
                + \eta\sqrt{v_t}\,\mathrm{d}W_t^{(2)}, \\
\mathrm{d}\langle W^{(1)},W^{(2)}\rangle_t &= \rho\,\mathrm{d}t,
\end{aligned}
\end{equation}
is two-dimensional and admits a Fokker-Planck operator on the joint grid
\((S_t,v_t)\).  
Quantum simulation can approximate this via:
\begin{itemize}
    \item tensor-product grids,
    \item discretised Laplacian and drift operators,
    \item controlled rotations encoding correlation \(\rho\).
\end{itemize}
Hamiltonian simulation for two-factor models has been demonstrated in
prototype quantum PDE solvers \citep{childs2021}.

\paragraph{Jump diffusion.}
A jump-diffusion process:
\begin{equation}
\mathrm{d}S_t = \mu S_t\,\mathrm{d}t + \sigma S_t\,\mathrm{d}W_t
  + S_{t^-}(J-1)\,\mathrm{d}N_t,
\end{equation}
has generator
\begin{equation}
Gf(s)
  = \mu s f'(s) 
    + \frac{1}{2}\sigma^2 s^2 f''(s)
    + \lambda \mathbb{E}\big[f(sJ)-f(s)\big].
\end{equation}
The jump operator is nonlocal but sparse, compatible with block-encoding
and Hamiltonian simulation schemes as in \citep{LowChuang2019}.

\paragraph{Regime-switching models.}
For models with discrete Markovian regimes:
\begin{equation}
\mathrm{d}S_t = \mu_{R_t} S_t\,\mathrm{d}t + \sigma_{R_t} S_t\,\mathrm{d}W_t,
\qquad
\mathbb{P}(R_{t+\Delta t}=j \mid R_t=i) = Q_{ij}\Delta t,
\end{equation}
the augmented state \((S_t,R_t)\) may be encoded in a quantum register with:
\begin{itemize}
    \item Hamiltonian terms for continuous \(S_t\) dynamics under regime \(i\),
    \item jump terms corresponding to the regime transitions \(Q\).
\end{itemize}
Quantum simulation of Markovian hybrid systems has been analysed by 
several works in physics-inspired quantum dynamics \citep{childs2021}.

\paragraph{Applications.}
Multi-period quantum dynamics provide a unified framework for stress testing
and tail-risk evaluation. They enable:
\begin{itemize}
    \item generation of extreme stress scenarios and shock trajectories,
    \item modelling of regime changes, volatility spikes, and liquidity evaporation,
    \item simulation of contagion and coupled risk-factor interactions,
    \item multi-horizon stress testing of portfolios under sequential shocks,
    \item fast evaluation of tail-sensitive metrics (VaR, CVaR) using full-path evolution.
\end{itemize}
By capturing the evolution of risk factors over time -- rather than sampling
from a static terminal distribution -- quantum dynamics complement the 
scenario-loading methods introduced in \Cref{sec:m3-quantum-sampling-of-risk-factors} and enrich the modelling
of extreme market behaviour.

\subsubsection{Scenario Stress Testing}
\label{sec:m3-scenario-stress-testing}

Stress testing evaluates portfolio resilience under extreme but plausible
market shocks. In a quantum setting, stress scenarios are implemented by
\emph{perturbing the Hamiltonian} governing the evolution of risk factors.
Let the baseline market dynamics be generated by a Hamiltonian \(H_0\)
(constructed as in \Cref{sec:m3-multi-period-evolution-via-quantum-dynamics}). A regulatory or internally defined stress
scenario is introduced via a perturbation term \(\Delta H\), giving the
stressed Hamiltonian:
\begin{equation}
    H_{\text{stress}} \;=\; H_0 \;+\; \Delta H.
    \label{eq:Hstress}
\end{equation}

\paragraph{Shock modelling in the Hamiltonian.}
A wide class of financial shocks can be expressed by modifying drift,
volatility, correlation, or jump parameters. For example, a volatility shock
in a one-factor diffusion may take the form:
\begin{equation}
    H_{\text{vol-shock}} 
    \;=\; H_0 \;+\; \lambda_\sigma\, \frac{\partial^2}{\partial S^2},
    \label{eq:volshock}
\end{equation}
where \(\lambda_\sigma>0\) scales the increase in instantaneous volatility.
Similarly, a correlation shock across factors \(i,j\) can be encoded as:
\begin{equation}
    H_{\text{corr-shock}} 
    \;=\; H_0 \;+\; \lambda_{\rho}\, S_i S_j,
    \label{eq:corrshock}
\end{equation}
representing higher co-movement or contagion between risk factors.

\paragraph{Quantum time evolution under stress.}
The stressed scenario is simulated by quantum time evolution:
\begin{equation}
    \ket{\psi(t)} 
    \;=\; e^{-i H_{\text{stress}} t} \ket{\psi(0)},
    \label{eq:stress-evolution}
\end{equation}
implemented via Trotter-Suzuki product formulas or variational simulation
(VQE/VQS based real-time evolution).  
The resulting state encodes the distribution of risk factors under the
specified shock for horizon \(t\), enabling tail-risk metrics or portfolio
valuation at time \(t\).

\paragraph{Regulator-defined vs.\ data-driven scenarios.}
This framework can encode several types of stress scenarios:

\begin{itemize}
    \item \textbf{Regulatory scenarios:}  
    Basel stress tests (e.g., severe recession, credit spread widening),
    PRA/BoE scenarios, ECB adverse scenarios.  
    Each corresponds to specific calibrated perturbations \(\Delta H\)
    on macro, credit, or market-factor Hamiltonians.

    \item \textbf{Historical scenarios:}  
    Events such as 2008 crisis volatility spikes, 2020 liquidity collapse, or
    multi-asset correlation breakdowns can be reproduced by fitting
    \(\Delta H\) to empirical shock parameters.

    \item \textbf{Data-driven shocks:}  
    Machine-learned change points or regime-switching clusters can define
    \(\Delta H\) terms that encode structural breaks or emergent contagion
    patterns.
\end{itemize}

\paragraph{Risk analytics from stressed evolution.}
Measurements performed on the evolved state \( \ket{\psi(t)} \) provide:
\begin{itemize}
    \item stressed loss distributions,
    \item tail metrics (VaR, CVaR) under the shock,
    \item propagation of contagion across correlated factors,
    \item multi-horizon stress paths for liquidity, volatility, or credit spreads,
    \item joint tail events for systemic-risk analysis.
\end{itemize}

These capabilities connect naturally with the multi-period dynamics of
\Cref{sec:m3-multi-period-evolution-via-quantum-dynamics} and allow quantum circuits to model complex stress propagation
patterns that are challenging for classical nested Monte Carlo engines.

\subsection{System-Level Risk Modelling}
\label{sec:m3-system-level-risk-modelling}

\subsubsection{Quantum Linear Algebra for Risk Aggregation}
\label{sec:m3-quantum-linear-algebra-for-risk-aggregation}

Risk aggregation is a central component of enterprise risk management, where
institutions must compute portfolio-level loss distributions, sensitivities,
and variance contributions across a high-dimensional set of risk factors.
Such tasks typically require repeated manipulation of large covariance matrices,
factor decompositions, or linear-system solves.  

Quantum linear algebra primitives - notably \emph{quantum principal component analysis} (QPCA) and \emph{quantum linear-system solvers} (QLS, e.g.\ HHL algorithm) - offer potential pathways to accelerate these computations by exploiting quantum state representations of covariance structures.

\paragraph{Quantum representation of covariance matrices.}
Let the covariance matrix of risk factors be
\begin{equation}
\Sigma \in \mathbb{R}^{d \times d},
\end{equation}
where \(d\) may range from a few dozen (trading books) to hundreds (enterprise-wide stress engines).  
Under amplitude encoding, one may construct a density operator
\begin{equation}
    \rho_\Sigma \;=\; \frac{\Sigma}{\mathrm{Tr}(\Sigma)},
    \label{eq:rhoSigma}
\end{equation}
which can be approximated through samples of the underlying factor distribution.
Eigenvalues of \(\rho_\Sigma\) correspond to normalised principal components of \(\Sigma\), enabling QPCA.

\paragraph{QPCA for extracting dominant risk components.}
QPCA (Lloyd \emph{et al.}, 2014) applies phase estimation to the density operator \(\rho_\Sigma\), yielding estimates of eigenvalues \(\lambda_i\) and eigenvectors \(\ket{v_i}\).  
In risk terms, this enables:

\begin{itemize}
    \item identification of dominant sources of volatility,
    \item factor reduction for high-dimensional stress testing,
    \item extraction of principal shocks for scenario generation.
\end{itemize}

Mathematically, principal components satisfy:
\begin{equation}
    \Sigma v_i = \lambda_i v_i, 
    \label{eq:pca-eig}
\end{equation}
and QPCA estimates \((\lambda_i, v_i)\) in time polylogarithmic in \(d\) under favourable conditions (e.g., low-rank structure).

\paragraph{QLS solvers for variance and sensitivity.}
Risk aggregation frequently requires computing the portfolio variance:
\begin{equation}
    \mathrm{Var}(P) 
    \;=\; w^\top \Sigma w,
    \label{eq:variance}
\end{equation}
where \( w \in \mathbb{R}^d \) is the portfolio weight vector.  
Other tasks involve solving systems of the form:
\begin{equation}
    \Sigma x = b,
    \label{eq:linear-system}
\end{equation}
e.g., for hedging, risk attribution, or regression against factor returns.

QLS solvers (HHL algorithm; Harrow–Hassidim–Lloyd, 2009) approximate
\begin{equation}
x \approx \Sigma^{-1} b
\end{equation}
in time polylogarithmic in \(d\), provided that \(\Sigma\) is well-conditioned and sparse (or efficiently block-encodable).  
When applicable, this offers potential acceleration relative to classical \(O(d^3)\) methods.

\paragraph{Applications to financial risk tasks.}
Quantum linear-algebra tools support:

\begin{itemize}
    \item \textbf{Fast portfolio variance estimation} via amplitude estimation applied to \eqref{eq:variance};
    \item \textbf{Factor decomposition and stress-shock design} using eigenvectors from QPCA;
    \item \textbf{Risk attribution} by computing marginal contributions \( \partial \mathrm{Var}(P) / \partial w_i \);
    \item \textbf{Liquidity-adjusted risk} using inverse-covariance solves for impact models;
    \item \textbf{Scenario compression} through low-rank quantum approximations of \(\Sigma\).
\end{itemize}

These quantum linear-algebra primitives complement the sampling and simulation
approaches of earlier sections by providing structural insights into risk-driver
hierarchies, factor correlations, and hedging sensitivities.

\subsubsection{Network-Based Exposure Structures}
\label{sec:m3-network-based-exposure-structures}

Counterparty credit exposures, funding dependencies, and interconnected balance-sheet
positions naturally form \emph{financial networks}.  
Systemic-risk events - such as default cascades, liquidity spirals, or fire sales - are
driven by the propagation of shocks across these networks.  
Quantum algorithms, particularly \textbf{quantum walks} and \textbf{Hamiltonian-based propagation},
provide a computational framework for modelling such contagion channels.

\paragraph{Network representation.}
Let the financial system be represented as a weighted directed graph
\(
G = (V,E)
\),
where nodes correspond to institutions and edges encode bilateral exposures.
The (possibly asymmetric) exposure matrix is
\begin{equation}
    W = (w_{ij})_{i,j=1}^{N},
    \label{eq:exposure-matrix}
\end{equation}
where \(w_{ij}\) measures the loss imposed on institution \(i\) if \(j\) defaults.
Normalising \(W\) yields a Markov-like transition matrix:
\begin{equation}
    P_{ij} = \frac{w_{ij}}{\sum_{k} w_{ik}},
    \label{eq:transition-matrix}
\end{equation}
which describes how shocks may propagate across the network.

\paragraph{Quantum walk formulation.}
Define the corresponding quantum-walk operator (Szegedy walk):
\begin{equation}
    U_P = (2\Pi_P - I)(2\Pi_Q - I),
    \label{eq:szegedy-walk}
\end{equation}
where
\(
\Pi_P
\)
and
\(
\Pi_Q
\)
are projectors derived from \(P\) and its transpose.
Quantum walks provide:
\begin{itemize}
    \item coherent exploration of exposure networks,
    \item accelerated mixing and probability propagation,
    \item efficient computation of network centrality and vulnerability.
\end{itemize}

\paragraph{Shock propagation.}
A market or credit shock initially affecting node \(i_0\) can be encoded as a basis state
\(
\ket{i_0}
\),
with the quantum walk spreading amplitude across the network:
\begin{equation}
    \ket{\psi_t} = U_P^{\,t} \ket{i_0}.
    \label{eq:quantum-walk-evolution}
\end{equation}
The probability of encountering a distressed institution after \(t\) iterations is then
\begin{equation}
\Pr(j \ \text{distressed at time } t)
= |\braket{j | \psi_t}|^2.
\end{equation}
This allows identification of critical contagion channels or highly vulnerable nodes.

\paragraph{Quantum Hamiltonian approach.}
Networks may also be encoded into Hamiltonians of the form:
\begin{equation}
    H = L_W = D_W - W,
    \label{eq:network-laplacian}
\end{equation}
where \(L_W\) is the weighted graph Laplacian and \(D_W\) is the diagonal degree matrix.  
The Schrödinger evolution
\begin{equation}
    \ket{\psi(t)} = e^{-iHt} \ket{\psi(0)}
    \label{eq:laplacian-evolution}
\end{equation}
simulates continuous-time propagation of stress or liquidity shocks, capturing dynamics akin to:
\begin{itemize}
    \item funding contagion,
    \item joint-default clustering,
    \item liquidity evaporation across intermediaries.
\end{itemize}

\paragraph{Applications to systemic-risk modelling.}
Quantum-walk or Hamiltonian-based propagation enables:
\begin{itemize}
    \item identification of systemic hubs with high contagion centrality,
    \item rapid detection of nodes that amplify shocks,
    \item simulation of multi-step default cascades,
    \item estimation of distributional outcomes under network stress,
    \item construction of scenario trees for macroprudential supervision.
\end{itemize}

These network-based quantum models complement the factor-based and dynamics-based
approaches of Sections~3.3 and~3.4.1, enabling richer modelling of interconnected
financial systems.

\subsubsection{Quantum Rare-Event and Contagion Simulation}
\label{sec:m3-quantum-rare-event-and-contagion-simulation}

Extreme market failures - such as joint defaults, liquidity freezes, or volatility explosions - 
are typically \emph{rare events}, making them difficult to study using classical Monte Carlo.
Quantum-enhanced sampling provides tools to boost the probability mass of these rare scenarios,
improving statistical efficiency.  
When combined with network-based propagation models (\Cref{sec:m3-network-based-exposure-structures}),
quantum techniques can simulate both the \emph{initiation} and the \emph{systemic spread}
of tail events.

\paragraph{Rare-event amplification via amplitude techniques.}
Let \(L\) denote the portfolio loss random variable with distribution \(p(l)\).
Define a threshold \(\tau\) corresponding to a regulatory or internal stress boundary.
The rare-event probability is:
\begin{equation}
    \pi_{\tau} = \Pr(L > \tau) = \sum_{l > \tau} p(l).
    \label{eq:rare-event-prob}
\end{equation}

In a quantum encoding where each loss value \(l\) is mapped to a basis state \(\ket{l}\),
a binary indicator function
\begin{equation}
    g(l) = 
    \begin{cases}
        1, & l > \tau,\\[4pt]
        0, & l \le \tau,
    \end{cases}
    \label{eq:indicator}
\end{equation}
is used to control an ancilla rotation, producing amplitudes proportional to
\(\sqrt{\pi_{\tau}}\).
Quantum amplitude amplification can then increase the visibility of these events:
\begin{equation}
    \ket{\psi_k}
    = Q^k A \ket{0},
    \label{eq:grover-k}
\end{equation}
where \(A\) prepares the loss distribution and \(Q\) is the Grover iterate.
This boosts the tail-event amplitude, enabling efficient sampling of rare losses.

\paragraph{Quantum-enhanced contagion propagation.}
Let the exposure network be encoded by the transition matrix \(P\) 
(cf.\ Eq.~\eqref{eq:transition-matrix}) or Laplacian Hamiltonian \(H\)
(cf.\ Eq.~\eqref{eq:network-laplacian}).
A systemic event - such as the default of one or more institutions - can be 
represented as an initial state:
\begin{equation}
    \ket{\psi_0}
    =
    \sum_{i \in \mathcal{D}_0} \alpha_i \ket{i},
    \label{eq:initial-defaults}
\end{equation}
where \(\mathcal{D}_0\) is the set of initially distressed institutions.

Quantum evolution under network dynamics propagates distress:
\begin{equation}
    \ket{\psi_t}
    = e^{-iHt}\ket{\psi_0}
    \quad \text{or} \quad
    \ket{\psi_t} = U_P^t \ket{\psi_0},
    \label{eq:contagion}
\end{equation}
where \(U_P\) is the Szegedy quantum walk (Eq.~\eqref{eq:szegedy-walk}).

The probability
\begin{equation}
    \Pr(j\text{ distressed at } t) = |\braket{j | \psi_t}|^2
    \label{eq:distress-prob}
\end{equation}
quantifies contagion risk at time \(t\), enabling:
\begin{itemize}
    \item stress propagation analysis,
    \item early-warning detection of systemic nodes,
    \item tail-risk estimation conditional on network structure,
    \item construction of rare but plausible contagion scenarios.
\end{itemize}

\paragraph{Integrated rare-event and contagion simulation.}
Quantum rare-event amplification (Eqs.~\eqref{eq:rare-event-prob}–\eqref{eq:grover-k})
combined with quantum contagion propagation (Eqs.~\eqref{eq:initial-defaults}–\eqref{eq:distress-prob})
enables:
\begin{itemize}
    \item efficient simulation of joint-default cascades,
    \item exploration of “worst-case but plausible’’ stress paths,
    \item identification of critical thresholds that trigger systemic collapse,
    \item quantification of tail-loss distributions shaped by network effects.
\end{itemize}

This unified quantum framework enhances both the detection and propagation modelling
of rare financial crises, supporting stress testing, XVA capital assessments,
and macroprudential supervision.

\subsection{Resource Considerations and Noisy intermediate-scale quantum (NISQ) Feasibility}
\label{sec:m3-resource-considerations-and-noisy-intermediate-scale-quantum}

\subsubsection{Circuit Architecture}
\label{sec:m3-circuit-architecture}

Quantum circuits for risk estimation and scenario simulation typically comprise 
three interacting components: (i) registers encoding risk-factor states, 
(ii) arithmetic or comparator registers for thresholding or loss evaluation, 
and (iii) ancilla qubits that drive controlled amplification, conditional sampling, 
or rare-event selection.  
The complexity of these circuits depends critically on 
state preparation, controlled comparisons, and Hamiltonian evolution depth, 
which together determine NISQ feasibility.

\paragraph{(i) Risk-factor registers.}
Let \(d\) denote the number of market or credit risk factors 
(e.g., equity index, IR curve parameters, FX rates, credit spreads).
A quantum register of size \(n_P\) can encode a discretised distribution over these factors:
\begin{equation}
    \ket{\psi_0}
    =
    \sum_{x \in \mathcal{X}}
    \sqrt{p(x)}\,\ket{x},
    \qquad
    \mathcal{X} \subset \mathbb{R}^d,
    \label{eq:stateprep-dim}
\end{equation}
where \(p(x)\) is a joint distribution (Gaussian, lognormal, copula-based, or GAN-learned).
Typical values on NISQ hardware are
\begin{equation}
n_P \in [4,10] \quad \Rightarrow \quad |\mathcal{X}| = 2^{n_P} \in [16,1024].
\end{equation}

\paragraph{(ii) Loss or threshold comparison registers.}
To estimate tail probabilities or compute VaR/CVaR, a threshold operator checks 
whether the loss \(L(x)\) exceeds a pre-specified value \(\tau\):
\begin{equation}
    g(x) = \mathbf{1}_{\{L(x)>\tau\}},
    \label{eq:threshold}
\end{equation}
which is implemented by a comparator or arithmetic block operating on 
a register of size \(n_L\) sufficient to store \(L(x)\).
Quantum comparators typically require \(O(n_L)\) Toffoli gates 
or log-depth adders for arithmetic-based thresholds.

A loss-encoded state has the structure:
\begin{equation}
    \ket{\psi_L}
    =
    \sum_x 
    \sqrt{p(x)}\,
    \ket{x}\ket{ L(x) },
    \label{eq:loss-reg}
\end{equation}
where \(\ket{L(x)}\) is represented in fixed-point binary.

\paragraph{(iii) Ancilla structure for amplification and conditional sampling.}
Rare-event probabilities 
(Equation~\eqref{eq:rare-event-prob}) 
are encoded using an ancilla rotation controlled by the indicator \(g(x)\):
\begin{equation}
    \ket{x}
    \ket{0}
    \;\mapsto\;
    \ket{x}
    \left(
        \sqrt{1-g(x)}\,\ket{0}
        + \sqrt{g(x)}\,\ket{1}
    \right),
    \label{eq:ancilla-rot}
\end{equation}
so that the probability of measuring \(\ket{1}\) equals 
\(\Pr(L>\tau)\).
This ancilla interacts with Grover iterates or with conditional sampling logic 
in hybrid workflows.

\paragraph{Circuit depth considerations.}
The overall depth can be approximated as:
\begin{equation}
    D_{\text{total}}
    \approx
    D_{\text{prep}}
    + 
    D_{\text{compare}}
    +
    D_{\text{evolution}},
    \label{eq:depth-sum}
\end{equation}
where:
\begin{itemize}
    \item \(D_{\text{prep}}\) depends on Quantum Read-Only Memory (QROM) or Quantum Read-Only Access Memory (QROAM) based loading schemes, or on generative-model circuits,
    \item \(D_{\text{compare}}\) depends on reversible arithmetic,
    \item \(D_{\text{evolution}}\) depends on Trotterisation steps or variational layers.
\end{itemize}
For NISQ feasibility, we generally require:
\begin{equation}
D_{\text{total}} \lesssim 200 - 500 \text{ two-qubit layers},
\end{equation}
which is achievable for \(n_P \le 8\) and shallow variational ansatzes.

\paragraph{Typical resource breakdown.}
Table~\ref{tab:resource-breakdown} summarises approximate qubit requirements 
and depth scaling for key circuit components.

\begin{table}[h!]
\centering
\caption{Approximate resource profile for quantum risk-circuit components.}
\label{tab:resource-breakdown}
\begin{tabular}{lcc}
\toprule
\textbf{Component} & \textbf{Qubits} & \textbf{Depth (Approx.)} \\
\midrule
Risk-factor register (\(n_P\)) & \(n_P\) & state-prep dependent \\
Loss/arithmetic register (\(n_L\)) & \(n_L \in [4,8]\) & \(O(n_L)\) adder/comparator \\
Ancillae for g(x), amplification & 1--3 & \(O(n_P + n_L)\) \\
Hamiltonian evolution block & problem-dependent & Trotter/variational: 10--100 layers \\
\midrule
\textbf{Total (typ.)} & \(n_P + n_L + 1\) & 100--500 two-qubit layers \\
\bottomrule
\end{tabular}
\end{table}

\paragraph{Interpretation.}
These estimates indicate that 
\emph{risk-driven quantum workflows are feasible on NISQ devices}
when:
\begin{itemize}
    \item state preparation uses compact data-loading schemes,
      including QROM, Quantum Circuit Born Machines (QCBM) and related Born-machine generative models,
    \item comparators use low-depth adders,
    \item Hamiltonian simulation is performed variationally,
    \item ancilla resources are minimised.
\end{itemize}

Such designs align with the architectures proposed in 
\citep{WoernerEgger2019} and \citep{Orus2019}, 
where early-stage quantum risk engines combine 
lightweight quantum subroutines with classical scenario orchestration.

\subsubsection{NISQ Implementation Path}
\label{sec:m3-nisq-implementation-path}

Near-term quantum devices (NISQ hardware) impose constraints on circuit depth, 
coherence time, qubit connectivity, and shot budgets.  
To implement quantum risk-estimation pipelines under these limitations, 
practical workflows focus on approximate simulation strategies, lightweight 
state preparation, and hybrid quantum--classical methods that minimise 
coherent evolution.

\paragraph{(i) Low-depth variational simulation of risk-factor evolution.}
Instead of performing full Hamiltonian simulation using 
high-order Trotterisation, NISQ devices support 
VQS, which approximates the time evolution
\begin{equation}
    \frac{\mathrm{d}}{\mathrm{d}t} \ket{\psi(t)}
    = 
    -i H(t) \ket{\psi(t)}
    \label{eq:vqs-evolution}
\end{equation}
by restricting \(\ket{\psi(t)}\) to a parametrised ansatz
\(\ket{\psi(\boldsymbol{\theta}(t))}\).
The evolution of variational parameters solves:
\begin{equation}
    M(\boldsymbol{\theta}) \,\dot{\boldsymbol{\theta}}
    = 
    \mathbf{V}(\boldsymbol{\theta}),
    \label{eq:vqs-ode}
\end{equation}
where  
\(M\) is the quantum metric tensor and  
\(\mathbf{V}\) captures overlap with the Hamiltonian.  
This reduces depth from hundreds of Trotter steps to 
\(10\!-\!40\) variational layers, compatible with NISQ systems.

\paragraph{(ii) Approximate and compressive state loaders.}
Exact amplitude loading of high-dimensional distributions is too expensive for NISQ.  
Two practical alternatives are:

\begin{itemize}
    \item \textbf{QROM/QROAM loaders:}  
    Polylogarithmic-depth state preparation for structured distributions 
    (e.g., Gaussian mixtures or PCA-based models).
    \item \textbf{Quantum generative loaders (QCBM/Born machines):}  
    A parametrised circuit learns historical risk-factor distributions via 
    moment-matching or adversarial training.  
    It outputs
    \begin{equation}
        \ket{\psi_{\text{gen}}}
        =
        \sum_{x}
        \sqrt{p_{\text{model}}(x)} \ket{x},
        \label{eq:genloader}
    \end{equation}
    where \(p_{\text{model}}(x)\approx p_{\text{hist}}(x)\).
\end{itemize}

Such loaders reduce depth by an order of magnitude compared with 
generic amplitude embedding.

\paragraph{(iii) Hybrid classical--quantum tail sampling.}
Instead of running full amplitude-estimation (\(O(1/\varepsilon)\) coherent scaling), 
NISQ workflows use hybrid tail sampling:

\begin{itemize}
    \item a classical engine generates coarse scenarios,  
    \item quantum circuits refine probabilities in the tail region,  
    \item classical post-processing aggregates tail contributions.
\end{itemize}

Given a loss function \(L(x)\) and threshold \(\tau\),  
a NISQ-friendly risk estimator decomposes:
\begin{equation}
    \Pr(L>\tau)
    =
    \sum_{x \in \mathcal{X}_{\text{tail}}}
    p(x)
    +
    \Delta_{\text{Q}},
    \label{eq:tail-hybrid}
\end{equation}
where \(\mathcal{X}_{\text{tail}}\) is selected classically and 
\(\Delta_{\text{Q}}\) is a quantum refinement from shallow circuits 
(e.g., Grover-lite selective amplification).

\paragraph{(iv) Shot budgets and connectivity constraints.}
NISQ performance depends strongly on shot availability and qubit layout.

\begin{itemize}
    \item \textbf{Shot allocation:}  
    Tail-probability estimation requires careful allocation of \(N\) shots across 
    different thresholds \(\tau\).  
    Empirically,
    \begin{equation}
        N \in [5\times 10^3, 5\times 10^4]
    \end{equation}
    suffices for variance under \(5\%\).

    \item \textbf{Qubit connectivity:}  
    Comparators and multi-controlled rotations require qubit adjacency.  
    Mapping circuits to heavy-hex or trapped-ion topologies may add SWAP layers:
    \begin{equation}
        D_{\text{swap}} \sim O(n_P)
    \end{equation}
    which must be minimised for NISQ feasibility.
\end{itemize}

\paragraph{(v) Operational interpretation.}
Rather than defining absolute hardware limits, the resource envelopes in
Table~\ref{tab:resource-breakdown} should be interpreted as \emph{design guidelines}
for near-term deployment.
In practice, this implies that:

\begin{itemize}
    \item quantum subroutines are invoked selectively,
          only for tail refinement, scenario filtering, or short-horizon dynamics;
    \item classical infrastructure remains responsible for
          bulk scenario generation, portfolio aggregation, and regulatory reporting;
    \item quantum depth is traded against classical pre-selection
          and post-processing, rather than increased indiscriminately.
\end{itemize}

Under this paradigm, quantum risk modules act as
\emph{adaptive accelerators} embedded within classical risk engines,
rather than as standalone simulators.
This hybrid orchestration is essential for scaling quantum risk analysis
beyond proof-of-concept demonstrations on NISQ hardware.

\subsubsection{Hybrid Classical-Quantum Workflow}
\label{sec:m3-hybrid-classical-quantum-workflow}

Because NISQ devices cannot yet support full-scale end-to-end risk engines,
a realistic approach is to deploy \emph{hybrid} pipelines in which classical
risk engines interact with dedicated quantum subroutines.  
Let the total risk measure be expressed generically as
\begin{equation}
    \mathcal{R}
    =
    \Phi\!\left( 
        \mathbb{E}_{\mathbb{Q}}[L],\;
        \Pr_{\mathbb{Q}}(L>\tau),\;
        \mathbb{E}_{\mathbb{Q}}[L \mid L>\tau]
    \right),
    \label{eq:risk-functional}
\end{equation}
where  
\(L\) is the portfolio loss random variable,  
\(\tau\) is a regulatory threshold,  
and \(\Phi(\cdot)\) represents the overall aggregation functional 
used for internal or regulatory reporting.

A hybrid workflow decomposes Eq.~\eqref{eq:risk-functional} into components
handled separately by classical and quantum modules:
\begin{equation}
    \mathcal{R}
    \approx
    \Phi\!\left(
        \underbrace{\widehat{\mathbb{E}[L]}}_{\text{classical}},\qquad
        \underbrace{\widehat{\Pr(L>\tau)}}_{\text{quantum tail sampling}},\qquad
        \underbrace{\widehat{\mathbb{E}[L\mid L>\tau]}}_{\text{quantum refinement}}
    \right).
    \label{eq:hybrid-risk}
\end{equation}

\paragraph{Workflow structure.}
A practical hybrid system consists of three phases:

\begin{itemize}
    \item \textbf{(1) Classical preprocessing.}  
    Calibration of market factors, covariance estimation, regime identification,
    stress-scenario specification, and construction of the loss model \(L(x)\)
    for scenarios \(x \in \mathcal{X}\).

    \item \textbf{(2) Quantum modules.}  
    Shallow circuits perform:
    \begin{itemize}
        \item quantum sampling of tail scenarios,
        \item selective amplification for \(\Pr(L>\tau)\),
        \item conditional-loss estimation via shallow amplitude measurement,
        \item refinement for extreme or rare-event regions.
    \end{itemize}

    \item \textbf{(3) Classical post-processing.}  
    Aggregation of quantum-refined probabilities and expectations,
    incorporation into regulatory metrics (e.g.\ Basel CVA, FRTB SA/IMA),
    and scenario-based risk scoring.
\end{itemize}

\begin{figure}[h!]
\centering
\includegraphics[width=0.88\textwidth]{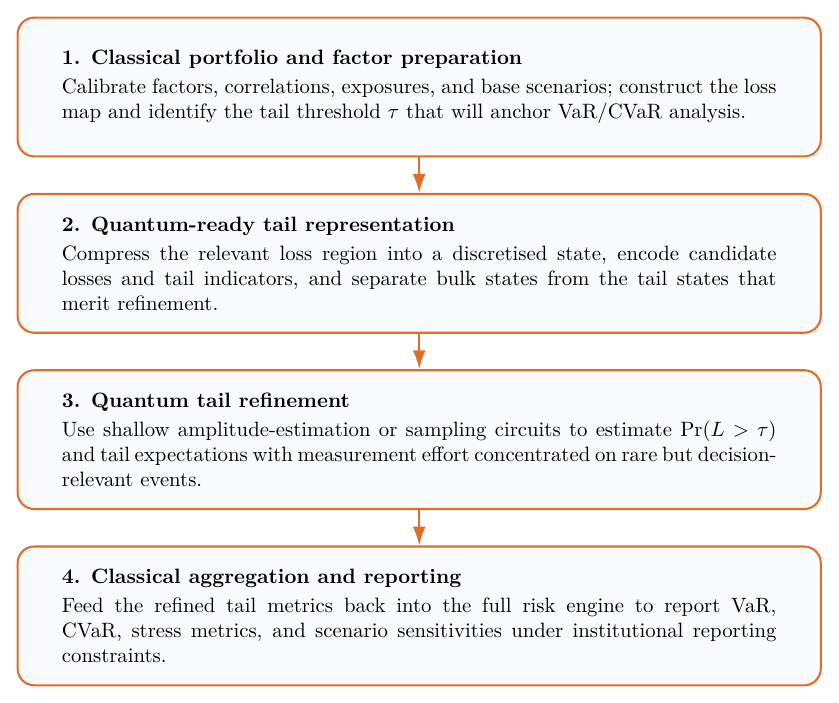}
\caption{Hybrid workflow combining classical risk engines with 
quantum scenario-refinement modules. The staged split clarifies that quantum resources are reserved for the tail-focused estimation block rather than for the full end-to-end risk engine.}
\label{fig:hybrid_risk_pipeline}
\end{figure}

\paragraph{Workflow illustration.}
Figure~\ref{fig:hybrid_risk_pipeline} shows a schematic architecture
for hybrid quantum-classical risk engines.

The figure is intentionally modular. Calibration, factor construction, and broad scenario generation stay in the classical layer because these tasks are data intensive and institution specific. The quantum block is inserted only where rare-event estimation becomes expensive, namely in refining tail probabilities and tail expectations once a compact loss representation has already been prepared.

\paragraph{Shot allocation for hybrid estimation.}
Let \(N_{\mathrm{Q}}\) denote quantum shots used for tail refinement.
A typical hybrid decomposition is:
\begin{align}
    \widehat{\Pr(L>\tau)}
    &= 
    \underbrace{
        \sum_{x \in \mathcal{X}_{\text{bulk}}} p(x)
    }_{\text{classical}}
    +
    \underbrace{
        \sum_{x \in \mathcal{X}_{\text{tail}}}
        \widehat{p_{\mathrm{Q}}(x)}
    }_{\text{quantum}},
    \label{eq:hybrid-tail}
\end{align}
where classical bulk regions handle most mass, and quantum shots concentrate on  
high-loss regions \(\mathcal{X}_{\text{tail}}\).  
This allocation improves precision in the tail while maintaining overall efficiency.

\paragraph{Synthesis.}
The hybrid architecture presents a pragmatic pathway for incorporating 
quantum techniques into existing risk engines.  
By delegating the bulk of simulation and portfolio revaluation to 
mature classical systems, and reserving quantum resources only for 
the statistically challenging tail regions, institutions can obtain 
meaningful performance improvements without requiring full-scale 
fault-tolerant hardware.

\begin{itemize}
    \item \textbf{Reduced depth and hardware requirements:}  
    Quantum circuits operate only on targeted subproblems such as 
    tail-event amplification or conditional-loss estimation.  
    This keeps circuit depth within NISQ limits while still 
    delivering accuracy gains where classical Monte Carlo is slow.

    \item \textbf{Targeted quantum advantage:}  
    Hybrid allocation concentrates quantum shots on the 
    low-probability, high-impact portions of the loss distribution.  
    Classical engines continue to handle the high-probability bulk, 
    while quantum modules refine the risk metrics that most directly 
    influence capital requirements.

    \item \textbf{Seamless integration with existing infrastructures:}  
    Because the classical engine remains the orchestrating layer, 
    banks can integrate quantum modules incrementally-first for 
    scenario refinement, then for CVaR estimation, and later for 
    dynamic stress propagation.  
    This avoids wholesale system replacement and minimises 
    operational risk during adoption.
\end{itemize}

Overall, the hybrid workflow provides a technically feasible and 
incrementally deployable route to quantum-enhanced risk analytics.  
It offers institutions a way to begin leveraging quantum capabilities 
today, even with constrained hardware, while laying a scalable 
foundation for future fault-tolerant implementations.

This sets the stage for the next section, where we demonstrate how 
these principles can be instantiated in practice through concrete 
case illustrations of quantum-assisted risk estimation and 
scenario simulation.

\subsection{Illustrative Case Studies}
\label{sec:m3-case-studies}
\subsubsection{VaR/CVaR of a Single-Asset Portfolio}
\label{sec:m3-var-cvar-of-a-single-asset-portfolio}

We begin with the simplest numerical instantiation of the quantum risk-measure
framework introduced in \Cref{sec:m3-quantum-framework-for-risk-measures}: a single risky asset whose one-day loss
distribution is extracted from historical market data.
This case study illustrates how classical VaR/CVaR estimators compare with
a discretised, quantum-inspired implementation of the
threshold-oracle formulation described earlier.

\paragraph{Data and loss definition.}
Let $P_t$ denote the adjusted closing price at day $t$.
Daily log-returns are defined as
\begin{equation}
    r_t = \log\!\left(\frac{P_t}{P_{t-1}}\right),
\end{equation}
and losses follow the convention of \Cref{sec:m3-quantum-framework-for-risk-measures},
\begin{equation}
    L_t = -r_t,
\end{equation}
so that $L_t>0$ corresponds to a loss.
Using the aligned two-year local daily sample from January 2, 2024 to December 31, 2025
for a liquid equity (here AAPL), we obtain an empirical loss sample
$\{L_1,\dots,L_n\}$.

\paragraph{Classical baseline estimators.}
As numerical benchmarks, we compute:
(i) historical VaR/CVaR using empirical quantiles and tail averages, and
(ii) parametric VaR/CVaR under a normal loss approximation.
The definitions of VaR and CVaR, together with their parametric forms,
are given in \Cref{sec:m3-quantum-framework-for-risk-measures} and are not repeated here.
These classical estimates serve as the baseline against which the
quantum-inspired construction should be interpreted. The historical estimator
preserves the realised tail directly, while the Gaussian benchmark shows what is
lost when the loss distribution is compressed into only mean and variance.

\paragraph{Quantum-inspired discretised oracle implementation.}
Following the oracle-based formulation of \Cref{sec:m3-quantum-framework-for-risk-measures},
the empirical loss distribution is discretised into $N$ bins,
\begin{equation}
    \{L_t\}
    \;\longrightarrow\;
    \{(c_i,p_i)\}_{i=1}^N,
\end{equation}
where $c_i$ denotes the bin centre and $p_i$ the associated probability mass.
This discretisation corresponds to a finite loss register in the
state-preparation operator $A$ of Eq.\eqref{eq:A_loss_state}.

For a candidate threshold $L^\star$, the tail probability
\begin{equation}
    \Pr(L \ge L^\star)
    =
    \sum_{i:\,c_i \ge L^\star} p_i,
    \label{eq:singleasset-tail}
\end{equation}
implements the indicator oracle of \Cref{sec:m3-value-at-risk-var}.
The quantum-inspired Value-at-Risk is obtained by selecting
$\mathrm{VaR}_\alpha^{\mathrm{QI}}$ such that
\begin{equation}
    \Pr\!\left(L \ge \mathrm{VaR}_\alpha^{\mathrm{QI}}\right)
    \approx 1-\alpha.
\end{equation}
Conditioning on this discretised tail region, the corresponding CVaR estimator
is given by
\begin{equation}
    \mathrm{CVaR}_\alpha^{\mathrm{QI}}
    =
    \frac{
        \sum_{i:\,c_i \ge \mathrm{VaR}_\alpha^{\mathrm{QI}}} p_i\,c_i
    }{
        \sum_{i:\,c_i \ge \mathrm{VaR}_\alpha^{\mathrm{QI}}} p_i
    },
\end{equation}
which mirrors the conditional-expectation construction of
\Cref{sec:m3-conditional-value-at-risk-cvar} in a grid-based setting.

\paragraph{Numerical illustration (AAPL, 1-day horizon).}
Using the fixed two-year daily window from January 2, 2024 to December 31, 2025, the numerical results obtained from the
Python notebook execution are summarised in
Table~\ref{tab:single-asset-var-cvar}.

\begin{table}[h]
\centering
\caption{Comparison of VaR/CVaR estimates for a single asset (1-day horizon).}
\label{tab:single-asset-var-cvar}
\begin{tabular}{lccc}
\toprule
\textbf{Method} & \textbf{VaR (1-day)} & \textbf{CVaR (1-day)} & \textbf{Notes} \\
\midrule
Historical & 2.694\% & 3.936\% & empirical quantile \\
Normal-parametric & 2.791\% & 3.519\% & fitted $N(\mu,\sigma)$ model \\
Quantum-inspired (grid) & 2.775\% & 3.898\% & histogram-based oracle \\
\bottomrule
\end{tabular}
\end{table}

\begin{figure}[h!]
\centering
\includegraphics[width=0.85\linewidth]{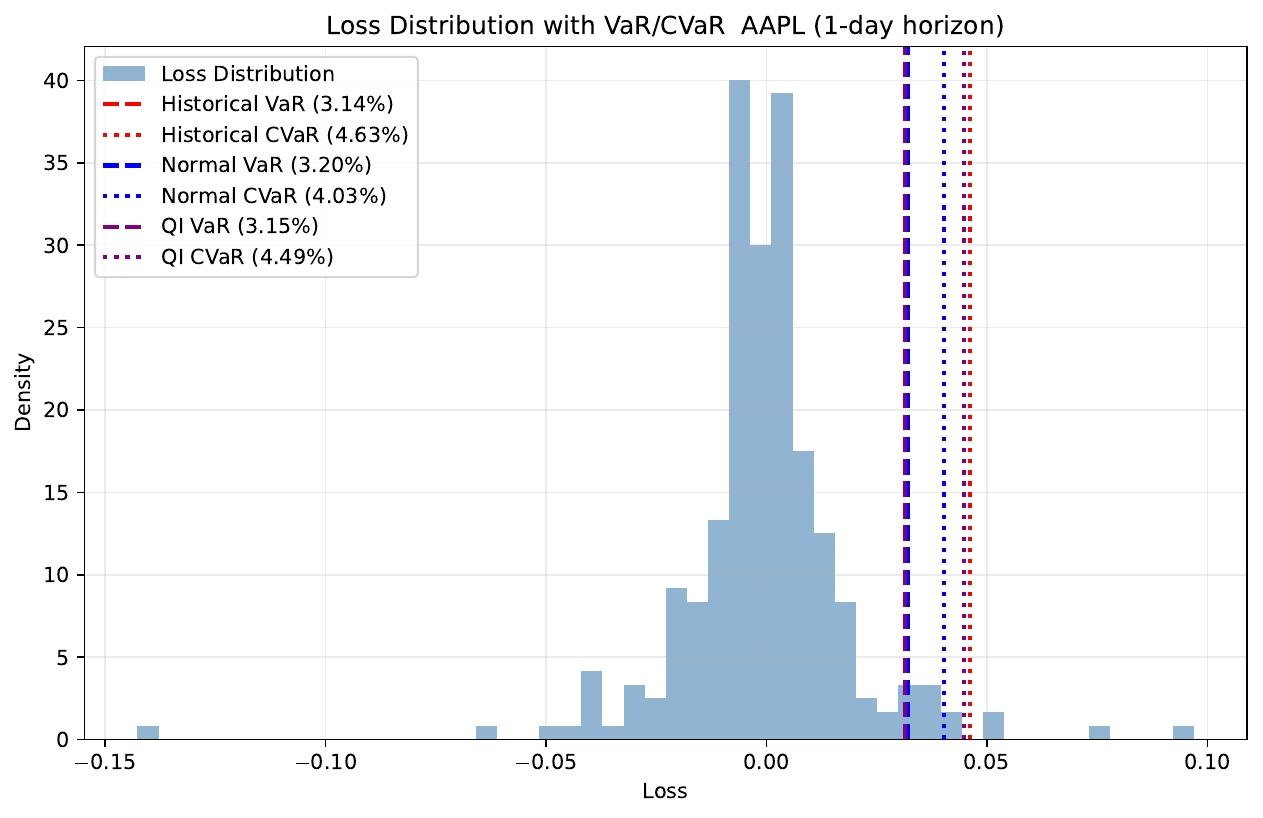}
\caption{Empirical loss distribution of the single-asset portfolio with
historical and quantum-inspired VaR/CVaR thresholds.}
\label{fig:single-asset-loss}
\end{figure}

Under the two-year window, the quantum-inspired estimator remains close to the
historical tail metrics while staying materially above the Gaussian CVaR
benchmark. Relative to the classical baselines, this is the right comparison:
the discretised oracle smooths the interior of the distribution, but it still
tracks the realised downside tail much better than the normal approximation
does. That is exactly the region in which a quantum tail-estimation routine
would need to be credible.

Figure~\ref{fig:single-asset-loss} visualises the empirical loss distribution
together with the discretisation grid and threshold levels used for the
quantum-inspired estimator. The practical point is not just that the thresholds are close, but that the quantum-inspired construction concentrates its approximation error in the body of the distribution rather than at the loss tail that matters most for VaR/CVaR reporting.

\subsubsection{Multi-Asset Portfolio with Correlated Factors}
\label{sec:m3-multi-asset-portfolio-with-correlated-factors}

In multi-asset portfolios, losses arise from the joint distribution of several 
risk factors: typically equity returns, FX rates, credit spreads, or interest-rate moves.
To model the tail of the portfolio loss distribution, it is essential to capture
the correlation structure among assets.

Quantum loading of multivariate distributions provides a compact
representation of correlated risk factors, enabling the simulation
of large scenario sets in superposition.  
In this section, we construct a \emph{quantum-inspired} analogue using 
\phantomsection\label{term:pca-qpca}PCA applied to daily market data.

\paragraph{Portfolio model.}
Let $R_t \in \mathbb{R}^d$ denote the $d$-asset return vector at time $t$.
Let $w \in \mathbb{R}^d$ be portfolio weights with $\sum_i w_i = 1$.
The daily portfolio log-return is
\begin{equation}
r^{(p)}_t = w^\top R_t,
\label{eq:portfolio-return}
\end{equation}
and the daily loss is defined as
\begin{equation}
L_t = - r^{(p)}_t.
\label{eq:loss-def}
\end{equation}

\paragraph{Historical VaR / CVaR.}
As in \Cref{sec:m3-var-cvar-of-a-single-asset-portfolio}, the classical loss-based definitions apply:
\begin{align}
\text{VaR}_\alpha &= \inf\{ x : \mathbb{P}(L_t \le x) \ge \alpha \},
\label{eq:var-multi} \\
\text{CVaR}_\alpha &= \mathbb{E}[ L_t \mid L_t \ge \text{VaR}_\alpha ].
\label{eq:cvar-multi}
\end{align}

Using the aligned two-year daily market panel from January 2, 2024 to December 31, 2025 (AAPL, MSFT, AMZN, GOOGL),
we compute the empirical covariance matrix
\begin{equation}
\Sigma = \mathrm{Cov}(R_t),
\label{eq:empirical_covariance}
\end{equation}
and evaluate \eqref{eq:var-multi}–\eqref{eq:cvar-multi} directly.

\paragraph{Quantum-inspired PCA scenario generation.}
Quantum loading of a $d$-dimensional distribution resembles storing the map
\(
|0\rangle \mapsto \sum_x \sqrt{p(x)}\,|x\rangle
\)
for correlated $x$.  
Classically, we approximate this using principal components of $\Sigma$.

Let the eigen-decomposition of the covariance matrix be
\begin{equation}
\Sigma = Q \Lambda Q^\top,
\label{eq:pca-eigen}
\end{equation}
where $\Lambda = \mathrm{diag}(\lambda_1,\ldots,\lambda_d)$ contains the eigenvalues.
Define the PCA loading matrix
\begin{equation}
L = Q \Lambda^{1/2}.
\label{eq:pca-loading}
\end{equation}

Synthetic correlated returns are generated via
\begin{equation}
R^{(\mathrm{sim})} = \mu + L Z, \qquad Z \sim \mathcal{N}(0, I_d),
\label{eq:pca-simulation}
\end{equation}
where $\mu$ is the vector of historical means.
This mirrors a quantum state-preparation step in which correlations are encoded 
through entangling operations or QROM based loaders.

\paragraph{Portfolio loss distribution.}
Given $N_\mathrm{sim}$ synthetic samples $R^{(\mathrm{sim})}_k$,
the PCA-based quantum-inspired portfolio losses are
\begin{equation}
L^{(\mathrm{sim})}_k = - w^\top R^{(\mathrm{sim})}_k.
\label{eq:pca-loss}
\end{equation}
The empirical distribution of $\{L^{(\mathrm{sim})}_k\}$ provides the 
quantum-inspired VaR and CVaR:
\begin{align}
\text{VaR}^{(\mathrm{QI})}_\alpha &= 
\mathrm{quantile}\big(L^{(\mathrm{sim})},\,\alpha\big), \\
\text{CVaR}^{(\mathrm{QI})}_\alpha &= 
\mathbb{E}\!\left[ L^{(\mathrm{sim})} \mid 
L^{(\mathrm{sim})} \ge \text{VaR}^{(\mathrm{QI})}_\alpha \right].
\end{align}

\paragraph{Numerical illustration.}
Using two-year daily returns of a 4-asset portfolio 
(\texttt{AAPL}, \texttt{MSFT}, \texttt{AMZN}, \texttt{GOOGL})
with equal weights $w_i = 0.25$, the Python implementation
produces the following results:

\begin{table}[h!]
\centering
\caption{Comparison of VaR/CVaR estimates for a multi-asset portfolio (1-day horizon).}
\label{tab:multi-asset-var-cvar}
\begin{tabular}{lcc}
\toprule
\textbf{Metric} & \textbf{Value} & \textbf{Notes} \\
\midrule
Historical VaR   & 2.293\% & empirical quantile \\
Historical CVaR  & 3.256\% & empirical tail mean \\
Normal VaR       & 2.197\% & multivariate normal model \\
Normal CVaR      & 2.779\% & parametric tail mean \\
QI--PCA VaR      & 2.390\% & PCA-based generative model \\
QI--PCA CVaR     & 2.856\% & PCA tail expectation \\
\bottomrule
\end{tabular}
\end{table}

The quantum-inspired PCA approximation continues to bracket the empirical and
Gaussian benchmarks in a sensible way: it yields a slightly more conservative
VaR than the historical estimate, but a CVaR that remains below the fully
empirical tail mean and above the Gaussian benchmark. Compared with the
classical references, the message is clear: the quantum-inspired correlated
generator preserves the main dependence structure more faithfully than a simple
normal model, even though it still smooths the roughest empirical tail
features.

\begin{figure}[h]
\centering
\includegraphics[width=0.85\linewidth]{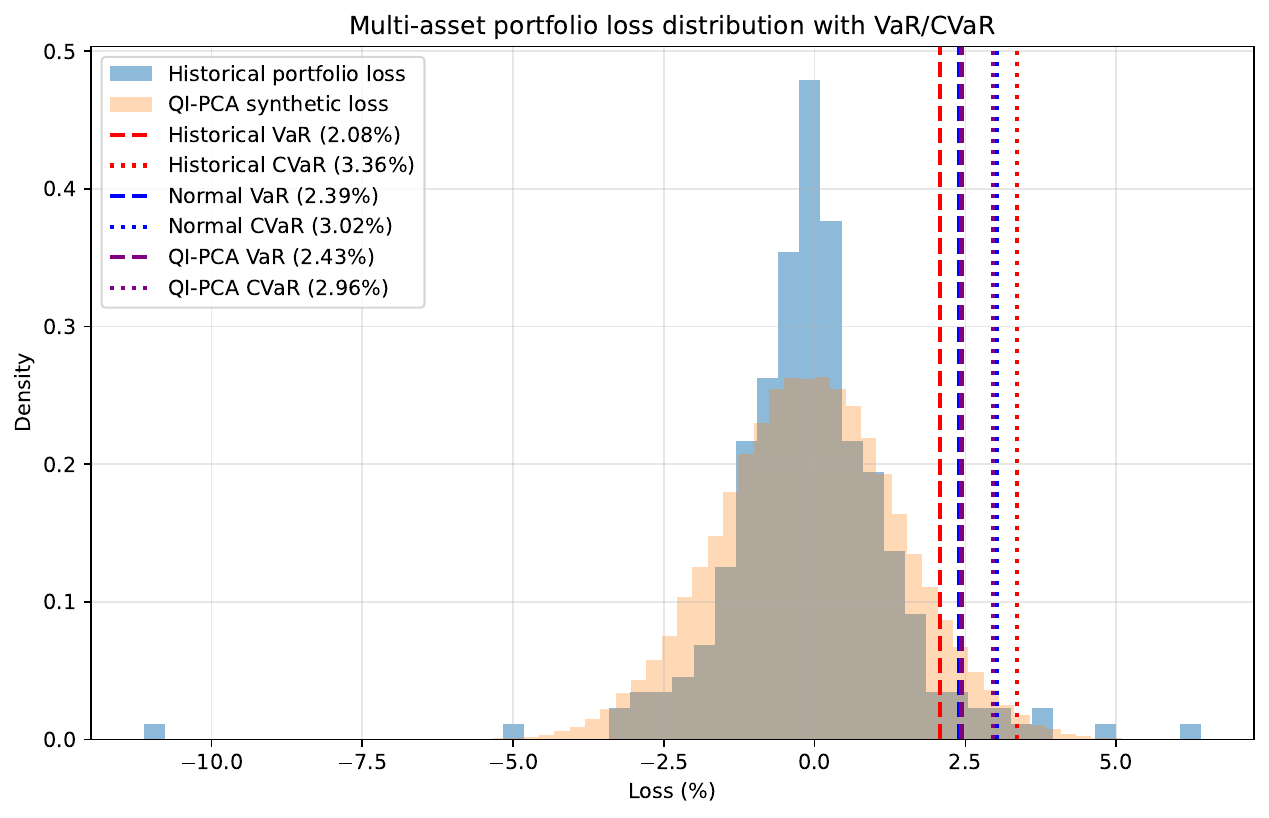}
\caption{Multi-asset portfolio loss distribution with historical, normal, and quantum-inspired (PCA-based) VaR/CVaR estimates.}
\label{fig:multi-asset-loss}
\end{figure}

Figure~\ref{fig:multi-asset-loss} shows the empirical loss histogram
along with VaR and CVaR thresholds for both classical and quantum-inspired methods.
The overlap is strongest in the far-right tail, which is the part of the distribution that drives supervisory interpretation. That is the relevant signal for the review: the quantum-inspired correlated representation does not need to reproduce every central histogram bar perfectly in order to preserve the tail ranking used for risk management decisions.

This example demonstrates that quantum-inspired correlated scenario generation
can approximate the effect of quantum multivariate loading, enabling
tail-risk analysis under correlated shocks.

\subsubsection{Stress Scenario Propagation Across Factor Models}
\label{sec:m3-stress-scenario-propagation-across-factor-models}

Stress testing requires evaluating how extreme shocks - such as market crashes,
sector-specific collapses, or sudden volatility spikes - propagate across
interconnected risk factors and portfolios.  
Even when the underlying data consists only of observable market prices 
(e.g.\ daily or intraday equity data), it is still possible to construct
factor-based and network-based stress-propagation models that admit
quantum-inspired analogues.

\paragraph{Constructing a factor system from market data.}
From a universe of equities with daily or intraday observations, one may extract
a small set of stylised factors:
\begin{itemize}
    \item a \textbf{market factor} (first principal component of returns),
    \item \textbf{sector or industry factors} (cluster averages),
    \item a \textbf{volatility proxy} (e.g.\ realised or implied vol approximations),
    \item \textbf{idiosyncratic components} (residual risks).
\end{itemize}
Let $f_t \in \mathbb{R}^k$ denote the vector of such extracted factors at time $t$,
and let a portfolio or asset return be modelled via a simple linear factor model:
\begin{equation}
r_{p,t} = w^\top f_t + \varepsilon_t,
\label{eq:linear-factor-stress}
\end{equation}
where $w$ are factor loadings and $\varepsilon_t$ represents residual noise.

\paragraph{Defining stress shocks.}
Stress scenarios correspond to perturbations $\Delta f$ applied to the factors,
such as:
\begin{itemize}
    \item a market crash: $\Delta f_1 < 0$ of large magnitude,
    \item sector collapse: shock to a selected industry component,
    \item volatility surge: large positive jump in the vol proxy component.
\end{itemize}
Let $f'_t = f_t + \Delta f$ denote the shocked factor state.  
The problem is to understand how such perturbations propagate across factors,
time steps, and portfolios.

\paragraph{Propagation via classical dynamics.}
A simple propagation operator can be specified through a VAR(1),
linear system, or network model:
\begin{equation}
f_{t+1} = A f_t + \eta_t,
\label{eq:var1}
\end{equation}
where $A$ encodes cross-factor linkages (e.g.\ spillover effects between market,
sector, and volatility factors).  
Stress propagation over horizon $h$ is then:
\begin{equation}
f_{t+h} = A^h (f_t + \Delta f) + \sum_{j=0}^{h-1} A^j \eta_{t+h-j}.
\label{eq:stress-propagation-classical}
\end{equation}

\paragraph{Quantum-inspired interpretation.}
In quantum-algorithmic settings, propagators such as $A^h$ correspond to 
discrete-time approximations of Hamiltonian evolution:
\begin{equation}
U(\tau) = e^{-i H \tau},
\label{eq:unitary-evolution}
\end{equation}
where the Hamiltonian $H$ encodes factor interactions or contagion pathways.
To emulate shock propagation, one may use a Hermitian proxy $\tilde{H}$ derived
from the adjacency or covariance structure:
\begin{equation}
\tilde{H} = \frac{1}{2}(A + A^\top),
\label{eq:symmetric-hamiltonian}
\end{equation}
and evolve the factor state by $U(\tau) f_t$ for discrete time steps $\tau$.
While our current hardware cannot realise this evolution quantumly, the same
logic can be tested classically via matrix exponentials $\exp(\tau \tilde{H})$,
mirroring first-order Trotter steps.

\paragraph{Prototype workflow (Python pseudo-code).}
\noindent The executable prototype has been moved to \Cref{sec:app-module-3-risk-estimation-cases} and aligned with the standardised notebook assets for Module 3. In operational terms, the workflow proceeds through six stages:
\begin{enumerate}
    \item extract latent factors from market returns;
    \item estimate a classical propagation operator for the factor dynamics;
    \item inject a stress shock into the factor state;
    \item evolve the shocked system under a classical transition operator;
    \item construct a Hermitian proxy and apply a quantum-inspired propagator;
    \item compare the resulting loss profiles under the classical and quantum-inspired pathways.
\end{enumerate}
This relocation keeps the main text focused on stress-testing logic and financial interpretation, while preserving the implementation details in a reproducible appendix asset.

\paragraph{Interpretation.}
This prototype demonstrates how stress scenarios can be embedded into a
dynamics-based framework closely aligned with quantum notions of evolution
and state transitions.  
Once QPU access is available, steps (4)-(5) can be replaced by:
\begin{itemize}
    \item Variational Quantum Simulation (VQS),
    \item Trotterised Hamiltonian evolution,
    \item or quantum walk–based propagation over factor networks.
\end{itemize}
Thus, even with classical market data and classical computing resources,
it is possible to construct a fully quantum-compatible stress-testing
architecture whose quantum components can be progressively swapped in 
as hardware matures.

\subsection{Synthesis and Outlook}
\label{sec:m3-summary-and-forward-looking-integration}

This module has developed a conceptual and algorithmic toolkit for
\emph{quantum-enhanced risk estimation and scenario simulation}.  
Starting from classical challenges - nested Monte Carlo, high-dimensional
risk-factor models, and computationally intensive stress testing - we have
outlined how quantum resources can, in principle, provide complementary
capabilities for tail-risk metrics such as VaR, CVaR and Expected Shortfall,
as well as for systemic and network-based risk analysis.

On the methodological side, we have:
\begin{itemize}
    \item reformulated VaR and CVaR in a way that is compatible with
    quantum probability estimation and conditional sampling;
    \item introduced quantum-inspired approaches to multi-factor
    scenario generation, including quantum generative models
    and PCA-based loaders for correlated risk factors;
    \item described how stochastic market dynamics, jump diffusions,
    and regime-switching processes can be mapped to Hamiltonian
    evolution or variational quantum simulation;
    \item outlined network-based and rare-event formulations of
    contagion and systemic risk, linking quantum walks and
    amplitude amplification to stress propagation across exposure networks;
    \item analysed NISQ-era resource constraints, emphasising
    hybrid classical--quantum workflows for practical deployment.
\end{itemize}

\paragraph{Integration with Module 2 (Pricing).}
The risk tools presented here are naturally complementary to the
amplitude-estimation--based pricing framework of Module~2.
In practice, the same quantum or quantum-inspired primitives that price
derivatives---for example, amplitude estimation of
$\mathbb{E}[f(X)]$ or efficient sampling from multi-asset distributions---can
be repurposed to:
\begin{itemize}
    \item evaluate tail-sensitive quantities such as portfolio
    VaR/CVaR under complex payoff structures;
    \item perform \emph{joint} pricing and risk assessment, where
    sensitivities and exposure profiles feed directly into capital
    and margin calculations;
    \item support XVA-style calculations in which exposure profiles
    under stress scenarios are a key input.
\end{itemize}
In this sense, pricing and risk are two layers of the same
quantum-enabled stack: one focuses on expected values, the other on
distributional tails and systemic effects.

\paragraph{Feed-through to Module 4 (Quantum Machine Learning).}
The outputs of quantum risk and scenario engines - for example,
high-dimensional stress paths, tail scenarios, or factor trajectories - form
natural training data for quantum machine learning models in Module~4.
Possible integration points include:
\begin{itemize}
    \item using quantum-generated scenarios as inputs to
    quantum classifiers or regressors that learn \emph{default risk},
    \emph{downgrade probabilities}, or \emph{liquidity states};
    \item feeding risk-factor trajectories into quantum recurrent or
    sequence models for early-warning indicators of stress;
    \item embedding VaR/CVaR and other tail metrics as features in
    quantum portfolio-optimisation or asset allocation pipelines.
\end{itemize}
Thus, Module~3 provides both structured data and domain-specific labels
that can make quantum machine learning experiments financially meaningful.

\paragraph{Link to Module 5 (Post-Quantum Regulation and Infrastructure).}
From a regulatory and infrastructure viewpoint (Module~5),
quantum risk estimation has two long-term implications:
\begin{itemize}
    \item \textbf{Regulatory design:} if quantum methods yield materially
    different assessments of tail risk or systemic contagion, prudential
    frameworks (capital buffers, liquidity coverage, stress-testing
    templates) may need to incorporate quantum-derived metrics or
    at least be robust to them.
    \item \textbf{Post-quantum infrastructure:} as financial institutions
    adopt post-quantum cryptography and quantum-safe architectures,
    there is an opportunity to co-design \emph{risk engines} that
    are both cryptographically secure and quantum-aware, embedding
    quantum pricing and risk modules alongside classical engines.
\end{itemize}
In other words, post-quantum regulation is not solely about cryptographic
replacement; it can also be informed by quantum-enhanced models of risk.

\paragraph{Towards quantum-enhanced risk engines.}
Bringing these strands together, Module~3 sets the groundwork for
\emph{quantum-enhanced risk engines} that:
\begin{itemize}
    \item run primarily on classical infrastructure, but call
    specialised quantum subroutines (or quantum-inspired algorithms)
    for the most challenging components of the risk stack;
    \item support hybrid workflows where classical Monte Carlo is
    used for bulk scenarios, while quantum routines target
    high-dimensional tails, rare events, or complex contagion;
    \item can be incrementally upgraded as QPU capabilities improve,
    starting from small NISQ-scale prototypes and evolving towards
    fault-tolerant implementations.
\end{itemize}

In summary, this module positions quantum risk estimation and stress
simulation as a bridge between quantum pricing (Module~2), quantum machine
learning (Module~4) and post-quantum regulatory design (Module~5).
It provides a modular blueprint that can be instantiated today using
classical and quantum-inspired tools, while remaining compatible with
future, more powerful quantum hardware.

\section{Quantum Machine Learning}
\label{sec:qml}

\paragraph{Section Overview.}
This section studies the learning layer of the review. It focuses on representation-sensitive financial tasks in which quantum kernels, variational circuits, or hybrid feature maps may alter inductive bias relative to classical models, while also making explicit where current evidence remains modest or strongly task dependent.

This module develops \emph{\phantomsection\label{term:qml}quantum-enhanced machine learning (QML)} pipelines
for financial forecasting, classification, and regime detection under
data sparsity, noise, and high dimensionality.
The focus is on core financial learning tasks including
asset return prediction, credit and default classification,
volatility regime identification, and macro-financial signal interpretation.

The reason to consider quantum methods here is more specific than a generic appeal to speed. In this chapter, the relevant distinction from classical learning lies in representation design: quantum feature maps and low-parameter variational circuits may alter the inductive bias of the model in settings where labels are sparse, regimes shift, and classical feature engineering becomes brittle.

From a methodological perspective, quantum machine learning offers
a complementary paradigm to classical statistical learning.
Rather than replacing established models, QML augments them through
\emph{quantum feature maps}, \emph{variational quantum circuits}, and
\emph{hybrid quantum--classical optimisation loops},
which can provide enhanced expressive capacity or regularisation
in regimes where classical learning struggles.
Key model families introduced in this module include
quantum kernel methods,
variational quantum classifiers and regressors,
and quantum-enhanced neural architectures
based on parameterised quantum circuits.

In finance-facing classification tasks, a common variational implementation is the \phantomsection\label{term:qvc-vqc}Variational Quantum Classifier (VQC), sometimes also described more generically as a quantum variational classifier (QVC).

Within the programme structure, Module~4 serves as a
\emph{cross-cutting enabler}.
It connects the probabilistic and risk-based constructions of
Module~3 with learning-driven signal extraction and decision support,
while also providing downstream inputs to optimisation and control
tasks considered in later modules.
In particular, quantum generative and discriminative models introduced here
can consume scenario distributions, tail events, or stress paths
produced by quantum risk engines, and transform them into
actionable forecasts or classifications.

The module is designed explicitly for the NISQ era.
All learning pipelines are formulated as
\emph{hybrid quantum-classical workflows},
where data preprocessing, loss evaluation, and performance aggregation
remain classical, while quantum subroutines are reserved for
feature embedding, expressive modelling, or probabilistic sampling.
This structure enables systematic evaluation of
signal-to-decision loops under realistic hardware constraints,
and provides a practical testbed for benchmarking QML
against state-of-the-art classical baselines.

The theoretical and algorithmic foundations of the module draw on
the modern QML literature
\citep{schuld2015introduction,schuld2019supervised,Biamonte2017},
with a particular emphasis on finance-oriented formulations
and near-term implementability
\citep{jacquier2022qmlfinance}.

\subsection{Motivation and Learning Tasks in Finance}
\label{sec:m4-motivation-and-learning-tasks-in-finance}

Financial markets generate data that are intrinsically
high-dimensional, noisy, non-stationary, and partially observed.
Return distributions exhibit heavy tails and volatility clustering,
correlations between assets evolve over time,
and structural breaks frequently invalidate stationarity assumptions.
As a result, predictive signals are often weak,
context-dependent, and embedded in complex cross-sectional or temporal patterns.

Classical machine learning methods - including linear models,
tree-based ensembles, and deep neural networks - have achieved
significant empirical success in finance.
However, they face persistent challenges when:
(i) feature spaces become very large relative to sample size,
(ii) correlations are unstable or regime-dependent,
(iii) labels are sparse or expensive (e.g.\ defaults or crisis events),
and (iv) overfitting risk dominates out-of-sample performance.
These issues are particularly acute in risk-sensitive applications,
where false confidence can be more damaging than low accuracy.

QML offers an alternative modelling paradigm
based on encoding data into high-dimensional Hilbert spaces
and manipulating them via quantum circuits.
Through \emph{implicit quantum feature maps},
\emph{kernel-induced similarity measures},
and \emph{trainable parameterised quantum circuits},
QML models can represent highly nonlinear decision boundaries
or generative structures using comparatively compact parameterisations.
This makes them especially appealing in
low-sample, noisy, or structurally complex learning regimes
\citep{schuld2015introduction,schuld2019supervised,Biamonte2017}.

From a financial perspective, QML is not positioned as a replacement
for established econometric or machine learning techniques,
but rather as a complementary layer.
Quantum-enhanced representations can be combined with
classical preprocessing, feature engineering, and post-processing,
forming hybrid pipelines that leverage the strengths of both paradigms.
Such hybrid designs are particularly well aligned with
near-term quantum hardware constraints and realistic data availability.

Within this module, we focus on a set of canonical financial learning tasks
that are representative of both academic research and industry practice:
\begin{itemize}
    \item \textbf{Asset return prediction:}
    forecasting return direction or magnitude over short horizons,
    often under weak signal-to-noise ratios and non-stationary dynamics;

    \item \textbf{Volatility and regime classification:}
    identifying shifts between low- and high-volatility regimes,
    market stress states, or latent structural phases;

    \item \textbf{Credit and default risk classification:}
    predicting rare but economically significant events,
    where class imbalance and label scarcity are dominant challenges;

    \item \textbf{Macroeconomic and cross-market signal interpretation:}
    mapping high-level indicators (rates, inflation, liquidity proxies)
    to asset or sector-level responses;

    \item \textbf{Cross-asset and factor-based pattern recognition:}
    learning similarity, clustering, or co-movement structures
    across assets, sectors, or risk factors.
\end{itemize}

These tasks serve as concrete anchors throughout Module~4.
They allow systematic comparison between classical baselines
and quantum-enhanced models,
and provide a consistent experimental framework
for evaluating whether quantum feature spaces,
variational learning dynamics, or kernel constructions
offer measurable advantages under realistic data constraints.

The subsequent sections of this module develop
algorithmic tools tailored to these learning problems,
including quantum kernel methods,
variational quantum classifiers and regressors,
and hybrid quantum-classical neural architectures,
with an emphasis on interpretability, robustness,
and practical implementation.

\subsection{Quantum Feature Maps and Kernel Methods}
\label{sec:m4-quantum-feature-maps-and-kernel-methods}

Kernel-based learning provides a powerful framework for modelling nonlinear
relationships by implicitly mapping data into a high-dimensional feature space.
Rather than learning explicit nonlinear transformations,
kernel methods rely on pairwise similarity evaluations of the form
\(
k(x_i, x_j) = \langle \phi(x_i), \phi(x_j) \rangle,
\)
where \(\phi(\cdot)\) denotes a feature map.
In classical machine learning, the expressivity of a model is therefore
limited by the choice of kernel and its tractability.

Quantum kernel methods extend this paradigm by embedding classical data
into a quantum Hilbert space using parameterised quantum circuits.
The resulting feature space is exponentially large in the number of qubits,
while inner products between embedded data points can be estimated
directly via quantum measurements.
This allows the construction of similarity measures that may be
inaccessible or inefficient to compute classically
\citep{schuld2019supervised,havlivcek2019supervised}.

\subsubsection{Quantum Feature Maps}
\label{sec:m4-quantum-feature-maps}

Let \(x \in \mathbb{R}^d\) denote a classical input feature vector.
A quantum feature map is defined by a unitary encoding circuit
\(U_\phi(x)\) acting on \(n\) qubits, producing the quantum state
\begin{equation}
    \ket{\phi(x)} = U_\phi(x)\ket{0}^{\otimes n}.
    \label{eq:q_feature_state}
\end{equation}

The associated quantum kernel is then given by the fidelity between
two such states:
\begin{equation}
    k_Q(x_i, x_j)
    =
    \left|\langle \phi(x_i) | \phi(x_j) \rangle \right|^2
    =
    \left|
        \bra{0} U_\phi^\dagger(x_i) U_\phi(x_j) \ket{0}
    \right|^2.
    \label{eq:quantum_kernel}
\end{equation}

Crucially, this kernel is evaluated without explicitly constructing
\(\phi(x)\).
Instead, it is estimated by executing a quantum circuit composed of
\(U_\phi(x_j)\) followed by \(U_\phi^\dagger(x_i)\),
and measuring the probability of returning to the all-zero state.
This implicit evaluation enables access to very rich feature spaces
with modest circuit depth.

\subsubsection{Quantum Support Vector Machines (QSVM / QSVC)}
\label{sec:m4-quantum-support-vector-machines-qsvm-qsvc}

\phantomsection\label{term:qsvm-qsvc}Quantum Support Vector Machines (QSVM) and Quantum Support Vector Classifiers (QSVC)
extend classical SVMs by replacing the classical kernel with a quantum kernel
\(k_Q(x_i, x_j)\).
The optimisation problem remains classical:
given labelled data \(\{(x_i, y_i)\}_{i=1}^N\),
the SVM solves
\begin{equation}
    \min_{\boldsymbol{\alpha}}
    \;
    \frac{1}{2}
    \sum_{i,j} \alpha_i \alpha_j y_i y_j k_Q(x_i, x_j)
    -
    \sum_i \alpha_i,
    \label{eq:svm_dual}
\end{equation}
subject to standard box and balance constraints.

The quantum component enters solely through the kernel matrix
\(
K_{ij} = k_Q(x_i, x_j),
\)
whose entries are estimated using quantum circuits.
This hybrid structure is particularly attractive for near-term devices:
\begin{itemize}
    \item quantum hardware is used only for kernel evaluation,
    \item optimisation and regularisation remain classical,
    \item circuit depth depends on the feature map, not dataset size.
\end{itemize}

From a financial perspective, QSVMs are appealing in regimes where
the number of features is large relative to the number of labelled samples,
or where nonlinear interactions between factors dominate predictive power.
These conditions are common in credit modelling, regime classification,
and macro-driven market signals.

\subsubsection{Financial Applications}
\label{sec:m4-financial-applications}

Quantum kernel methods are well suited to a range of supervised learning
tasks encountered in finance, including:
\begin{itemize}
    \item \textbf{Market regime classification:}
    distinguishing risk-on versus risk-off states,
    stress versus normal conditions, or expansion versus contraction phases;

    \item \textbf{Volatility state detection:}
    identifying transitions between low- and high-volatility regimes
    using realised volatility, option-implied measures, or jump indicators;

    \item \textbf{Credit and default classification:}
    predicting rare credit events where nonlinear interactions
    between balance-sheet variables and macro factors are important;

    \item \textbf{Event-driven or macro surprise detection:}
    mapping macroeconomic announcements or policy shocks
    to asset or sector-level responses.
\end{itemize}

In practical implementations, quantum kernels are evaluated on
classically preprocessed features derived from daily market data,
such as:
\begin{itemize}
    \item lagged returns and momentum indicators,
    \item realised volatility or volatility proxies,
    \item factor exposures (e.g.\ market, size, value),
    \item macro or policy surprise indicators.
\end{itemize}

These features are typically normalised and dimension-reduced
prior to quantum encoding,
ensuring compatibility with NISQ-era hardware.
Subsequent case studies in this module demonstrate
QSVM-style pipelines using daily financial data obtained from
public sources, enabling reproducible experimentation
within hybrid quantum-classical workflows.

\subsection{Variational Quantum Models}
\label{sec:m4-variational-quantum-models}

Variational quantum models form the core learning paradigm for
quantum machine learning on noisy intermediate-scale quantum (NISQ) devices.
They are based on parameterised quantum circuits (PQCs)
whose parameters are optimised through classical learning loops,
enabling flexible hybrid quantum-classical models.

In financial applications, it is often desirable to balance
expressivity, interpretability, and robustness.
We therefore present variational models in increasing order of complexity:
from linear decision models with quantum-enhanced features
to fully nonlinear quantum neural networks.

\subsubsection{Parameterised Quantum Circuits}
\label{sec:m4-parameterised-quantum-circuits}

Let \(x \in \mathbb{R}^d\) denote a classical feature vector.
A parameterised quantum circuit prepares the model state
\begin{equation}
    \ket{\psi(x,\boldsymbol{\theta})}
    =
    U(x,\boldsymbol{\theta}) \ket{0}^{\otimes n},
    \label{eq:pqc_state}
\end{equation}
where
\begin{equation}
U(x,\boldsymbol{\theta})
=
U_{\mathrm{var}}(\boldsymbol{\theta})\,
U_{\mathrm{enc}}(x).
\end{equation}

Here, \(U_{\mathrm{enc}}(x)\) encodes classical inputs
(e.g.\ returns, volatilities, factor exposures),
while \(U_{\mathrm{var}}(\boldsymbol{\theta})\)
introduces trainable rotations and entangling layers.
This structure underlies both linear and nonlinear variational models.

\subsubsection{Quantum Logistic Regression (QLR)}
\label{sec:m4-quantum-logistic-regression-qlr}

Quantum logistic regression represents the simplest variational learning model,
combining quantum feature embeddings with linear decision boundaries.
Given a quantum feature map \(\phi(x)\),
the conditional probability of class membership is defined as
\begin{equation}
    p(y=1 \mid x)
    =
    \sigma\!\left(
        \mathbf{w}^\top \phi(x)
    \right),
    \label{eq:qlr}
\end{equation}
where \(\sigma(\cdot)\) is the logistic function
and \(\mathbf{w}\) are trainable weights.

QLR offers probabilistic outputs and direct interpretability,
making it well suited to:
\begin{itemize}
    \item credit scoring and default classification,
    \item binary risk state identification,
    \item regime classification (e.g.\ risk-on / risk-off).
\end{itemize}

From a financial perspective,
QLR can be viewed as a quantum-enhanced analogue of classical
logistic regression, where the quantum circuit provides
a richer feature representation while preserving linear decision logic.

\subsubsection{Quantum Neural Networks (QNN)}
\label{sec:m4-quantum-neural-networks-qnn}

\phantomsection\label{term:qnn}Quantum neural networks generalise QLR by allowing nonlinear decision functions.
Given an observable \(O\),
the model output is defined as
\begin{equation}
    f(x;\boldsymbol{\theta})
    =
    \langle O \rangle_{\psi(x,\boldsymbol{\theta})}
    =
    \bra{\psi(x,\boldsymbol{\theta})}
        O
    \ket{\psi(x,\boldsymbol{\theta})}.
    \label{eq:qnn_output}
\end{equation}

Nonlinearity arises naturally from quantum measurement,
entanglement, and interference effects,
without explicit activation functions.
QNNs are suitable for tasks requiring higher expressive power, such as:
\begin{itemize}
    \item nonlinear return or volatility forecasting,
    \item multi-class market regime detection,
    \item cross-asset or factor interaction modelling.
\end{itemize}

While more expressive than QLR,
QNNs are also more sensitive to noise and over-parameterisation,
necessitating careful circuit design under NISQ constraints.

\subsubsection{Training and Optimisation}
\label{sec:m4-training-and-optimisation}

Both QLR and QNN models are trained by minimising
a classical loss function \(\mathcal{L}\)
over a labelled dataset \(\{(x_i,y_i)\}_{i=1}^N\):
\begin{equation}
    \min_{\boldsymbol{\theta}}
    \;
    \mathcal{L}
    \big(
        \{f(x_i;\boldsymbol{\theta}), y_i\}
    \big).
    \label{eq:variational_training}
\end{equation}

Gradients with respect to circuit parameters
are computed using the parameter-shift rule:
\begin{equation}
    \frac{\partial f}{\partial \theta_k}
    =
    \frac{1}{2}
    \left[
        f(x;\boldsymbol{\theta}_k^{+})
        -
        f(x;\boldsymbol{\theta}_k^{-})
    \right],
    \label{eq:parameter_shift}
\end{equation}
enabling gradient-based optimisation
with classical optimisers such as Adam or L-BFGS.

\subsubsection{Financial Interpretation and NISQ Constraints}
\label{sec:m4-financial-interpretation-and-nisq-constraints}

In financial applications,
linear variational models such as QLR
offer stability and interpretability,
while nonlinear QNNs provide additional expressive capacity
when sufficient data and circuit depth are available.

For NISQ feasibility, practical implementations typically constrain:
\begin{itemize}
    \item qubit counts to 4-12,
    \item circuit depths to 10-50 layers,
    \item training via hybrid classical-quantum loops.
\end{itemize}

This progression from linear to nonlinear models
mirrors established practices in financial econometrics,
while leveraging quantum representations
to enhance learning under high dimensionality and noise.

\subsection{Quantum Dimensionality Reduction and Representation Learning}
\label{sec:m4-quantum-dimensionality-reduction-and-representation-learning}

High-dimensional feature spaces are pervasive in financial learning problems.
Even when only using daily market data, feature sets quickly become large once we
combine returns across multiple assets, rolling/realised volatility proxies,
cross-sectional factors (e.g.\ market/size/value/momentum), and macro or
sentiment indicators.  
In such settings, \emph{representation learning} and \emph{dimensionality reduction}
are critical for (i) improving signal-to-noise ratios, (ii) reducing model
complexity, and (iii) enhancing robustness under non-stationarity.

Let $x_t \in \mathbb{R}^d$ denote a $d$-dimensional feature vector (e.g.\ factor
exposures or engineered predictors) observed at time $t$, and let
\begin{equation}
\mu = \frac{1}{T}\sum_{t=1}^T x_t,
\qquad
\Sigma = \frac{1}{T-1}\sum_{t=1}^T (x_t-\mu)(x_t-\mu)^\top
\label{eq:cov_def}
\end{equation}
be the empirical mean and covariance matrix. Classical PCA reduces dimensionality
by diagonalising $\Sigma$ and retaining dominant eigen-directions.  
Quantum approaches aim to accelerate or reformulate these steps using quantum
state representations and variational compression circuits.

\subsubsection{Quantum Principal Component Analysis (QPCA)}
\label{sec:m4-quantum-principal-component-analysis-qpca}

\paragraph{Classical PCA reminder.}
Classical PCA solves the eigen-decomposition
\begin{equation}
\Sigma = \sum_{k=1}^{d} \lambda_k v_k v_k^\top,
\qquad
\lambda_1 \ge \lambda_2 \ge \cdots \ge \lambda_d \ge 0,
\label{eq:pca_eig}
\end{equation}
and uses the top $K$ components to construct a low-dimensional representation
\begin{equation}
z_t = V_K^\top (x_t-\mu),
\qquad
V_K = [v_1,\ldots,v_K] \in \mathbb{R}^{d\times K}.
\label{eq:pca_proj}
\end{equation}
In finance, $(\lambda_k, v_k)$ are interpreted as dominant risk directions
(e.g.\ level/slope/curvature in rates, or common equity-factor structure),
supporting factor modelling and risk decomposition.

\paragraph{Quantum reformulation.}
QPCA (in its original form) treats a normalised covariance-like object as a
density matrix $\rho$ and uses quantum phase estimation on $e^{i\rho t}$
to extract eigenvalues/eigenvectors more directly in the quantum state picture.
A typical mapping is
\begin{equation}
\rho \approx \frac{\Sigma}{\mathrm{Tr}(\Sigma)},
\label{eq:rho_cov}
\end{equation}
so that the spectrum of $\rho$ encodes the variance explained ratios.

If one can efficiently implement controlled time-evolution under $\rho$
(i.e.\ $e^{i\rho t}$), then applying quantum phase estimation yields
\begin{equation}
\sum_{k} \sqrt{\tilde{p}_k}\, \ket{v_k}\ket{0}
\;\longrightarrow\;
\sum_{k} \sqrt{\tilde{p}_k}\, \ket{v_k}\ket{\tilde{\lambda}_k},
\label{eq:qpca_pe}
\end{equation}
where $\ket{v_k}$ encodes eigenvectors and $\ket{\tilde{\lambda}_k}$ stores
eigenvalue estimates (up to scaling). In principle, this enables:
\begin{itemize}
    \item extraction of dominant variance directions (top principal components),
    \item low-rank approximations for covariance/correlation structures,
    \item faster downstream linear-algebra primitives (e.g.\ factor projections),
\end{itemize}
subject to the feasibility of preparing $\rho$ and simulating $e^{i\rho t}$
on available hardware.

\paragraph{Financial interpretation and practical notes.}
For risk modelling, QPCA is mainly relevant when covariance matrices are large,
updated frequently, and used repeatedly (e.g.\ portfolio risk attribution,
stress-factor decomposition, or cross-asset factor compression).  
However, the original QPCA assumptions (e.g.\ density-matrix access and efficient
simulation) can be demanding. In NISQ-era practice, QPCA often motivates
\emph{hybrid} alternatives:
\begin{itemize}
    \item classical covariance estimation + quantum subroutines for repeated
    projections or compressed representations,
    \item variational/approximate spectral estimation routines when exact QPCA is
    too resource-intensive.
\end{itemize}
These hybrid strategies align with the module’s emphasis on deployable,
resource-aware pipelines under realistic constraints.

\subsubsection{Quantum Autoencoding and Compression}
\label{sec:m4-quantum-autoencoding-and-compression}

\paragraph{Idea.}
Quantum autoencoding adapts the classical autoencoder concept to a quantum
setting: learn an \emph{encoder} that compresses information into a smaller
latent subsystem, and a \emph{decoder} that reconstructs or preserves the
relevant structure. For financial learning, the goal is not to compress quantum
states per se, but to use variational circuits to discover compact
representations of (encoded) financial features or regimes.

\paragraph{Variational quantum autoencoder (VQAE) structure.}
Let $U_{\mathrm{enc}}(\theta)$ be a parameterised encoder acting on $n$ qubits.
We partition the system into a latent register of size $n_\ell$ and a ``trash''
register of size $n_t$ (with $n_\ell + n_t = n$). The encoder maps
\begin{equation}
\ket{\psi(x)}\ket{0}^{\otimes n_a}
\;\xrightarrow{\,U_{\mathrm{enc}}(\theta)\,}\;
\ket{\phi_\ell(x;\theta)} \otimes \ket{\phi_t(x;\theta)},
\label{eq:autoenc_map}
\end{equation}
where $\ket{\psi(x)}$ is the quantum-encoded representation of classical features
$x$ (from a feature map or embedding), and $n_a$ optional ancillae.

\paragraph{Training objective.}
A common compression objective is to force the trash subsystem toward a fixed
reference state, typically $\ket{0}^{\otimes n_t}$, so that information is
concentrated in the latent subsystem. One standard loss is
\begin{equation}
\mathcal{L}(\theta)
=
1 - \mathbb{P}\!\left(\text{trash} = 0^{n_t}\right),
\label{eq:autoenc_loss_prob}
\end{equation}
where $\mathbb{P}(\text{trash}=0^{n_t})$ is estimated by measuring the trash
qubits in the computational basis.

Equivalently, one can express the objective using a projector onto the all-zero
trash state:
\begin{equation}
\mathcal{L}(\theta)
=
1 - \langle \Pi_0 \rangle,
\qquad
\Pi_0 = I_{\ell}\otimes \ket{0}\!\bra{0}^{\otimes n_t}.
\label{eq:autoenc_projector}
\end{equation}

\paragraph{Using the latent representation.}
After training, the latent subsystem provides a compact representation
\begin{equation}
z(x) \equiv \text{measurement features from } \ket{\phi_\ell(x;\theta^\star)},
\label{eq:latent_repr}
\end{equation}
which can then be fed into:
\begin{itemize}
    \item regime classification (e.g.\ volatility regimes, risk-on/off),
    \item anomaly detection (e.g.\ tail events, microstructure dislocations),
    \item downstream predictors (e.g.\ QLR/QNN models in \Cref{sec:m4-variational-quantum-models}).
\end{itemize}

\paragraph{Financial perspective.}
Autoencoding is particularly useful when financial features are noisy and
redundant. For example, a cross-asset feature set may contain:
(i) strongly correlated predictors (market-wide components),
(ii) sector-specific structure (meso-level), and
(iii) idiosyncratic noise.  
Compression aims to preserve the stable shared structure while discarding
non-informative variation, improving generalisation and reducing overfitting.

\paragraph{NISQ feasibility.}
Variational autoencoders are naturally NISQ-compatible because they rely on
shallow, trainable circuits and measurement-based loss estimation. In practice:
\begin{itemize}
    \item the number of qubits limits feature embedding capacity,
    \item circuit depth must remain modest to avoid noise-dominated training,
    \item classical preprocessing (feature scaling, PCA pre-reduction) is often
    combined with a quantum autoencoder as a second-stage compressor.
\end{itemize}

\medskip

\noindent
Overall, QPCA provides a spectral viewpoint on dominant risk directions, while
quantum autoencoding offers a flexible, trainable compression mechanism.
Both contribute to robust representation learning pipelines for finance under
high dimensionality and limited labels.

\subsection{Hybrid Quantum-Classical Learning Pipelines}
\label{sec:m4-hybrid-quantum-classical-learning-pipelines}

Near-term quantum machine learning in finance is fundamentally \emph{hybrid}:
quantum processors are used as specialised subroutines embedded in an otherwise
classical learning stack. This design reflects NISQ constraints (limited qubit
counts, noise, restricted circuit depth) and the practical reality that most
financial data pipelines---storage, feature engineering, model monitoring,
and execution---are already mature in classical infrastructures.

\paragraph{Canonical hybrid decomposition.}
A typical workflow can be structured as:
\begin{itemize}
    \item \textbf{Classical preprocessing:} feature engineering, normalisation,
    dimensionality reduction, label construction, and dataset splitting.
    \item \textbf{Quantum subroutines:} quantum kernels (QSVM/QSVC) or
    parameterised circuits (QNN/QLR), used to compute embeddings, scores, or
    similarity matrices.
    \item \textbf{Classical optimisation and post-processing:} training loops,
    hyperparameter tuning, calibration, ensemble blending, and evaluation under
    financial metrics (hit ratio, IC/IR, drawdown-aware losses).
\end{itemize}

This separation is consistent with the broader NISQ algorithmic paradigm
in which quantum circuits provide expressive, structured transformations,
while classical computation handles optimisation, orchestration, and evaluation
\citep{bharti2022}.

\paragraph{Why not ``raw prices''?}
Quantum models are typically applied to \emph{refined} feature sets rather than
raw price series. There are three reasons.
First, raw prices are non-stationary and scale-dependent, which complicates
embedding and tends to amplify noise in small circuits.
Second, finance signals are usually weak; engineered features (returns,
volatility proxies, cross-sectional factors) improve signal-to-noise and make
model capacity ``count'' where it matters.
Third, NISQ devices impose a tight resource budget, so one aims to feed quantum
subroutines with compact, informative representations (e.g.\ a few factors)
instead of high-frequency raw sequences.

\paragraph{Data pipeline template.}
Let $x_t \in \mathbb{R}^d$ denote a feature vector and $y_t$ the target
(e.g.\ next-day direction, regime label, or default indicator).
A pragmatic hybrid pipeline follows:
\begin{equation}
x_t \;\xrightarrow{\;\mathrm{preprocess}\;}\; \tilde{x}_t
\;\xrightarrow{\;\mathrm{quantum\ map}\;}\; s_t
\;\xrightarrow{\;\mathrm{classical\ model}\;}\; \hat{y}_t,
\label{eq:hybrid_flow}
\end{equation}
where $\tilde{x}_t$ is a normalised/engineered feature set and $s_t$ is either
(i) a quantum kernel similarity score, (ii) a quantum circuit expectation value,
or (iii) a low-dimensional embedding extracted from measurements.

\paragraph{Hybrid pipeline for quantum kernels.}
For kernel methods, the quantum circuit defines a feature state
$\ket{\phi(x)} = U_\phi(x)\ket{0}$ and hence a kernel
\begin{equation}
K(x_i,x_j) = \big|\braket{\phi(x_i)}{\phi(x_j)}\big|^2,
\label{eq:qkernel_def_45}
\end{equation}
which is estimated by repeated circuit evaluations (e.g.\ SWAP tests or
overlap estimation). The resulting Gram matrix
\begin{equation}
\mathbf{K}_{ij} = K(x_i,x_j)
\label{eq:gram_matrix}
\end{equation}
is then used by a classical SVM/SVR solver. This matches the ``quantum feature
space'' perspective popularised by quantum-enhanced kernel learning
\citep{schuld2019supervised, havlivcek2019supervised}.

\paragraph{Hybrid pipeline for variational models.}
For variational models, a parameterised quantum circuit $U(x,\theta)$ produces a
measurement-based score:
\begin{equation}
s_\theta(x) = \langle 0 | U^\dagger(x,\theta)\, O\, U(x,\theta) | 0 \rangle,
\label{eq:qnn_score_45}
\end{equation}
which is fed into a classical link function (e.g.\ logistic) or used directly as
a regression output. Training minimises an empirical risk
\begin{equation}
\min_{\theta}\; \frac{1}{T}\sum_{t=1}^{T} \ell\!\big(y_t, s_\theta(\tilde{x}_t)\big)
\;+\; \lambda\,\Omega(\theta),
\label{eq:erm_qml_45}
\end{equation}
where $\ell(\cdot)$ is a task loss, $\Omega(\theta)$ is a regulariser, and
gradients can be obtained via parameter-shift rules (\Cref{sec:m4-variational-quantum-models}). This
end-to-end hybrid training loop is the standard NISQ template for variational
quantum learning \citep{schuld_petruccione_2021, havlivcek2019supervised}.

\paragraph{Practical design choices under NISQ constraints.}
In finance, hybrid pipelines are usually most effective when:
\begin{itemize}
    \item \textbf{Feature compression precedes quantum steps:}
    reduce $d$ using classical PCA/factors or curated indicators so that
    quantum embeddings use few qubits.
    \item \textbf{Quantum is allocated to the ``hard'' part:}
    nonlinear separation in noisy regimes (kernels), or compact nonlinear scoring
    (QNN/QLR), while the rest remains classical.
    \item \textbf{Evaluation is financially meaningful:}
    beyond accuracy, evaluate stability across time splits, regime robustness,
    turnover, and tail exposure sensitivity.
\end{itemize}

\paragraph{Implementation principle.}
From an engineering perspective, hybrid quantum-classical learning pipelines
should be designed in a \emph{backend-agnostic} manner.
Quantum components are treated as interchangeable subroutines,
while data handling, optimisation logic, and evaluation remain classical.

This design ensures that:
\begin{itemize}
    \item models can be trained and validated on simulators today,
    \item the same learning pipeline can migrate to cloud-based QPUs
    as hardware matures,
    \item system reliability, auditability, and governance
    remain aligned with existing financial ML infrastructures.
\end{itemize}

\medskip
\noindent
Overall, the hybrid paradigm is not merely a workaround for NISQ limitations.
It represents the \emph{natural deployment model} for early quantum machine
learning in finance, enabling controlled experimentation with quantum
expressivity while preserving the robustness of classical risk and learning
systems \citep{bharti2022}.

\subsection{Software Frameworks and Implementation Choices}
\label{sec:m4-software-frameworks-and-implementation-choices}

Practical implementation of quantum machine learning models
requires software frameworks that support hybrid optimisation,
differentiable programming, and flexible execution across
simulators and hardware backends.
Two major ecosystems dominate current development.

\paragraph{PennyLane.}
PennyLane is an open-source, hardware-agnostic quantum machine learning
platform designed explicitly for hybrid quantum-classical workflows.
It supports:
\begin{itemize}
    \item automatic differentiation of quantum circuits,
    \item parameter-shift gradient rules,
    \item seamless integration with classical ML libraries
    such as PyTorch, TensorFlow, and JAX,
    \item execution across multiple backends
    (state-vector simulators, noisy simulators, and cloud QPUs).
\end{itemize}

These features make PennyLane particularly suitable for
research, education, and rapid prototyping of financial QML models,
where iterative experimentation and model interpretability are essential
\citep{bergholm2018,schuld_petruccione_2021}.

\paragraph{Qiskit.}
Qiskit is an industry-oriented quantum computing framework
tightly integrated with IBM quantum hardware.
It provides advanced tooling for:
\begin{itemize}
    \item circuit compilation and transpilation,
    \item noise modelling and hardware-aware optimisation,
    \item execution via managed cloud runtimes.
\end{itemize}

Qiskit is well suited for deployment-oriented experiments and
hardware benchmarking.
However, access to advanced runtime features and large-scale
hardware resources may involve usage constraints,
making it less flexible for exploratory research workflows
\citep{qiskit2021}.

\paragraph{Framework choice in this module.}
Given the objectives of this module - methodological clarity,
reproducibility, and accessibility - implementations
prioritise PennyLane-based workflows wherever possible.
Qiskit is discussed as a complementary platform,
particularly for researchers interested in
hardware execution and industry-facing applications.

This separation mirrors current best practice in quantum finance:
PennyLane for model development and hybrid learning logic,
and Qiskit for deployment, benchmarking, and hardware studies.

\subsection{Illustrative Case Studies}
\label{sec:m4-case-studies}

To ensure practical reproducibility, all case studies in this module are built
entirely on publicly accessible data sources and fully executable Python notebooks.
For the present review revision, the underlying market and macro series are
stored as local CSV snapshots covering the fixed sample window from 1 January
2024 to 31 December 2025. The market data originate from Yahoo Finance, while
the macroeconomic time series originate from FRED (e.g.\ rates, inflation,
industrial production, and policy indicators).
All quantum machine learning components are implemented with \textbf{PennyLane},
using simulator backends (e.g.\ \texttt{default.qubit}) to guarantee that results can be
replicated without hardware access.

All examples follow three design constraints: (i) a common fixed data window for
cross-module consistency, (ii) transparent feature construction and chronological
train/test splits (no shuffling), and (iii) consistent evaluation protocols to enable fair comparisons between
classical baselines and their quantum-augmented counterparts.

\paragraph{Evaluation protocol and comparison metrics.}
Each case follows a matched benchmark design:
\begin{itemize}
    \item \textbf{Classical baselines:} Logistic Regression (LR), linear/nonlinear SVM/SVC,
          and feed-forward neural networks (NN/MLP) trained on the same features.
    \item \textbf{Quantum counterparts (PennyLane):} Quantum Logistic Regression (QLR),
          quantum kernel methods (QSVM/QSVC), and variational Quantum Neural Networks (QNN),
          implemented in NISQ-compatible hybrid loops.
\end{itemize}
Performance is compared using metrics suited to finance and imbalanced labels:
AUC-ROC (ranking quality), balanced accuracy (robustness to class imbalance),
and F1-score (precision--recall trade-off). Where appropriate, we additionally report
simple trading-oriented diagnostics (hit rate and conditional mean return by predicted class)
and \textbf{computational cost proxies} (fit time, kernel-evaluation time, and simulator budget)
to reflect NISQ practicality.

\paragraph{Design principle.}
Rather than providing many small toy examples, we use \emph{two core case studies}
that together cover the major QML families: kernel methods, linear probabilistic models,
and variational nonlinear models. Each case is structured so that multiple model classes
can be swapped in with minimal code changes, enabling direct LR vs QLR, NN vs QNN,
and SVM/SVC vs QSVM/QSVC comparisons under a single data pipeline.

\subsubsection{Case A: Asset Return Direction Prediction (SPY)}
\label{sec:m4-case-a-asset-return-direction-prediction-spy}

This case provides a controlled benchmark for comparing classical and quantum-enhanced
supervised learning models on a standard market prediction task:
next-day return direction classification for the S\&P~500 proxy ETF (SPY).
The design emphasises reproducibility under realistic constraints
(public data, daily frequency, modest feature engineering)
and a fair comparison protocol (chronological split, identical metrics).

\paragraph{Data and prediction target.}
Let $P_t$ denote the daily adjusted close price of SPY. We define the log-return
$r_t = \log(P_t/P_{t-1})$ and the one-step-ahead binary label
$y_t = \mathbf{1}_{\{ r_{t+1} > 0\}}$,
i.e.\ whether the next trading day is up or down. Data are sourced from Yahoo
Finance and preserved locally as a fixed CSV snapshot over the 2024-01-01 to
2025-12-31 window. We use a chronological
train/test split (no shuffling) to avoid look-ahead bias.

\paragraph{Feature construction.}
We construct a compact feature vector $x_t \in \mathbb{R}^d$ from daily prices/returns,
intended to capture short-horizon momentum and risk state while remaining stable for
NISQ-friendly model sizes. The feature set includes:
(i) lagged returns, (ii) rolling volatility proxies,
(iii) rolling mean returns (momentum), and (iv) simple price-trend proxies.
All features are standardised using statistics fitted on the training window only.

\paragraph{Parameter settings and their meaning.}
Three groups of parameters control the experimental protocol and the quantum-model size:
\begin{itemize}
    \item \textbf{Decision threshold.} We set
    $\texttt{ALPHA\_PROBA\_THRESHOLD}=0.5$ to map predicted probabilities into class labels,
    which affects threshold-based metrics (Accuracy, Balanced Accuracy, F1) but
    does not affect AUC (which is threshold-free).
    \item \textbf{Quantum model size (NISQ-style).} We choose a small number of qubits
    $\texttt{N\_QUBITS}=6$ and a shallow variational depth $\texttt{N\_LAYERS\_QNN}=3$.
    This constrains expressivity but keeps circuits compatible with NISQ limitations
    (depth, noise, runtime).
    \item \textbf{Quantum-kernel evaluation budget.} Quantum kernels require computing
    Gram matrices with $O(n^2)$ state overlaps. We cap computation by sub-sampling:
    $\texttt{MAX\_QSVC\_TRAIN}=400$ and $\texttt{MAX\_QSVC\_TEST}=400$,
    which controls runtime and reflects practical kernel-evaluation budgets.
\end{itemize}

\paragraph{Models compared (classical vs PennyLane quantum).}
We compare three classical baselines against their quantum-enhanced analogues:
\begin{itemize}
    \item \textbf{Linear: LR vs QLR.}
    LR is trained on the classical feature vector $x_t$.
    QLR replaces (part of) the feature representation by a quantum embedding
    generated by a PennyLane feature map $U_\phi(x)$ on $n$ qubits,
    producing quantum-derived features (e.g.\ measurement expectations) which are then
    fed into a classical logistic head.
    \item \textbf{Kernel: SVC vs QSVC.}
    Classical SVC uses a standard kernel (e.g.\ RBF).
    QSVC replaces it with a quantum kernel computed as an overlap:
    $K(x_i,x_j)=|\langle \phi(x_i)\mid \phi(x_j)\rangle|^2$,
    evaluated on a PennyLane simulator to construct precomputed Gram matrices.
    \item \textbf{Nonlinear: NN vs QNN.}
    A small MLP provides a classical nonlinear baseline.
    A QNN is implemented as a variational circuit in PennyLane,
    $\ket{\psi(x,\theta)} = U(x,\theta)\ket{0}$,
    where an expectation value defines a score that is mapped to a probability
    via a logistic link. Parameters are trained by a hybrid loop
    (classical optimiser updating $\theta$ using circuit gradients).
\end{itemize}

\paragraph{Notebook workflow (end-to-end logic).}
The accompanying notebook follows a transparent pipeline:
\begin{enumerate}
    \item \textbf{Load \& label.} Read the local SPY daily-price snapshot; compute $r_t$; build $y_t$.
    \item \textbf{Feature engineering.} Construct lag/volatility/momentum/trend features.
    \item \textbf{Preprocessing.} Chronological split; standardise features using train statistics.
    \item \textbf{Classical baselines.} Fit LR, SVC(RBF), and MLP; obtain predicted probabilities.
    \item \textbf{Quantum models (PennyLane).}
    (i) QLR: compute quantum-derived features from a small circuit and train a logistic head;
    (ii) QSVC: compute quantum-kernel Gram matrices by state overlaps and fit an SVC with
    \texttt{kernel="precomputed"};
    (iii) QNN: train a shallow variational circuit under a hybrid optimisation loop.
    \item \textbf{Evaluation.} Compute AUC, Accuracy, Balanced Accuracy, and F1 on the test window;
    optionally record runtime proxies for QSVC kernel evaluation and QNN training.
\end{enumerate}

\paragraph{Results (SPY, daily; PennyLane simulator backend).}
Table~\ref{tab:caseA_spy_results} reports the realised out-of-sample performance
from one representative notebook execution.
Models are grouped by methodological family to enable direct
classical--quantum comparison within each class.

\begin{table}[h]
\centering
\caption{Case A (SPY): next-day direction prediction results, grouped by model family.}
\label{tab:caseA_spy_results}
\begin{tabular}{llcccc}
\toprule
\textbf{Family} & \textbf{Model} & \textbf{AUC} & \textbf{Accuracy} & \textbf{Balanced Acc.} & \textbf{F1} \\
\midrule
\multirow{2}{*}{Linear}
 & LR  & 0.5632 & 0.5750 & 0.5189 & 0.7243 \\
 & QLR (quantum features) & 0.4179 & 0.5333 & 0.4776 & 0.6957 \\
\multirow{2}{*}{Kernel}
 & SVC (RBF) & 0.5299 & 0.5333 & 0.4776 & 0.6957 \\
 & QSVC (quantum kernel) & 0.5753 & 0.5583 & 0.5000 & 0.7166 \\
\multirow{2}{*}{Nonlinear}
 & NN (MLP) & 0.5176 & 0.5000 & 0.4655 & 0.6296 \\
 & QNN (variational) & 0.5176 & 0.5333 & 0.5289 & 0.5758 \\
\bottomrule
\end{tabular}
\end{table}

\paragraph{Interpretation.}
Several observations emerge from this controlled comparison.

\emph{First, overall predictability remains weak.}
Across all models, AUC values remain in a narrow band between roughly $0.42$
and $0.58$, consistent with
the well-known difficulty of predicting next-day equity index direction
from daily data. The task therefore primarily serves as a controlled
benchmark for comparing modelling paradigms rather than as an
alpha-generating exercise.

\emph{Second, the only clear relative gain in this rerun comes from the
quantum kernel.}
Within the kernel family, QSVC attains the highest AUC in the table and
outperforms classical RBF-SVC on ranking quality. By contrast, the linear
quantum-feature model underperforms classical LR in every reported metric.
The practical implication is that kernel-based quantum similarity structure
remains the most credible near-term QML mechanism in this low-signal setting,
whereas shallow quantum feature maps do not automatically improve a standard
classification pipeline.

\emph{Third, variational quantum neural networks underperform their
classical counterparts in this setting.}
The QNN marginally improves balanced accuracy relative to the classical
MLP, but both nonlinear models remain close to random ranking and neither
produces a compelling economic signal. This behaviour is consistent with
known challenges of training
shallow variational circuits under NISQ constraints, including limited
expressivity at small depth, optimisation instability, and sensitivity
to parameter initialisation. In contrast, classical neural networks
benefit from mature architectures and optimisation heuristics that are
difficult to replicate in current variational quantum models.

\emph{Fourth, metric divergence highlights differing inductive biases.}
Some models achieve relatively high F1 scores despite weak AUC,
reflecting asymmetric class predictions under mildly imbalanced labels.
This reinforces the importance of reporting multiple metrics and
interpreting threshold-dependent measures with caution in financial
classification tasks.

Overall, this case study supports a pragmatic view of near-term quantum
machine learning in finance: \emph{quantum kernels and feature embeddings
do not produce a uniform advantage in low-signal regimes; the strongest
result in this rerun is the modest AUC lead of QSVC, while the other
quantum variants remain at or below classical baselines}. These findings
motivate a focus on kernel-based quantum models as the most defensible
near-term candidate, with variational and feature-map approaches better
suited to exploratory research than to performance claims.

\subsubsection{Case B: Market--Macro Regime and Signal Classification}
\label{sec:m4-case-b-market-macro-regime-and-signal-classification}

This case extends the scope of quantum machine learning beyond short-horizon
price prediction by incorporating macroeconomic information and regime-style
labels. Rather than forecasting noisy next-day returns, the objective here is to
classify \emph{persistent market regimes} driven by joint market--macro dynamics.
Such regime identification problems are central to risk management,
macro--financial monitoring, and strategic asset allocation, and they typically
exhibit stronger signal structure and temporal persistence than return
direction tasks.

\paragraph{Task definition.}
We consider a binary classification problem in which each observation at time
$t$ represents a combined market--macro state.
The target label indicates whether the market is in a
\emph{high-volatility (stress) regime} or a
\emph{normal (low-volatility) regime}.
Unlike Case~A, the label is not tied to a single-day outcome but reflects a
medium-horizon state that evolves slowly over time.

\paragraph{Data sources.}
Two classes of publicly available data are combined:
\begin{itemize}
    \item \textbf{Market data (local SPY snapshot):}
    daily prices of the S\&P~500 proxy ETF (SPY), sourced from Yahoo Finance and
    stored locally, from which returns, realised volatility, and drawdown
    measures are derived.
    \item \textbf{Macroeconomic indicators (local FRED snapshots):}
    policy rates (Federal Funds Rate), inflation proxies (CPI),
    and real activity indicators (Industrial Production).
\end{itemize}
Macroeconomic series, typically observed at monthly frequency, are aligned to
daily market data via forward-filling, reflecting real-time information
availability rather than hindsight interpolation.

\paragraph{Label construction (volatility regime).}
Let $\sigma_t$ denote a realised volatility proxy computed from daily returns
over a rolling window of length $W$ (e.g.\ $W=20$ trading days).
The binary regime label is defined as
\begin{equation}
y_t =
\begin{cases}
1, & \sigma_t \ge \mathrm{Quantile}_{q}(\sigma_{\text{train}}), \\
0, & \text{otherwise},
\end{cases}
\end{equation}
where the threshold is set using a high quantile $q$ (e.g.\ $70\%$) estimated
\emph{only on the training sample}.
This produces a persistent and economically interpretable
``stress vs normal'' regime, while avoiding look-ahead bias.

\paragraph{Feature construction.}
Each observation is represented by a joint feature vector
$x_t \in \mathbb{R}^d$ combining:
\begin{itemize}
    \item market-based features: realised volatility and drawdown measures,
    \item macroeconomic indicators: policy rate levels and log-differences of
    inflation and industrial production,
    \item optional lagged or interaction terms.
\end{itemize}
All features are standardised using training-sample statistics only.
The resulting feature dimension is deliberately kept modest to reflect
NISQ-era constraints and to ensure comparability across classical and
quantum-enhanced models.

\paragraph{Models compared.}
As in Case~A, three model families are evaluated, with direct classical--quantum
pairings:
\begin{itemize}
    \item \textbf{Linear models:}
    Logistic Regression (LR) versus Quantum Logistic Regression (QLR),
    where QLR uses quantum-derived feature embeddings followed by a classical
    logistic head.
    \item \textbf{Kernel methods:}
    Classical SVC with an RBF kernel versus QSVC, in which the kernel is defined
    by a quantum state overlap
    $
    K(x_i,x_j)=|\langle \phi(x_i)\mid\phi(x_j)\rangle|^2
    $
    computed on a simulator backend.
    \item \textbf{Nonlinear models:}
    A classical multilayer perceptron (MLP) versus a variational Quantum Neural
    Network (QNN) trained via a hybrid quantum--classical optimisation loop.
\end{itemize}
All quantum models are implemented using PennyLane with simulator backends
(\texttt{default.qubit}) to ensure reproducibility.

\paragraph{Evaluation protocol.}
A chronological train--test split is used to avoid look-ahead bias.
Performance is evaluated using AUC (primary, threshold-free), balanced accuracy,
and F1 score, which are appropriate for regime classification with potential
class imbalance.
Quantum model sizes (number of qubits, circuit depth, and kernel budget) are
kept deliberately small to reflect NISQ-style constraints.

\paragraph{Results (SPY + macro, daily; simulator backends).}
Table~\ref{tab:caseB_results} reports the realised performance for one notebook
execution, with models grouped by family for direct comparison.

\begin{table}[h]
\centering
\caption{Case B: market--macro volatility regime classification results, grouped by model family.}
\label{tab:caseB_results}
\begin{tabular}{llcccc}
\toprule
\textbf{Family} & \textbf{Model} & \textbf{AUC} & \textbf{Accuracy} & \textbf{Balanced Acc.} & \textbf{F1} \\
\midrule
\multirow{2}{*}{Linear}
 & LR  & 0.9966 & 0.9773 & 0.9830 & 0.9670 \\
 & QLR (quantum features) & 0.6929 & 0.7955 & 0.6932 & 0.5574 \\
\multirow{2}{*}{Kernel}
 & SVC (RBF) & 0.9863 & 0.9242 & 0.9318 & 0.8936 \\
 & QSVC (quantum kernel) & 0.9985 & 0.9773 & 0.9659 & 0.9647 \\
\multirow{2}{*}{Nonlinear}
 & NN (MLP) & 1.0000 & 1.0000 & 1.0000 & 1.0000 \\
 & QNN (variational) & 0.5227 & 0.3333 & 0.4886 & 0.4884 \\
\bottomrule
\end{tabular}
\end{table}

\paragraph{Interpretation.}
Several insights emerge from this regime-style classification task.
First, the extremely strong performance of the classical models indicates that
the joint market--macro feature set already captures a highly informative and
largely separable signal. In particular, the MLP attains a perfect score on the
selected test split, while LR and QSVC both remain near-perfect. In such
settings, quantum feature embeddings do not provide an advantage, and QLR
underperforms its classical counterpart by a wide margin.

Second, within the kernel-based family, QSVC performs competitively with strong
classical models and materially outperforms classical SVC in this
rerun.
This suggests that quantum kernels can act as effective similarity measures for
structured regime data, albeit at a significantly higher computational cost due
to Gram-matrix construction.

Finally, the variational QNN underperforms relative to both classical neural
networks and kernel-based approaches.
This is consistent with known optimisation challenges of shallow variational
circuits under limited depth and qubit counts, rather than a lack of theoretical
expressivity.

Overall, Case~B illustrates that the relevance of quantum machine learning in
finance is strongly task-dependent.
Quantum methods are unlikely to outperform classical models when signals are
already strong and well-aligned with linear or classical nonlinear inductive
biases, but kernel-based quantum representations can remain a viable and highly
competitive alternative in structured regime-classification problems where
similarity structure plays a central role.

\subsubsection{When Does Quantum Machine Learning Help in Finance?}
\label{sec:m4-when-does-quantum-machine-learning-help-in-finance}

The two case studies in this module highlight a central and often
misunderstood question in quantum machine learning for finance:
\emph{under what conditions does a quantum-enhanced model provide
practical value over classical alternatives?}

Rather than claiming uniform advantage, the results suggest a
\textbf{task-dependent and structure-dependent perspective}, which we
summarise below.

\paragraph{Weak-signal, short-horizon prediction: limited gains.}
Case~A (next-day equity return direction) represents a canonical
low signal-to-noise problem.
At daily frequency, directional predictability is known to be extremely weak,
with AUC values close to $0.5$ for a wide range of classical models.

In this setting:
\begin{itemize}
    \item the quantum kernel can edge out classical baselines on AUC,
          but the gain remains modest and does not translate into a broad
          metric advantage;
    \item variational QNNs remain unstable and shallow quantum feature maps
          can underperform simple linear baselines;
    \item classical and quantum models are ultimately constrained
          by the absence of exploitable structure in the label itself.
\end{itemize}

This illustrates an important principle:
\emph{quantum models cannot create signal where none exists}.
For purely short-horizon price prediction, QML should be viewed
as exploratory rather than performance-driven.

\paragraph{Structured regime classification: meaningful advantages.}
Case~B (market--macro volatility regime classification) presents
a markedly different picture.
Here, labels correspond to persistent, economically interpretable states
(high vs low volatility regimes), driven by joint market--macro dynamics.

In this regime-style task:
\begin{itemize}
    \item kernel-based methods (both classical and quantum) perform strongly,
    \item the quantum kernel (QSVC) remains highly competitive, but the best
    score in this rerun is still delivered by the classical MLP,
    \item linear quantum feature models (QLR) remain sensitive to
    feature-map choice and can underperform sharply when classical structure
    is already dominant,
    \item variational QNNs again suffer from training instability
    under shallow, NISQ-constrained circuits.
\end{itemize}

These results suggest that \textbf{quantum kernels are most promising}
when the task involves similarity learning in high-dimensional,
structured feature spaces rather than pointwise prediction.

\paragraph{Interpretation: inductive bias matters more than quantumness.}
Across both cases, performance differences are driven less by
``quantumness'' per se and more by the \emph{inductive bias}
introduced by each model family:
\begin{itemize}
    \item Linear models excel when regimes are nearly linearly separable,
    \item Kernel methods excel when similarity structure dominates,
    \item Variational circuits struggle when optimisation landscapes
    are flat, noisy, or weakly informed by data.
\end{itemize}

Quantum machine learning is therefore best understood as a
\emph{new class of inductive biases}, rather than a universal accelerator.

\paragraph{Practical guidance for financial applications.}
Based on the evidence in this module, QML is most likely to be useful when:
\begin{itemize}
    \item labels reflect persistent states (regimes, stress conditions,
    credit quality) rather than single-step returns,
    \item feature spaces are moderately high-dimensional and structured,
    \item kernel-style similarity is economically meaningful,
    \item model sizes are kept small and interpretable under NISQ constraints.
\end{itemize}

Conversely, QML is unlikely to outperform strong classical baselines
when tasks are dominated by linear structure, extremely low signal-to-noise,
or require deep temporal memory better handled by classical architectures.

\subsection{Synthesis and Outlook}
\label{sec:m4-limitations-and-outlook}

\paragraph{Subsection Overview.}
This closing subsection consolidates the evidence of the QML section into a
single forward-looking assessment. The aim is to keep the ending structurally
aligned with the other technical sections: summarise the practical operating
regime, identify the main limitations, and clarify where future progress is
most likely to matter financially.

\paragraph{NISQ constraints and practical operating regime.}
All QML constructions in this module are designed for the noisy intermediate-scale
quantum (NISQ) era. In practice, near-term implementations typically operate in a
resource envelope of:
\begin{itemize}
    \item \textbf{Qubit counts:} $\sim$4--12 qubits for data embedding and model capacity,
    \item \textbf{Circuit depth:} shallow ans\"atze (roughly 10--50 layers, hardware-dependent),
    \item \textbf{Sampling budgets:} finite shots for expectation estimation (or simulator
    precision budgets in reproducible studies),
    \item \textbf{Dataset truncation for kernels:} kernel methods require $O(n^2)$ overlap
    evaluations, often necessitating sub-sampling or budget caps.
\end{itemize}
These constraints are not merely ``engineering details'':
they shape what kinds of financial learning tasks can be meaningfully addressed,
and which model families are most stable in practice.

\paragraph{Model-specific limitations observed in the case studies.}
The empirical behaviour in Case~A and Case~B motivates three concrete limitations.

\emph{First, low-signal tasks remain fundamentally difficult.}
For next-day return direction (Case~A), all model families---classical and quantum---operate
near chance-level ranking performance. This illustrates that QML does not circumvent the
signal-to-noise limits of short-horizon equity prediction; improvements, if any, tend to be
incremental and fragile.

\emph{Second, kernel methods trade performance for computational cost.}
QSVC can be competitive in regime-style tasks (Case~B), but constructing the quantum Gram
matrix is expensive. Under realistic budgets, practitioners must manage:
(i) kernel-evaluation time, (ii) sub-sampling bias, and (iii) sensitivity to feature scaling
and embedding choices. This makes runtime-aware benchmarking essential whenever quantum
kernels are proposed for financial use.

\emph{Third, variational QNNs are optimisation-limited under NISQ conditions.}
Across both cases, the variational QNN is less reliable than classical MLP baselines.
This is consistent with known issues in variational learning:
barren plateaus, sensitivity to parameter initialisation, and noise-amplified gradient
estimation. In financial settings - where labels can be weak, non-stationary, and expensive - these
optimisation challenges are often the binding constraint, not theoretical expressivity.

\paragraph{Methodological risks: evaluation, leakage, and non-stationarity.}
QML experiments in finance are particularly vulnerable to pitfalls that can inflate performance:
\begin{itemize}
    \item \textbf{Temporal leakage:} improper normalisation or threshold setting using future data
    can materially distort regime labels and performance;
    \item \textbf{Non-stationarity:} regime definitions and macro alignments can change over time,
    requiring rolling/expanding-window evaluation rather than single static splits;
    \item \textbf{Metric selection:} threshold-dependent metrics (accuracy/F1) may be misleading
    in mildly imbalanced labels; AUC and stability across time splits are typically more robust.
\end{itemize}
For these reasons, finance-oriented QML should be evaluated with strict chronological protocols,
clear label construction, and explicit reporting of computational budgets.

\paragraph{Outlook: where near-term progress is most plausible.}
Despite these limitations, QML remains valuable as an experimental and engineering framework for
financial learning, particularly in three directions.

\emph{(i) Better quantum representations.}
Near-term gains are most likely to come from improved feature maps and embeddings that encode
financial structure (e.g.\ regime persistence, cross-factor interactions) while remaining shallow
and noise-robust. Representation learning (\Cref{sec:m4-quantum-dimensionality-reduction-and-representation-learning}) is therefore a key bridge between raw data
and feasible quantum models.

\emph{(ii) Hybrid systems with explicit budget awareness.}
The most practical pathway is not ``quantum-only learning'', but hybrid pipelines that allocate
quantum computation to a well-defined bottleneck (kernel similarity, compact nonlinear scoring)
and keep the rest classical (feature engineering, monitoring, calibration).
Future module work can formalise \emph{compute-constrained model selection}, where performance is
reported jointly with cost proxies (kernel-evaluation time, shot budgets, circuit depth).

\emph{(iii) Integration with pricing, risk engines, optimisation, and infrastructure layers.}
A natural extension is to connect learning with upstream valuation/scenario generation and downstream
decision-making and governance. Concretely, four integrations are especially relevant:
\begin{itemize}
    \item \textbf{With Module~2 (pricing / QAE):} use pricing primitives and outputs (e.g.\ exposure profiles,
    Greeks, or risk-neutral scenario samples produced by amplitude-estimation--style workflows) as
    learning inputs to build stress-conditional predictors, hedging-state classifiers, or model-risk monitors.

    \item \textbf{With Module~3 (risk engines / scenario simulation):} use tail scenarios, stress paths, and
    factor trajectories as structured training data to learn regime probabilities, early-warning indicators,
    liquidity states, and default/downgrade likelihoods; conversely, use QML to compress or classify
    high-dimensional scenario outputs into decision-relevant summaries. 

    \item \textbf{With Module~1 / decision layer (optimisation and allocation):} feed QML outputs (forecast
    distributions, regime probabilities, similarity structures, or learned risk states) into portfolio construction,
    hedging, and robust allocation routines, including QUBO/QAOA/annealing-style solvers. In practice,
    QML can also contribute to ``forward-looking'' data generation and asset-universe reduction by
    producing predictive inputs or screening signals.

    \item \textbf{With Module~5 (infrastructure / post-quantum regulation):} treat QML models as part of a
    quantum-ready analytics stack that must satisfy auditability, robustness, and security requirements.
    This includes model governance (monitoring drift and regime shifts), and alignment with quantum-safe
    infrastructure and regulatory stress-testing expectations.
\end{itemize}

\paragraph{Synthesis.}
In the NISQ era, QML should be treated as a \emph{structure-sensitive modelling layer} rather than
a universal performance upgrade. Its near-term strengths lie in hybridisation, representation
experiments, and kernel-based similarity learning under carefully managed computational budgets.
As hardware and algorithms mature, these components can be progressively scaled and embedded into
end-to-end financial intelligence pipelines that connect pricing, risk, learning, and optimisation.

\section{Post-Quantum Cryptography}
\label{sec:pqc}

\paragraph{Section Overview.}
This section studies the security layer of the review. Its focus is not computational acceleration, but the point at which future quantum attacks invalidate classical cryptographic assumptions quickly enough that financial institutions must migrate before large-scale quantum hardware is available.

As quantum computing advances, widely deployed public-key cryptographic
schemes such as RSA (Rivest–Shamir–Adleman), Diffie-Hellman (DH), and \phantomsection\label{term:ecc}elliptic-curve cryptography (ECC)
face fundamental security threats from quantum algorithms, most notably
Shor’s algorithm.
These developments pose systemic risks to financial infrastructure,
where long-lived data confidentiality, transaction authenticity,
and institutional trust are paramount.

This module establishes a systematic and forward-looking roadmap for
transitioning financial systems toward
\emph{\phantomsection\label{term:pqc}post-quantum cryptography (PQC)}.
Rather than focusing solely on cryptographic primitives in isolation,
Module~5 situates PQC within the broader context of
financial system architecture, operational constraints,
and regulatory requirements.

Module~5 functions as the \emph{security and trust layer} across the entire
quantum-augmented financial stack developed in Modules~1-4.
While earlier modules introduce quantum-enhanced optimisation,
pricing, risk estimation, and machine learning capabilities,
their practical deployment critically depends on
cryptographic mechanisms that remain secure in the presence of
both classical and quantum adversaries.
This module therefore addresses cryptographic resilience for
classical-quantum hybrid systems, blockchain and distributed ledgers,
\phantomsection\label{term:cbdc}central bank digital currency (CBDC) architectures,
interbank messaging protocols, and \phantomsection\label{term:defi}DeFi smart contract platforms.

Beyond algorithmic security, the module emphasises
migration strategy, interoperability, and governance.
It connects post-quantum cryptographic theory with
financial system design choices, implementation trade-offs,
and compliance considerations, providing a realistic and
incremental path from classical cryptographic infrastructures
to quantum-resilient security primitives.

The distinction from the earlier modules is direct: here the quantum question is not where quantum hardware may accelerate a task, but where future quantum attacks invalidate classical security assumptions quickly enough that financial infrastructure must migrate in advance.

In this sense, Module~5 does not merely ``append'' security
to quantum-enabled finance, but defines the conditions under which
the capabilities introduced in previous modules
can be deployed, trusted, and sustained in a
post-quantum world.

\subsection{Motivation: Quantum Threats to Financial Infrastructure}
\label{sec:m5-motivation-quantum-threats-to-financial-infrastructure}

Modern financial systems rely critically on public-key cryptography
as the foundation for confidentiality, authentication, integrity,
and non-repudiation.
Protocols such as TLS (Transport Layer Security) for secure communication, PKI-based (Public Key Infrastructure) identity
and certificate infrastructures, digital signature schemes,
and cryptographic hash functions are deeply embedded in
banking systems, capital markets, payment rails, and distributed ledgers.

At the infrastructure level, these cryptographic primitives underpin:
(i) secure interbank messaging and settlement,
(ii) authentication of market participants and institutions,
(iii) legal enforceability of digital contracts and records, and
(iv) consensus and validation mechanisms in blockchain-based systems.
As a result, cryptographic failure does not merely imply data leakage,
but threatens systemic trust, legal finality, and financial stability.

Quantum algorithms-most notably Shor’s algorithm for integer factorisation
and discrete logarithms-render widely deployed public-key schemes such as
RSA, DSA (Digital Signature Algorithm), and ECC fundamentally insecure
once sufficiently large, fault-tolerant quantum computers become available see
\citep{shor1994algorithms}.
Since RSA and ECC form the backbone of most contemporary financial
authentication and signature infrastructures,
their cryptanalytic vulnerability constitutes a
\emph{structural, not incremental, risk}.

From a financial perspective, the threat is amplified by two characteristics
of cryptographic infrastructure.
First, cryptographic systems are deeply embedded and slow to replace,
often constrained by legacy software, hardware security modules (HSMs),
regulatory certification, and cross-border interoperability.
Second, financial data and records typically have long economic,
legal, and regulatory lifetimes, far exceeding the expected timeline
for the arrival of large-scale quantum computers.

\paragraph{Harvest-now, decrypt-later risk.}
A particularly severe and often underestimated threat is the
\emph{\phantomsection\label{term:hndl}harvest-now, decrypt-later} (HNDL) attack model.
Even in the absence of practical quantum computers today,
adversaries can intercept and store encrypted financial communications,
transaction records, and identity exchanges,
with the intention of decrypting them retroactively
once quantum capabilities mature.

This threat model is especially critical for:
\begin{itemize}
    \item long-term financial contracts and legal agreements,
    \item settlement and custody records,
    \item digital identity credentials and KYC data,
    \item regulatory archives and supervisory disclosures,
    \item blockchain transaction histories and smart contract states.
\end{itemize}

In these contexts, confidentiality must be preserved not merely for
months or years, but often for decades.
Consequently, cryptographic systems that are secure only against
classical adversaries already fail to meet the forward-security
requirements of financial institutions and regulators.

\paragraph{Implications for financial system design.}
The quantum threat to cryptography therefore represents a
\emph{time-delayed systemic risk}.
Unlike market or credit risks, it does not manifest gradually through
price signals, but materialises abruptly once cryptographic assumptions
collapse.
This creates a discontinuous risk profile, where assets, records,
or infrastructures may transition from secure to fully compromised
within a short technological window.

For this reason, post-quantum cryptography is not a niche technical upgrade,
but a strategic requirement for the long-term resilience of
financial infrastructure.
It motivates proactive migration strategies that can be deployed
well before quantum computers reach cryptographically relevant scale,
ensuring continuity of trust, compliance, and legal certainty
in a post-quantum environment.

\subsection{Classical and Post-Quantum Cryptographic Primitives}
\label{sec:m5-classical-and-post-quantum-cryptographic-primitives}

\subsubsection{Symmetric Cryptography}
\label{sec:m5-symmetric-cryptography}

Symmetric cryptographic primitives are comparatively robust against
quantum adversaries.
Unlike public-key systems, which are broken outright by Shor’s algorithm,
symmetric encryption and message authentication schemes are affected
primarily by Grover’s search algorithm, which provides only a
\emph{quadratic} speed-up over classical brute-force attacks
\citep{grover1996fast}.

In a classical setting, exhaustive key search against a symmetric cipher
with key length $k$ requires $O(2^{k})$ operations.
Grover’s algorithm reduces this complexity to $O(2^{k/2})$ in the quantum
setting.
While this represents a meaningful asymptotic improvement, it does not
fundamentally compromise symmetric cryptography.
Instead, it implies a straightforward and well-understood mitigation:
\emph{doubling the effective security margin by increasing key sizes}.

For example, \phantomsection\label{term:aes}AES-128 (Advanced Encryption Standard-128), which offers roughly $2^{128}$ classical security,
is reduced to approximately $2^{64}$ security against a quantum adversary,
which is no longer considered sufficient for long-term financial data
protection.
In contrast, AES-256 retains an effective quantum security level of
approximately $2^{128}$ operations under Grover-style attacks,
placing it well beyond foreseeable practical attack capabilities.
As a result, AES-256 is widely regarded as quantum-resistant in practice
and is recommended for post-quantum transition strategies
\citep{bernstein2009understanding,nistpqc2016}.

A similar argument applies to modern stream ciphers and authenticated
encryption schemes such as ChaCha20-Poly1305.
Their security ultimately reduces to key search and collision resistance,
both of which degrade only quadratically under quantum attacks.
By selecting sufficiently large keys and hash outputs,
their effective security remains robust in a post-quantum threat model.

\paragraph{Implications of Grover’s algorithm for financial systems.}
From an infrastructure perspective, the impact of Grover’s algorithm is
significant but manageable.
Symmetric cryptography remains viable as the backbone for:
(i) bulk data encryption,
(ii) secure communication channels,
(iii) storage encryption for financial records, and
(iv) integrity protection via message authentication codes (MACs).

However, Grover’s algorithm does impose practical constraints that are
particularly relevant for financial systems:
\begin{itemize}
    \item \textbf{Key length requirements:}
    legacy deployments using 112-bit or 128-bit symmetric keys may no longer
    meet long-horizon confidentiality requirements.
    \item \textbf{Latency and throughput trade-offs:}
    larger keys marginally increase computational and hardware costs,
    which must be accounted for in high-throughput systems such as
    payment rails and high-frequency trading infrastructure.
    \item \textbf{Hardware security modules (HSMs):}
    symmetric cryptographic acceleration and key management must be
    updated to support longer keys and post-quantum-compliant configurations.
\end{itemize}

Crucially, these adaptations do not require new cryptographic primitives,
but rather disciplined parameter choices and infrastructure upgrades.
This distinguishes symmetric cryptography from public-key cryptography,
where entirely new mathematical constructions are required.

\paragraph{Role in post-quantum architectures.}
In post-quantum financial architectures, symmetric cryptography
continues to play a central role.
Most post-quantum cryptographic schemes rely on symmetric primitives
internally-for example, in hash-based constructions, key derivation,
and hybrid encryption protocols.
As a result, the quantum resilience of symmetric cryptography forms
the stable foundation upon which post-quantum public-key systems are built.

In summary, while Grover’s algorithm necessitates careful reassessment
of security parameters, symmetric cryptography remains
\emph{fundamentally compatible with the post-quantum era}.
Its continued viability provides an anchor of continuity in the broader
cryptographic migration of financial infrastructure.

\subsubsection{Public-Key Cryptography and Quantum Vulnerability}
\label{sec:m5-public-key-cryptography-and-quantum-vulnerability}

In contrast to symmetric cryptography, public-key cryptosystems based on
integer factorisation or discrete logarithms are \emph{fundamentally broken}
by quantum computation.
The decisive result is Shor’s algorithm, which solves both the
integer factorisation problem and the discrete logarithm problem in
polynomial time on a fault-tolerant quantum computer
\citep{shor1994algorithms}.

This represents not a gradual degradation of security, but a
\emph{structural break} in cryptographic assumptions:
once sufficiently large quantum computers exist, the core hardness
assumptions underlying widely deployed public-key systems collapse entirely.

\paragraph{Affected cryptographic primitives.}
Shor’s algorithm directly compromises the following classes of primitives:
\begin{itemize}
    \item \textbf{RSA-based encryption and signatures,}
    whose security relies on the hardness of integer factorisation;
    \item \textbf{Diffie-Hellman (DH) and Elliptic-Curve Diffie-Hellman (ECDH),}
    whose security relies on the discrete logarithm problem;
    \item \textbf{Digital Signature Algorithm (DSA) and ECDSA,}
    which underpin authentication and non-repudiation in financial systems;
    \item \textbf{Public Key Infrastructure (PKI),}
    including X.509 certificates and certificate authorities,
    which inherit the vulnerability of the underlying signature schemes.
\end{itemize}

Unlike Grover’s algorithm, which merely reduces effective key strength,
Shor’s algorithm enables an attacker to recover private keys
\emph{efficiently and deterministically}.
As a result, increasing key sizes does not restore security:
no feasible parameter choice within these cryptosystems
can provide post-quantum protection.

\paragraph{Systemic implications for financial infrastructure.}
Public-key cryptography is deeply embedded in the trust fabric of
modern financial systems.
It underpins:
\begin{itemize}
    \item \textbf{Secure communication protocols} such as TLS,
    used in online banking, trading platforms, and payment systems;
    \item \textbf{Interbank messaging networks} (e.g.\ SWIFT),
    where authentication and non-repudiation are legally binding;
    \item \textbf{Blockchain and distributed ledger systems,}
    where digital signatures authorise asset transfers and smart contract execution;
    \item \textbf{Digital identity and access control frameworks,}
    including customer authentication and institutional credentialing.
\end{itemize}

The quantum vulnerability of public-key cryptography therefore constitutes
a \emph{systemic, cross-layer risk}.
Once private keys can be extracted, an adversary can:
(i) impersonate financial institutions,
(ii) forge legally binding signatures,
(iii) retroactively invalidate transaction authenticity,
and (iv) undermine trust in historical records and audit trails.

\paragraph{Irreversibility and long-horizon exposure.}
A particularly severe feature of this vulnerability is its irreversibility.
Any data, transaction, or signature protected today by RSA, DH, or ECC
may become retrospectively insecure once quantum decryption becomes feasible.
This is especially problematic for:
\begin{itemize}
    \item long-dated financial contracts,
    \item settlement and clearing records,
    \item regulatory filings and compliance archives,
    \item blockchain ledgers with permanent public availability.
\end{itemize}

In these contexts, cryptographic failure is not merely a technical issue,
but a legal, regulatory, and systemic stability concern.

\paragraph{Why migration is unavoidable.}
The impact of Shor’s algorithm implies that
public-key cryptography based on factorisation or discrete logarithms
has no viable long-term future in a quantum-capable world.
Unlike symmetric schemes, these systems cannot be salvaged
by parameter tuning or incremental upgrades.
Instead, they must be replaced by cryptographic primitives
based on fundamentally different hardness assumptions,
such as lattice problems, hash-based constructions, or code-based systems.

This observation motivates the development and standardisation of
\emph{post-quantum public-key cryptography},
which forms the core focus of the subsequent sections of this module.

\subsubsection{Post-Quantum Cryptography}
\label{sec:m5-post-quantum-cryptography}

Post-quantum cryptography (PQC) refers to cryptographic schemes that are designed
to remain secure against both classical and quantum adversaries,
\emph{without relying on quantum communication channels or quantum hardware}.
Unlike quantum key distribution (QKD), PQC is fully software-based and can be
deployed on existing digital infrastructure, making it the primary practical
response to quantum cryptographic threats in large-scale financial systems.

The security of PQC schemes is based on computational problems for which
no efficient classical or quantum algorithms are currently known.
Crucially, these problems are \emph{not reducible} to integer factorisation
or discrete logarithms, and therefore are not vulnerable to Shor’s algorithm.

\paragraph{Security assumptions and problem families.}
Leading PQC constructions are built on several hardness assumptions:
\begin{itemize}
    \item \textbf{Lattice-based problems}, such as Learning With Errors (LWE)
    and Module-LWE, which appear resistant to both classical and quantum attacks;
    \item \textbf{Hash-based constructions}, relying solely on the security of
    cryptographic hash functions and minimal algebraic structure;
    \item \textbf{Code-based cryptography}, based on the hardness of decoding
    random linear error-correcting codes;
    \item \textbf{Multivariate polynomial systems}, though these have seen
    mixed security outcomes and are currently less favoured.
\end{itemize}

Among these, lattice-based and hash-based schemes currently dominate
standardisation and deployment discussions due to their balance between
security, efficiency, and implementability.

\paragraph{NIST Post-Quantum Cryptography standardisation.}
To provide a coordinated migration path away from quantum-vulnerable
public-key systems, the U.S. National Institute of Standards and Technology (NIST)
launched a multi-round Post-Quantum Cryptography standardisation process.
After extensive cryptanalysis and performance evaluation, NIST selected
a small set of primary algorithms for standardisation.

\paragraph{NIST-selected PQC primitives.}
The leading candidates include:
\begin{itemize}
    \item \textbf{CRYSTALS-Kyber:}
    a lattice-based key encapsulation mechanism (KEM) designed to replace
    RSA and Diffie-Hellman for secure key exchange;
    \item \textbf{CRYSTALS-Dilithium:}
    a lattice-based digital signature scheme intended to replace ECDSA and RSA
    signatures in authentication and non-repudiation applications;
    \item \textbf{FALCON:}
    a lattice-based signature scheme offering significantly smaller signatures
    at the cost of more complex implementation requirements;
    \item \textbf{SPHINCS+:}
    a stateless hash-based signature scheme with very conservative security
    assumptions, trading efficiency and signature size for long-term robustness.
\end{itemize}

These algorithms collectively cover the two core public-key functionalities
required by financial systems: \emph{key establishment} and
\emph{digital signatures}.

\paragraph{Design trade-offs and financial relevance.}
From a financial-infrastructure perspective, PQC adoption involves
important trade-offs:
\begin{itemize}
    \item \textbf{Security margin vs efficiency:}
    lattice-based schemes offer efficient performance but rely on newer
    assumptions, while hash-based schemes offer minimal assumptions at higher cost;
    \item \textbf{Key and signature sizes:}
    PQC primitives typically involve larger keys and signatures than ECC,
    affecting bandwidth, storage, and ledger scalability;
    \item \textbf{Implementation complexity:}
    some schemes (e.g.\ FALCON) require careful numerical handling,
    raising operational and audit concerns in regulated environments.
\end{itemize}

These considerations are particularly relevant for:
interbank messaging protocols,
high-throughput payment systems,
blockchain and distributed ledgers,
and long-lived regulatory records.

\paragraph{Why PQC is the dominant migration path for finance.}
Unlike QKD, which requires new physical infrastructure and trusted channels,
PQC can be integrated into existing protocols such as TLS, PKI,
and blockchain signature schemes through software upgrades.
This makes PQC the \emph{only scalable and regulator-compatible option}
for protecting financial infrastructure against quantum adversaries
over the next several decades.

As a result, PQC is now viewed not as an optional enhancement,
but as a necessary foundation for maintaining trust, legal validity,
and systemic stability in a quantum-capable future.

\subsection{PQC in Financial Systems}
\label{sec:m5-pqc-in-financial-systems}

\subsubsection{Banking and Interbank Infrastructure}
\label{sec:m5-banking-and-interbank-infrastructure}

Modern banking and interbank infrastructures rely pervasively on public-key
cryptography to ensure confidentiality, authentication, integrity, and
non-repudiation across a wide range of mission-critical processes.
These cryptographic primitives are embedded not only in external-facing
communications, but also deep within internal operational and settlement layers.

Core use cases include:
\begin{itemize}
    \item secure interbank messaging and settlement coordination
    (e.g.\ SWIFT-like messaging systems and RTGS (Real-time Gross Settlement) infrastructures),
    \item authentication and key exchange in domestic and cross-border
    payment rails,
    \item TLS for APIs, cloud-based banking services,
    and data exchange between financial institutions and third-party providers.
\end{itemize}

Public-key cryptography underpins trust relationships across institutional
boundaries. As a result, quantum vulnerability at this layer represents a
\emph{systemic risk}: compromise of cryptographic primitives does not affect
individual transactions in isolation, but undermines the integrity of the
entire financial communication fabric.

\paragraph{Quantum risk amplification in interbank systems.}
Interbank infrastructures are particularly exposed to quantum threats for three
reasons.
First, messages and credentials often have long legal and operational lifetimes,
making them susceptible to harvest-now, decrypt-later attacks.
Second, trust is transitive: compromise of a single institution’s cryptographic
keys can cascade across correspondent banking networks.
Third, interbank systems operate under strict availability and reliability
requirements, limiting the feasibility of disruptive cryptographic upgrades.

These characteristics imply that post-quantum transition in banking cannot be
treated as a simple algorithmic replacement problem, but must be approached as
a coordinated infrastructure-level transformation.

\paragraph{Hybrid cryptographic migration strategies.}
Given these constraints, migration toward post-quantum cryptography is expected
to proceed through \emph{hybrid cryptographic deployments}, in which classical
and post-quantum algorithms coexist during a transitional period.
In a hybrid scheme, security relies on the conjunction of two primitives:
a classical algorithm (e.g.\ RSA or ECDH) and a post-quantum algorithm
(e.g.\ CRYSTALS-Kyber).

Formally, a hybrid key exchange produces a session key
\begin{equation}
K = \mathrm{KDF}(K_{\mathrm{classical}} \,\|\, K_{\mathrm{PQC}}),
\end{equation}
ensuring that confidentiality is preserved as long as at least one component
remains secure.
This approach provides backward compatibility with existing infrastructure
while offering forward security against future quantum adversaries.

\paragraph{Operational and regulatory considerations.}
From an operational perspective, hybrid deployments introduce additional
complexity:
larger key sizes, increased handshake latency, and expanded certificate formats.
However, these costs are generally manageable within modern banking networks and
are outweighed by the systemic risk of delayed migration.

From a regulatory standpoint, hybrid cryptography aligns well with prudential
risk-management principles.
It enables phased adoption, continuous validation, and auditability, all of
which are essential for regulated financial institutions.
As a result, supervisory bodies increasingly view hybrid PQC deployment as the
most realistic near-term pathway toward quantum-resilient banking infrastructure.

\paragraph{Implications for financial system stability.}
The transition to post-quantum cryptography in banking and interbank systems is
not merely a technical upgrade.
It directly affects legal certainty, operational continuity, and systemic trust.
Failure to migrate in a timely and coordinated manner risks creating
cryptographic single points of failure that could be exploited once
quantum-capable adversaries emerge.

Consequently, PQC adoption in interbank infrastructure should be understood as a
core component of financial stability policy, rather than a discretionary
cybersecurity enhancement.

\subsubsection{Blockchain and Distributed Ledger Systems}
\label{sec:m5-blockchain-and-distributed-ledger-systems}

Most existing public and permissioned blockchains rely on elliptic-curve-based
digital signature schemes, such as ECDSA or EdDSA (Edwards-curve Digital Signature Algorithm), to authenticate transactions
and enforce ownership of on-chain assets.
These schemes are secure against classical adversaries but become fundamentally
vulnerable in the presence of sufficiently powerful quantum computers.

At the protocol level, digital signatures serve as the primary cryptographic
primitive that links private keys to transaction validity.
As a result, quantum attacks on signature schemes directly threaten the
security assumptions underlying distributed ledger systems.

This raises a set of critical and interrelated questions:
\begin{itemize}
    \item \textbf{Ledger immutability under post-quantum adversaries:}
    whether historical blocks remain trustworthy if signature schemes are no
    longer cryptographically binding;
    \item \textbf{Key rotation and address migration:}
    how users and institutions can safely migrate from quantum-vulnerable keys
    to post-quantum credentials without disrupting ledger continuity;
    \item \textbf{Long-term validity of historical transactions:}
    whether past transactions remain legally and economically valid once the
    cryptographic assumptions under which they were created are broken.
\end{itemize}

\paragraph{Quantum attacks and the signature layer.}
In classical blockchain security models, ledger immutability is often described
as an emergent property of hash chaining and distributed consensus.
However, this immutability implicitly assumes that transaction signatures cannot
be forged.
Quantum algorithms - in particular Shor’s algorithm - invalidate this assumption
by enabling efficient recovery of private keys from public keys for ECDSA- and
EdDSA-based systems.

Once a private key is recovered, an adversary can:
(i) forge transactions,
(ii) retroactively spend previously controlled outputs, and
(iii) impersonate legitimate users or smart contracts.
This represents a \emph{structural break} in the security model of classical
blockchains, rather than a marginal degradation.

\paragraph{Exposure asymmetry and dormant keys.}
An important nuance is that not all blockchain addresses are equally exposed.
Addresses whose public keys have never been revealed on-chain
(e.g.\ pay-to-hash constructions) remain partially protected until a transaction
is spent.
In contrast, addresses with publicly visible keys - including reused addresses,
smart contracts, and custodial wallets - are immediately vulnerable once
quantum capabilities mature.

This asymmetry creates a race condition between defenders and attackers:
users must migrate assets to post-quantum-safe addresses before quantum
adversaries can exploit exposed public keys.

\paragraph{Migration challenges and governance constraints.}
Unlike traditional IT systems, blockchains lack a central authority capable of
mandating cryptographic upgrades.
Transitioning to post-quantum signatures therefore requires:
\begin{itemize}
    \item protocol-level changes (soft forks or hard forks),
    \item coordinated upgrades by validators, miners, and node operators,
    \item wallet- and application-level support for new signature schemes.
\end{itemize}

These technical challenges are compounded by governance and coordination issues,
particularly in large public networks where stakeholder incentives are
heterogeneous.
Delayed or fragmented migration increases systemic risk by leaving high-value
assets exposed for extended periods.

\paragraph{Post-quantum strategies for distributed ledgers.}
Several approaches have been proposed to enhance quantum resilience in
blockchain systems:
\begin{itemize}
    \item \textbf{Post-quantum signature replacement:}
    adopting lattice-based or hash-based signature schemes for transaction
    authentication;
    \item \textbf{Hybrid signature schemes:}
    requiring both classical and post-quantum signatures during a transitional
    phase to ensure backward compatibility;
    \item \textbf{Account abstraction and upgradable keys:}
    decoupling account identity from a fixed cryptographic primitive to allow
    flexible key rotation;
    \item \textbf{Checkpointing and cryptographic finality:}
    anchoring ledger states to external, quantum-resistant trust anchors
    (e.g.\ hash-based commitments or notarisation layers).
\end{itemize}

Each approach involves trade-offs between security, performance, and complexity,
and no single solution dominates across all blockchain architectures.

\paragraph{Financial transmission channels.}
For financial-market uses of distributed ledgers, the impact of quantum-vulnerable
signatures is broader than retail wallet theft. It directly affects tokenised
securities, stablecoins, collateral vaults, custody arrangements, and
smart-contract-controlled settlement instructions. If the key controlling a
tokenised bond, fund share, or collateral pool can be forged, the resulting
failure is not merely cryptographic: it can disrupt ownership records, margin
flows, segregation of client assets, and the enforceability of settlement
finality. In permissioned financial ledgers, the same logic extends to node
authentication, validator governance, and the evidentiary status of transaction
records.

\paragraph{System-wide implications for market structure.}
Post-quantum migration in blockchain finance also has market-design
consequences. Larger signatures and more complex verification rules can reduce
throughput, increase gas or validation costs, and force redesign of wallet,
custody, and smart-contract interfaces. But the opposite side of that trade-off
is strategically important: a credible post-quantum migration path can make
tokenised assets, regulated stablecoins, and on-chain settlement systems more
compatible with supervisory expectations around auditability, operational
resilience, and legal certainty. The blockchain question is therefore not only
how to preserve cryptographic security, but how to preserve market confidence as
digital-asset infrastructures mature.

\paragraph{Implications for financial and regulatory use cases.}
For financial applications of distributed ledgers - including tokenised assets,
stablecoins, and regulated DeFi protocols - quantum vulnerability at the
signature layer has direct legal and economic implications.
If transaction authenticity can no longer be cryptographically guaranteed,
questions arise regarding asset ownership, settlement finality, and auditability.

Consequently, post-quantum migration for blockchain systems is not merely a
technical concern, but a prerequisite for maintaining trust, compliance, and
long-term viability in quantum-aware financial infrastructures.

\subsubsection{Central Bank Digital Currencies (CBDCs)}
\label{sec:m5-central-bank-digital-currencies-cbdcs}

Central Bank Digital Currencies (CBDCs) operate at the intersection of
monetary sovereignty, national security, and large-scale financial
infrastructure.
Unlike permissionless blockchains, CBDC architectures are designed to support
long-lived public trust, state-level adversaries, and legally binding finality
of transactions.
As a result, post-quantum security requirements in CBDCs are significantly more
stringent than in most private-sector or decentralised systems.

CBDCs are expected to remain operational over decades and to safeguard
financial records whose confidentiality, authenticity, and integrity must be
preserved across generational time horizons.
This makes them particularly vulnerable to
\emph{harvest-now, decrypt-later} attacks and retroactive cryptographic
compromise once quantum capabilities mature.

\paragraph{Threat model and design constraints.}
The CBDC threat model differs fundamentally from that of commercial payment
systems or public blockchains.
It explicitly includes:
\begin{itemize}
    \item well-resourced state-level adversaries,
    \item long-term archival exposure of transaction and identity data,
    \item legal and regulatory requirements for auditability and non-repudiation,
    \item high systemic cost of cryptographic failure.
\end{itemize}

Consequently, CBDC security design must assume that classical public-key
cryptography will eventually be broken, and that cryptographic agility and
quantum resilience are not optional features but core system requirements.

\paragraph{Integration points for post-quantum cryptography.}
Post-quantum cryptographic primitives can be integrated into CBDC architectures
across multiple layers:
\begin{itemize}
    \item \textbf{Digital identity and wallet authentication:}
    replacing or augmenting classical PKI-based credentials with
    lattice-based or hash-based authentication schemes for citizens,
    financial institutions, and service providers;
    \item \textbf{Transaction signing and validation:}
    adopting post-quantum digital signature schemes to ensure that
    transaction authorisation remains secure against quantum adversaries;
    \item \textbf{Secure communication and key establishment:}
    deploying post-quantum key encapsulation mechanisms (KEMs) to protect
    network-level confidentiality and integrity;
    \item \textbf{Audit trails and regulatory oversight:}
    ensuring that supervisory access, forensic analysis, and compliance
    mechanisms remain verifiable and tamper-resistant in a post-quantum world.
\end{itemize}

\paragraph{Hybrid and layered cryptographic architectures.}
Given the scale and criticality of CBDC systems, most realistic migration paths
involve \emph{hybrid cryptographic designs}, in which classical and
post-quantum primitives coexist during extended transition periods.
Hybrid signatures and dual key-establishment mechanisms provide defence in depth,
ensuring that system security does not hinge on a single cryptographic assumption.

Such layered architectures also enable gradual migration of user wallets,
intermediaries, and backend systems, reducing operational risk and allowing
for staged regulatory approval.

\paragraph{Privacy, programmability, and post-quantum trade-offs.}
CBDC designs often balance competing objectives, including user privacy,
traceability, and programmability.
Post-quantum cryptographic schemes introduce new trade-offs in this context:
\begin{itemize}
    \item lattice-based schemes typically involve larger key and signature sizes,
    affecting bandwidth and storage requirements;
    \item hash-based signatures offer strong security assurances but impose
    state-management or one-time-use constraints;
    \item zero-knowledge and selective-disclosure mechanisms must be re-evaluated
    for quantum resistance.
\end{itemize}

These trade-offs must be assessed not only from a technical perspective, but
also in terms of user experience, legal enforceability, and interoperability
with existing financial infrastructure.

\paragraph{Strategic and regulatory implications.}
For central banks, post-quantum security is inseparable from monetary and
financial stability.
A cryptographic failure in a CBDC system would have consequences far beyond
technical disruption, potentially undermining public confidence in the currency
itself.

As a result, PQC adoption in CBDCs is increasingly viewed as a matter of
strategic foresight rather than speculative preparation.
Early integration of quantum-resilient primitives enables central banks to
future-proof digital currency systems while maintaining continuity with
existing regulatory frameworks and international standards.

In this sense, CBDCs represent one of the most compelling and high-impact
deployment domains for post-quantum cryptography in finance.

\subsubsection{DeFi and Smart Contracts}
\label{sec:m5-defi-and-smart-contracts}

Decentralised finance (DeFi) systems rely almost entirely on cryptographic
primitives to replace traditional trust intermediaries.
Public-key cryptography underpins user authentication, transaction
authorisation, smart contract execution, and on-chain governance.
As a result, the advent of quantum-capable adversaries poses fundamental
design challenges for DeFi protocols.

Unlike centrally governed financial systems, DeFi ecosystems are
permissionless, globally distributed, and highly path-dependent.
Once deployed, smart contracts are often immutable, and governance upgrades
are constrained by coordination frictions and incentive misalignment.
This makes post-quantum cryptographic migration in DeFi both technically and
institutionally complex.

\paragraph{Cryptographic dependence in DeFi.}
At a minimum, DeFi protocols depend on public-key cryptography for:
\begin{itemize}
    \item user authentication and transaction signatures,
    \item contract-level access control and role management,
    \item oracle authentication and cross-chain messaging,
    \item decentralised governance and voting mechanisms.
\end{itemize}

Most existing systems rely on ECDSA or EdDSA signatures, inheriting the same
quantum vulnerabilities as the underlying blockchain platforms.
A successful quantum attack against private keys would enable
fund theft, governance capture, and irreversible protocol compromise.

\paragraph{Post-quantum design trade-offs.}
Integrating post-quantum cryptography into DeFi introduces a distinct set of
trade-offs that differ from those in banking or CBDC contexts:
\begin{itemize}
    \item \textbf{Signature size and gas costs:}
    lattice- and hash-based signatures are significantly larger than ECDSA,
    increasing on-chain storage requirements and transaction fees;
    \item \textbf{Backward compatibility:}
    existing contracts and user wallets are tightly coupled to classical
    cryptographic assumptions, making retrofitting PQC non-trivial;
    \item \textbf{State and key management:}
    some PQC schemes (e.g.\ hash-based signatures) impose stateful or
    one-time-use constraints that are difficult to reconcile with
    smart contract execution models.
\end{itemize}

These constraints imply that naive substitution of cryptographic primitives
is unlikely to be viable for large-scale DeFi systems.

\paragraph{Hybrid and staged migration strategies.}
A more realistic approach involves \emph{hybrid cryptographic designs}, where
classical and post-quantum primitives coexist.
Examples include:
\begin{itemize}
    \item dual-signature schemes requiring both ECDSA and PQC-based signatures,
    \item contract-level opt-in mechanisms for post-quantum authentication,
    \item off-chain or layer-2 aggregation of post-quantum proofs to reduce
    on-chain verification costs.
\end{itemize}

Such strategies allow DeFi protocols to preserve backward compatibility while
gradually increasing quantum resilience.

\paragraph{Governance and protocol evolution.}
DeFi governance itself is a critical attack surface in a post-quantum setting.
Token-based voting, multisignature treasuries, and timelocked upgrades all rely
on classical cryptographic authentication.
Quantum attacks could enable adversaries to capture governance power and push
malicious protocol upgrades.

Post-quantum secure governance therefore requires:
\begin{itemize}
    \item quantum-resilient voting and delegation mechanisms,
    \item secure key rotation and recovery processes,
    \item cryptographic auditability of historical governance actions.
\end{itemize}

These considerations elevate PQC from a purely technical concern to a core
component of DeFi protocol design and economic security.

\paragraph{Implications for programmability and composability.}
DeFi’s defining features - composability and permissionless programmability - 
further complicate PQC adoption.
Protocols are interconnected through shared contracts, standards, and
assumptions.
Introducing post-quantum primitives in one protocol can have cascading effects
on others.

As a result, PQC adoption in DeFi is likely to proceed unevenly, driven by
systemic risk awareness, high-value protocol exposure, and regulatory pressure,
rather than by immediate performance considerations.

In summary, post-quantum cryptography represents both a technical challenge and
a governance stress test for decentralised finance.
Its successful integration will require not only cryptographic innovation,
but also careful coordination across protocol design, economic incentives,
and community governance structures.

\subsection{Migration Strategies, Hybrid Cryptography, and Crypto-Agility}
\label{sec:m5-migration-strategies-hybrid-cryptography-and-crypto-agility}

Given the uncertainty surrounding the timeline, scale, and capabilities of
fault-tolerant quantum computers, an abrupt transition from classical
cryptography to post-quantum cryptography is neither feasible nor desirable
for financial systems.
Instead, realistic deployment paths rely on \emph{hybrid and transitional
architectures} that balance security, performance, interoperability, and
regulatory compliance.

\paragraph{Rationale for hybrid cryptographic designs.}
Hybrid cryptographic schemes combine classical and post-quantum primitives
within a single protocol or security layer, ensuring that the system remains
secure as long as \emph{at least one} component is uncompromised.
This approach mitigates both premature migration risks and delayed-response
risks associated with quantum uncertainty.

Formally, a hybrid scheme enforces security under the assumption:
\begin{equation}
\text{Security} = \text{Classical secure} \;\lor\; \text{Post-quantum secure}.
\end{equation}
Such constructions are particularly attractive in financial infrastructure,
where risk tolerance is low and backward compatibility is essential.

\paragraph{Hybrid deployment patterns.}
Several hybrid patterns are emerging as practical migration mechanisms:
\begin{itemize}
    \item \textbf{Hybrid key exchange:}
    combining classical Diffie-Hellman or ECDH with lattice-based KEMs
    (e.g.\ Kyber) to derive session keys;
    \item \textbf{Dual-signature schemes:}
    requiring both classical (e.g.\ ECDSA) and post-quantum signatures
    for transaction authorisation or governance actions;
    \item \textbf{Layered security models:}
    deploying PQC at transport or application layers
    while maintaining classical cryptography at legacy interfaces;
    \item \textbf{Off-chain and aggregation techniques:}
    performing post-quantum verification off-chain or in batch form
    to reduce on-chain or real-time computational overhead.
\end{itemize}

These patterns allow gradual adoption without disrupting
existing operational workflows or contractual obligations.

\paragraph{Crypto-agility as a design principle.}
Beyond specific algorithm choices, long-term resilience requires
\emph{crypto-agility}: the ability to rapidly replace or upgrade
cryptographic primitives without redesigning the entire system.
Crypto-agility is especially critical in finance, where systems must remain
secure over decades while threat models evolve.

Key enablers of crypto-agility include:
\begin{itemize}
    \item abstraction layers separating cryptographic primitives
    from business logic,
    \item modular key management and certificate infrastructures,
    \item algorithm-agnostic protocol negotiation mechanisms,
    \item governance processes for emergency cryptographic upgrades.
\end{itemize}

From this perspective, post-quantum migration is not a one-time event,
but an ongoing risk-management capability.

\paragraph{Governance and regulatory coordination.}
Hybrid cryptographic architectures raise governance and compliance questions
that extend beyond technical implementation.
Financial institutions operate under strict regulatory regimes governing
auditability, data retention, and operational resilience.

Effective PQC migration therefore requires:
\begin{itemize}
    \item alignment with regulatory timelines and supervisory guidance,
    \item standardised migration playbooks across institutions,
    \item certification and assurance frameworks for PQC implementations,
    \item coordinated industry adoption to avoid fragmentation.
\end{itemize}

International bodies and standard-setting organisations play a central role
in reducing coordination risk and ensuring interoperability across borders.

\paragraph{Interaction with quantum-enabled financial systems.}
In a quantum augmented financial stack, cryptographic migration cannot be
considered in isolation.
Hybrid PQC architectures must coexist with quantum optimisation,
quantum risk engines, and quantum machine learning components introduced
in earlier modules.

In particular:
\begin{itemize}
    \item cryptographic trust anchors must protect quantum-generated outputs,
    \item audit trails must remain verifiable under post-quantum assumptions,
    \item secure interfaces between classical and quantum subsystems
    must be explicitly defined.
\end{itemize}

This positions post-quantum cryptography as a \emph{horizontal security layer}
that enables, rather than constrains, the deployment of quantum-enhanced
financial analytics and decision systems.

\paragraph{Outlook.}
Hybrid cryptography and crypto-agility provide a pragmatic bridge between
today’s classical infrastructure and a quantum-capable future.
They allow financial systems to hedge cryptographic risk in much the same way
they hedge market or credit risk: through diversification, redundancy,
and forward-looking governance.

As quantum technologies mature, these transitional architectures will form the
foundation upon which fully post-quantum-native financial systems can be built,
without sacrificing stability, trust, or legal continuity.

\subsection{Beyond Cryptography: System-Level Quantum Safety}
\label{sec:m5-beyond-cryptography-system-level-quantum-safety}

While cryptographic primitives are a foundational component of quantum
resilience, \emph{cryptography alone does not guarantee system security}.
Financial infrastructures are complex socio-technical systems, and their
robustness against quantum-era threats depends equally on system architecture,
operational controls, and governance mechanisms.

A truly quantum-resilient financial architecture therefore requires a
\emph{defence-in-depth} approach, in which post-quantum cryptography is
embedded within a broader system-level security framework.

\paragraph{Security beyond algorithms.}
Even in a post-quantum world, many of the most effective attacks against
financial systems are unlikely to be purely cryptanalytic.
Instead, adversaries will continue to exploit:
misconfigured access controls, credential compromise, insider threats,
software vulnerabilities, and behavioural anomalies.

As a result, several security layers remain largely unaffected by quantum
algorithms and continue to play a critical protective role:
\begin{itemize}
    \item \textbf{Access control and identity management:}
    role-based and attribute-based access control (RBAC/ABAC),
    hardware security modules (HSMs), and secure key custody;
    \item \textbf{Multi-factor authentication (MFA):}
    combining cryptographic credentials with biometric, behavioural,
    or hardware-based authentication factors;
    \item \textbf{Intrusion detection and anomaly monitoring:}
    network- and application-level monitoring, behavioural baselining,
    and real-time alerting;
    \item \textbf{Secure audit logs and compliance tooling:}
    tamper-evident logging, regulatory reporting, and forensic traceability.
\end{itemize}

These mechanisms provide resilience against a broad class of operational and
adversarial risks that are orthogonal to quantum computation.

\paragraph{Quantum safety as a system property.}
From a systems perspective, quantum safety should be understood as an
\emph{emergent property} rather than a feature of any single cryptographic
algorithm.
A system may employ post-quantum cryptography yet remain insecure if
key management, access policies, or monitoring layers are weak.

Conversely, a well-designed system with strong identity controls,
segmentation, and anomaly detection can significantly mitigate damage even if
certain cryptographic components become vulnerable.
This mirrors long-standing principles in financial risk management, where
controls, redundancy, and monitoring reduce tail risks rather than eliminate
them.

\paragraph{Role of monitoring and learning-based defences.}
As quantum technologies evolve, system-level monitoring becomes even more
important.
Advanced intrusion detection and fraud monitoring systems increasingly rely on
machine learning models to detect deviations from normal behaviour.

In this context, quantum machine learning methods introduced in Module~4
can serve as complementary tools for:
\begin{itemize}
    \item detecting anomalous transaction patterns,
    \item identifying regime shifts in network or system behaviour,
    \item monitoring model drift and emerging attack signatures.
\end{itemize}

Crucially, these learning-based defences operate independently of the
cryptographic hardness assumptions threatened by quantum algorithms.

\paragraph{Auditability, compliance, and legal resilience.}
Financial systems are subject not only to technical security requirements but
also to legal and regulatory obligations.
Audit trails, record immutability, and compliance tooling must remain reliable
over long time horizons, often spanning decades.

Post-quantum system safety therefore requires:
\begin{itemize}
    \item cryptographically protected audit logs with post-quantum assurances,
    \item verifiable access histories and change management records,
    \item regulatory frameworks that recognise hybrid and transitional security
    architectures.
\end{itemize}

These requirements reinforce the importance of system-level design choices
that extend beyond algorithm selection.

\paragraph{Implications for quantum-augmented finance.}
In a quantum-augmented financial stack, cryptography, learning, optimisation,
and risk engines coexist.
System-level quantum safety ensures that:
\begin{itemize}
    \item outputs from quantum risk or optimisation modules remain trustworthy,
    \item decision-making pipelines are auditable and explainable,
    \item failures or compromises in one layer do not cascade system-wide.
\end{itemize}

Post-quantum cryptography thus functions as a necessary but insufficient
condition for security.
Only when combined with robust system controls, monitoring, and governance does
it enable a resilient transition toward quantum-era financial infrastructure.

\paragraph{Synthesis.}
Beyond cryptographic migration, quantum safety in finance is ultimately a
systems engineering and governance challenge.
By embedding PQC within layered security architectures, financial institutions
can manage quantum risk in a manner consistent with existing principles of
operational resilience, regulatory compliance, and long-term trust.

\subsection{Integration Across the Quantum Finance Stack}
\label{sec:m5-integration-across-the-quantum-finance-stack}

Post-quantum cryptography (PQC) functions as a horizontal security and trust layer
across the entire quantum finance stack developed in Modules~1--4.
While each module targets a distinct computational capability - optimisation,
pricing, risk estimation, and machine learning - they all rely on secure
communication, authenticated execution, and long-term data integrity.

From a systems perspective, PQC does not replace the functionality of these
modules; rather, it enables their safe deployment in adversarial, regulated,
and long-lived financial environments.

\paragraph{Integration with Module~1 (Quantum Optimisation).}
Quantum optimisation workflows involve sensitive inputs (return forecasts,
covariance matrices, constraints) and high-value outputs (portfolio allocations,
capital deployment decisions).
PQC ensures:
\begin{itemize}
    \item secure transmission of optimisation inputs between data providers,
    quantum backends, and classical controllers,
    \item authentication and integrity of optimisation results returned by
    quantum or hybrid solvers,
    \item tamper-resistant audit trails for regulatory and compliance review.
\end{itemize}
This is particularly relevant in hybrid architectures, where optimisation logic
is split across classical infrastructure and remote quantum hardware.

\paragraph{Integration with Module~2 (Quantum Pricing).}
Quantum amplitude estimation and related pricing algorithms often operate on
confidential market data, proprietary models, and legally binding contract parameters.
PQC supports:
\begin{itemize}
    \item confidentiality of pricing inputs and payoff structures,
    \item integrity of pricing outputs used for valuation, margining, or settlement,
    \item long-term protection of pricing records against retrospective decryption
    under future quantum adversaries.
\end{itemize}
This is critical for derivatives with long maturities and for regulatory
record-keeping obligations extending over decades.

\paragraph{Integration with Module~3 (Quantum Risk and Scenario Analysis).}
Risk engines generate stress scenarios, tail distributions, and systemic risk
signals that directly inform capital buffers and supervisory decisions.
PQC enables:
\begin{itemize}
    \item secure storage and transmission of stress scenarios and tail-risk paths,
    \item protection of institution-specific risk exposures from strategic leakage,
    \item cryptographic integrity of scenario libraries shared across institutions
    or regulators.
\end{itemize}
In this sense, PQC complements quantum risk estimation by ensuring that
risk intelligence remains trustworthy and non-manipulable.

\paragraph{Integration with Module~4 (Quantum Machine Learning).}
Quantum and hybrid machine-learning pipelines depend on high-quality datasets,
trained models, and decision outputs.
PQC provides safeguards for:
\begin{itemize}
    \item confidentiality of training data and feature representations,
    \item authenticity and version control of trained models,
    \item secure dissemination of regime signals, forecasts, or similarity structures
    into downstream decision layers.
\end{itemize}
This is especially important where QML outputs feed directly into trading,
risk management, or automated control systems.

\paragraph{System-level perspective.}
Taken together, PQC enables quantum finance modules to operate as part of a
coherent, production-grade infrastructure rather than isolated experimental tools.
It ensures that advances in optimisation, pricing, risk, and learning are not
undermined by cryptographic fragility.

In this sense, Module~5 completes the quantum finance stack:
it anchors computational innovation within a security framework that is
compatible with adversarial environments, regulatory scrutiny, and the
long time horizons characteristic of financial systems.

\subsection{Illustrative Case Studies}
\label{sec:m5-case-studies}

To complement the conceptual and architectural discussion,
this module may optionally include a small number of focused case illustrations.
These cases are not intended as full implementation blueprints,
but as concrete demonstrations of how post-quantum cryptography
can be integrated into existing financial systems under realistic constraints. 

Each case is deliberately scoped at the system and governance level, reflecting the fact that post-quantum security challenges in finance are primarily organisational and architectural, rather than purely cryptographic. The emphasis is on \emph{migration logic}, \emph{design trade-offs},
and \emph{operational feasibility}, rather than cryptographic novelty. 

\paragraph{Case 1: Post-Quantum Migration of Blockchain Signature Schemes.}
Most public blockchains rely on elliptic-curve-based signatures
(e.g.\ ECDSA or EdDSA) for transaction authentication.
This case illustrates a staged migration strategy in which:
\begin{itemize}
    \item post-quantum signature schemes (e.g.\ Dilithium or SPHINCS+)
    are introduced alongside existing elliptic-curve signatures,
    \item addresses or accounts are gradually migrated to PQC-secured formats,
    \item legacy signatures remain verifiable to preserve historical
    ledger validity.
\end{itemize}
The case highlights trade-offs between signature size, verification cost,
and backward compatibility, as well as governance challenges
in coordinating network-wide cryptographic upgrades.

\paragraph{Case 2: Hybrid TLS and PQC Deployment in Interbank Communication.}
Secure interbank messaging and API-based financial services rely heavily on
TLS and PKI infrastructures.
This case examines a hybrid TLS configuration in which:
\begin{itemize}
    \item classical key exchange (e.g.\ ECDHE) is combined with a
    post-quantum key encapsulation mechanism (e.g.\ CRYSTALS-Kyber),
    \item session keys remain secure even if one cryptographic primitive
    is later compromised,
    \item certificate authorities and key management systems are adapted
    to support PQC-enabled credentials.
\end{itemize}
The illustration demonstrates how PQC can be deployed incrementally
without disrupting existing operational workflows or regulatory compliance.

\paragraph{Case 3: Quantum-Resilient Audit Trails for DeFi Protocols.}
DeFi protocols generate on-chain records that serve as both execution logs
and audit evidence.
This case considers the design of quantum-resilient audit trails by:
\begin{itemize}
    \item anchoring smart-contract events to hash-based or lattice-based
    signature schemes,
    \item combining cryptographic commitments with off-chain compliance logs,
    \item ensuring long-term verifiability of governance decisions,
    liquidations, and protocol upgrades.
\end{itemize}
The focus is on preserving transparency and accountability
under future quantum adversaries, rather than on immediate performance gains.

\paragraph{Pedagogical and practical role of the cases.}
These illustrations serve three purposes within the module:
\begin{itemize}
    \item to translate abstract PQC concepts into concrete financial settings,
    \item to expose system-level trade-offs faced by practitioners,
    \item to provide a foundation for further technical or policy-oriented study.
\end{itemize}

Importantly, the case studies reinforce a central message of this module:
\emph{post-quantum security is a migration problem, not a single algorithmic switch}.
Effective adoption depends as much on governance, interoperability,
and operational design as on cryptographic primitives themselves.

\subsection{Implementation Constraints}
\label{sec:m5-limitations-and-outlook}

Despite rapid progress in post-quantum cryptography,
significant challenges remain before quantum-resilient security
can be fully embedded into global financial infrastructure.

\paragraph{Current limitations.}
Several constraints shape the near- to medium-term outlook:
\begin{itemize}
    \item \textbf{Cryptanalytic uncertainty.}
    While leading PQC schemes are designed to resist known quantum attacks,
    their long-term robustness depends on assumptions about lattice,
    code-based, or hash-based hardness that continue to be actively studied.
    Unlike RSA or ECC, which benefited from decades of cryptanalytic scrutiny,
    many PQC primitives have comparatively shorter security track records.

    \item \textbf{Performance and resource overhead.}
    Post-quantum schemes typically involve larger keys, signatures,
    and ciphertexts, leading to higher computational cost,
    increased bandwidth usage, and greater storage requirements.
    These overheads are particularly relevant in high-throughput
    financial systems, low-latency trading environments,
    and resource-constrained blockchain platforms.

    \item \textbf{Migration and operational complexity.}
    Financial infrastructure is deeply layered and long-lived.
    Migrating cryptographic primitives requires coordinated changes across
    software, hardware, governance frameworks, legal contracts,
    and regulatory reporting systems.
    Legacy dependencies and backward-compatibility requirements
    significantly slow the pace of transition.
\end{itemize}

These limitations imply that post-quantum security is not a
single technological upgrade, but a multi-year systems-engineering
and governance challenge.

\paragraph{Risk-based perspective.}
From a financial standpoint, the transition to PQC should be understood
through a risk-management lens rather than a deterministic timeline.
Quantum threats represent a \emph{low-probability but high-impact}
systemic risk.
As with market, credit, or operational risks, the appropriate response
is not to wait for certainty, but to reduce exposure proactively.

In this sense, early PQC adoption acts as a form of
\emph{cryptographic risk hedging}:
even if large-scale fault-tolerant quantum computers arrive later
than anticipated, the cost of preparedness is outweighed by the
reduction in tail risk associated with catastrophic cryptographic failure.

\paragraph{Outlook: toward crypto-agile financial systems.}
Looking forward, the most important structural development is not
any single post-quantum algorithm, but the emergence of
\emph{crypto-agile system design}.
Hybrid cryptography, algorithm agility, and modular security architectures
enable financial systems to adapt as both quantum capabilities
and cryptographic knowledge evolve.

Rather than aiming for a once-and-for-all migration,
future financial infrastructure is likely to:
\begin{itemize}
    \item support multiple cryptographic primitives in parallel,
    \item allow rapid algorithm substitution without system-wide disruption,
    \item integrate cryptographic governance into broader
    risk and compliance frameworks.
\end{itemize}

As quantum technologies mature,
these transitional architectures will form the foundation upon which
fully post-quantum-native financial systems can be built.
Crucially, this evolution can occur without sacrificing
operational stability, legal continuity, or public trust.

\paragraph{Position within the quantum finance stack.}
Within the broader programme, Module~5 completes the architectural picture.
While Modules~1--4 focus on computation, optimisation, pricing,
risk estimation, and learning under quantum enhancement,
post-quantum cryptography ensures that the resulting systems
remain trustworthy, auditable, and legally robust in a quantum-capable world.

In this sense, post-quantum cryptography is not merely a defensive add-on,
but a prerequisite for sustainable quantum-augmented finance.
It anchors innovation in a security framework capable of
supporting the long time horizons, systemic importance,
and regulatory scrutiny characteristic of modern financial systems.

\subsection{Synthesis and Outlook}
\label{sec:m5-concluding-remarks}

Module~5 completes the programme by establishing the cryptographic
and security foundations required for quantum-augmented finance.
While the preceding modules focus on computational advantage,
expressivity, and decision intelligence,
post-quantum cryptography ensures that these capabilities
can be deployed without undermining trust, legal certainty,
or systemic stability.

Taken together, Modules~1--4 demonstrate how quantum technologies
can enhance optimisation, pricing, risk estimation, and learning
within financial systems.
Module~5 provides the indispensable counterbalance:
it addresses the security externalities introduced by quantum computation
and defines a principled migration path away from cryptographic primitives
that will eventually become obsolete.

From a systems perspective, post-quantum cryptography transforms
quantum finance from a collection of algorithmic advances
into a coherent and deployable infrastructure.
It ensures that quantum-enhanced workflows remain:
(i) cryptographically sound over long time horizons,
(ii) compatible with regulatory and legal requirements,
and (iii) resilient against both classical and quantum adversaries.

Importantly, this module reframes security as a design constraint
rather than an afterthought.
By embedding cryptographic agility, hybrid defences,
and governance-aware security models at the architectural level,
financial systems can evolve toward quantum capability
without repeated disruptive overhauls.

In this sense, Module~5 does not merely conclude the programme;
it anchors it.
It ensures that gains in speed, precision, and learning power
translate into sustainable financial innovation,
aligned with the long-lived nature of contracts, institutions,
and public trust.

Collectively, the five modules outline a coherent architecture
for quantum-augmented financial infrastructure:
one that is computationally powerful, analytically expressive,
and secure by design in a post-quantum world.

\section{Integrated Perspective and Research Agenda}
\label{sec:integration}
Taken together, the five modules suggest that quantum finance should not be framed as a single algorithmic claim, but as a layered transition in financial computation. Optimisation, pricing, and risk estimation supply the decision and valuation machinery; machine learning adds a structure-sensitive modelling layer; and post-quantum cryptography ensures that these gains can be embedded in long-lived institutional systems without creating unacceptable security fragility.

A clear next editorial step is to strengthen the bridges between modules with more shared examples. The current revision already moves in that direction by introducing a reproducible U.S.-equity optimisation benchmark, a consolidated computational appendix, and a common author-year reference framework. A natural next step is therefore to propagate the same level of empirical integration across pricing, risk, learning, and security so that one financial setting can anchor the full review more explicitly.

As a first integrated draft, however, this manuscript now provides a stable base file for exactly that process: the original arguments are preserved, the review reads as one document rather than six separate papers, and the LaTeX structure is set up for iterative refinement without modifying the original module files.

\clearpage
\appendix
\section{Computational Appendix and Reproducibility Assets}
\label{app:computational-assets}

This appendix records the scope and validation status of the executable materials supporting the review. To keep the manuscript stylistically consistent, no inline code listings are reproduced here. All executable content is instead referenced through curated package names and filenames only. \Cref{tab:appendix-asset-packages,tab:appendix-validation-status} summarise the package structure and rerun status, while the public notebook repository associated with the review is \CodeRepo.

\subsection{Asset Package Structure}
\label{sec:app-asset-package-structure}

\begin{table}[H]
\centering
\small
\caption{Canonical asset packages supporting the review.}
\label{tab:appendix-asset-packages}
\begin{tabular}{p{0.11\linewidth}p{0.22\linewidth}p{0.27\linewidth}p{0.28\linewidth}}
\toprule
\textbf{Module} & \textbf{GitHub package} & \textbf{Public notebook} & \textbf{Scope} \\
\midrule
Module 1 & \repodir{module_01_optimisation} & \repofile{module_01_optimisation/module_01_classical_vs_qaoa_portfolio.ipynb}{module_01_classical_vs_qaoa_portfolio.ipynb} & Local classical-versus-QAOA portfolio benchmark, supporting data extracts, summary tables, and manuscript figure. \\
Module 2 & \repodir{module_02_pricing} & \repofile{module_02_pricing/module_02_asian_option_qae.ipynb}{module_02_asian_option_qae.ipynb} & Asian-option pricing notebook together with simulated pricing data exports, the circuit schematic, and the workflow diagram used in the chapter. \\
Module 3 & \repodir{module_03_risk} & \repofile{module_03_risk/module_03_single_asset_var_cvar.ipynb}{module_03_single_asset_var_cvar.ipynb} and \repofile{module_03_risk/module_03_multi_asset_var_cvar.ipynb}{module_03_multi_asset_var_cvar.ipynb} & Single-asset and multi-asset VaR/CVaR notebooks, local market-data extracts, refreshed summary tables, a data manifest, the stress-workflow diagram, and updated loss-distribution figures. \\
Module 4 & \repodir{module_04_qml} & \repofile{module_04_qml/module_04_case_a_quantum_classification.ipynb}{module_04_case_a_quantum_classification.ipynb} and \repofile{module_04_qml/module_04_case_b_macro_regime_learning.ipynb}{module_04_case_b_macro_regime_learning.ipynb} & Two notebook-based learning cases covering SPY direction classification and macro-regime classification, together with bundled local SPY/FRED snapshots, data manifests, and executed results tables. \\
Module 5 & Not applicable & Not applicable & Reference-driven policy and architecture chapter; no standalone executable notebook is maintained in the current version. \\
\bottomrule
\end{tabular}
\end{table}

The legacy notebook holdings are still retained for continuity with the earlier combined draft, but the chapter-level packages listed above are now the cleanest route for later auditing and revision.

\subsection{Execution and Validation Status}
\label{sec:app-execution-and-validation-status}

\begin{table}[H]
\centering
\small
\caption{Validation status of the public computational assets on 26 March 2026.}
\label{tab:appendix-validation-status}
\begin{tabular}{p{0.16\linewidth}p{0.13\linewidth}p{0.25\linewidth}p{0.34\linewidth}}
\toprule
\textbf{Asset} & \textbf{Status} & \textbf{Public notebook} & \textbf{Validation note} \\
\midrule
Module 1 benchmark & Executed & \repofile{module_01_optimisation/module_01_classical_vs_qaoa_portfolio.ipynb}{module_01_classical_vs_qaoa_portfolio.ipynb} & The notebook was rerun on the fixed 2024-01-01 to 2025-12-31 window; classical search selected \texttt{AMGN / COST / GOOGL}, while QAOA returned the nearby portfolio \texttt{COST / CSX / GOOGL}. \\
Module 2 pricing case & Executed & \repofile{module_02_pricing/module_02_asian_option_qae.ipynb}{module_02_asian_option_qae.ipynb} & The notebook ran successfully and reproduced the hybrid QAE workflow. The executed outputs recorded \(V_{\mathrm{MC}}=5.1866\), exact QAE price \(=5.1603\), and shot-based QAE price \(=5.1071\). \\
Module 3 single-asset case & Executed & \repofile{module_03_risk/module_03_single_asset_var_cvar.ipynb}{module_03_single_asset_var_cvar.ipynb} & The notebook ran successfully; the manuscript table and figure were refreshed to the executed VaR/CVaR outputs. \\
Module 3 multi-asset case & Executed & \repofile{module_03_risk/module_03_multi_asset_var_cvar.ipynb}{module_03_multi_asset_var_cvar.ipynb} & The notebook ran successfully; the manuscript table and figure were refreshed to the executed multi-asset VaR/CVaR outputs. \\
Module 4 Case A & Executed & \repofile{module_04_qml/module_04_case_a_quantum_classification.ipynb}{module_04_case_a_quantum_classification.ipynb} & The notebook was rerun successfully on the fixed 2024-01-01 to 2025-12-31 window; the manuscript table now reflects the executed SPY classification outputs. \\
Module 4 Case B & Executed & \repofile{module_04_qml/module_04_case_b_macro_regime_learning.ipynb}{module_04_case_b_macro_regime_learning.ipynb} & The notebook was rerun successfully on the fixed 2024-01-01 to 2025-12-31 window; the regime-classification table in the manuscript now reflects the executed outputs. \\
Module 5 & Not applicable & Not applicable & This chapter is reference-driven and architectural rather than notebook-driven in the current review. \\
\bottomrule
\end{tabular}
\end{table}

\subsection{Module 1: Local Optimisation Benchmark}
\label{sec:app-module-1-local-optimisation-benchmark}

The executable companion to \Cref{sec:module1-local-case} is curated in package \repodir{module_01_optimisation}. The principal notebook is \repofile{module_01_optimisation/module_01_classical_vs_qaoa_portfolio.ipynb}{module_01_classical_vs_qaoa_portfolio.ipynb}. The six raw input price files sit beside the generated summary tables \repofile{module_01_optimisation/module_01_qaoa_portfolio_summary.csv}{module_01_qaoa_portfolio_summary.csv} and \repofile{module_01_optimisation/module_01_qaoa_frontier.csv}{module_01_qaoa_frontier.csv}. The manuscript figure is \assetfile{module_01_classical_vs_qaoa_portfolio_case.pdf}. The public repository entry point is \CodeRepo.

This benchmark is intentionally small enough to permit exact classical enumeration. Its purpose is therefore methodological validation rather than a claim of near-term quantum advantage. The value of the case is that the financial problem, the QUBO mapping, the classical benchmark, and the QAOA result are all auditable from local project assets.

\subsection{Module 2: Asian-Option Pricing Case}
\label{sec:app-module-2-asian-option-pricing-case}

The executable companion to \Cref{sec:m2-case-study-quantum-pricing-of-an-asian-option} is curated in package \repodir{module_02_pricing}. The canonical notebook is \repofile{module_02_pricing/module_02_asian_option_qae.ipynb}{module_02_asian_option_qae.ipynb}. The pricing chapter also uses two workflow figures, \assetfile{asian_qae_circuit.pdf} and \assetfile{hybrid_qae_pipeline.pdf}. To make the simulation inputs auditable, the local asset package now also includes \repofile{module_02_pricing/module_02_asian_option_histogram.csv}{module_02_asian_option_histogram.csv}, \repofile{module_02_pricing/module_02_asian_option_pricing_summary.csv}{module_02_asian_option_pricing_summary.csv}, and \repofile{module_02_pricing/module_02_asian_option_parameters.json}{module_02_asian_option_parameters.json}. The public repository entry point is \CodeRepo.

The executed notebook confirms the logic of the hybrid QAE construction described in the main text. In the local execution recorded above, the Monte Carlo benchmark price was \(5.1866\), the exact amplitude-based reconstruction produced \(5.1603\), and the shot-based estimate produced \(5.1071\). The small gap is consistent with finite-shot noise and does not alter the qualitative interpretation in the chapter.

\subsection{Module 3: Risk Estimation Cases}
\label{sec:app-module-3-risk-estimation-cases}

The executable companions to the case studies in \Cref{sec:m3-case-studies} are curated in package \repodir{module_03_risk}. The single-asset workflow is filed as \repofile{module_03_risk/module_03_single_asset_var_cvar.ipynb}{module_03_single_asset_var_cvar.ipynb}, and the multi-asset workflow is filed as \repofile{module_03_risk/module_03_multi_asset_var_cvar.ipynb}{module_03_multi_asset_var_cvar.ipynb}. The corresponding figures are packaged together with the workflow diagram \assetfile{hybrid_risk_pipeline.pdf}. The local data folder now also preserves the copied market-data inputs \repofile{module_03_risk/AAPL_2024-01-01_2025-12-31_daily.csv}{AAPL_2024-01-01_2025-12-31_daily.csv}, \repofile{module_03_risk/MSFT_2024-01-01_2025-12-31_daily.csv}{MSFT_2024-01-01_2025-12-31_daily.csv}, \repofile{module_03_risk/GOOGL_2024-01-01_2025-12-31_daily.csv}{GOOGL_2024-01-01_2025-12-31_daily.csv}, and \repofile{module_03_risk/AMZN_2024-01-01_2025-12-31_daily.csv}{AMZN_2024-01-01_2025-12-31_daily.csv}, together with \repofile{module_03_risk/module_03_single_asset_var_cvar_summary.csv}{module_03_single_asset_var_cvar_summary.csv}, \repofile{module_03_risk/module_03_multi_asset_var_cvar_summary.csv}{module_03_multi_asset_var_cvar_summary.csv}, and \repofile{module_03_risk/module_03_risk_data_manifest.json}{module_03_risk_data_manifest.json}. The public repository entry point is \CodeRepo.

The single-asset case was rerun locally and now matches the refreshed manuscript table: Historical VaR/CVaR \(=\) \(2.694\%\) / \(3.936\%\), normal-parametric VaR/CVaR \(=\) \(2.791\%\) / \(3.519\%\), and quantum-inspired grid VaR/CVaR \(=\) \(2.775\%\) / \(3.898\%\). The multi-asset case was also rerun locally and the manuscript table was updated accordingly: historical VaR/CVaR \(=\) \(2.293\%\) / \(3.256\%\), normal VaR/CVaR \(=\) \(2.197\%\) / \(2.779\%\), and quantum-inspired PCA VaR/CVaR \(=\) \(2.390\%\) / \(2.856\%\).

\subsection{Module 4: Quantum Machine Learning Cases}
\label{sec:app-module-4-quantum-machine-learning-cases}

The two learning cases discussed in \Cref{sec:m4-case-studies} are curated in package \repodir{module_04_qml}. Case A is the SPY direction-classification notebook \repofile{module_04_qml/module_04_case_a_quantum_classification.ipynb}{module_04_case_a_quantum_classification.ipynb}; Case B is the macro-regime notebook \repofile{module_04_qml/module_04_case_b_macro_regime_learning.ipynb}{module_04_case_b_macro_regime_learning.ipynb}. The local asset package now preserves the shared SPY market snapshot \repofile{module_04_qml/SPY_2024-01-01_2025-12-31_daily.csv}{SPY_2024-01-01_2025-12-31_daily.csv}, the macro snapshots \repofile{module_04_qml/FEDFUNDS_2024-01-01_2025-12-31.csv}{FEDFUNDS_2024-01-01_2025-12-31.csv}, \repofile{module_04_qml/CPIAUCSL_2024-01-01_2025-12-31.csv}{CPIAUCSL_2024-01-01_2025-12-31.csv}, and \repofile{module_04_qml/INDPRO_2024-01-01_2025-12-31.csv}{INDPRO_2024-01-01_2025-12-31.csv}, together with \repofile{module_04_qml/module_04_case_a_data_manifest.json}{module_04_case_a_data_manifest.json}, \repofile{module_04_qml/module_04_case_b_data_manifest.json}{module_04_case_b_data_manifest.json}, \repofile{module_04_qml/module_04_case_a_results.csv}{module_04_case_a_results.csv}, \repofile{module_04_qml/module_04_case_b_results.csv}{module_04_case_b_results.csv}, and \repofile{module_04_qml/README.md}{module_04_qml/README.md}. The public repository entry point is \CodeRepo.

Both QML cases were rerun locally during the present revision cycle on the fixed two-year window. In Case A, the quantum-kernel classifier delivered the strongest AUC in a weak-signal next-day prediction task, while the remaining quantum variants stayed at or below their classical counterparts. In Case B, the strongest overall model remained the classical multilayer perceptron, but QSVC also remained near-perfect and materially outperformed both classical SVC and the variational QNN on the selected test configuration.

\subsection{Module 5: Policy and Architecture Sources}
\label{sec:app-module-5-policy-and-architecture-sources}

Module~5 is not tied to a standalone notebook in the current version of the review. Its reproducibility requirement is therefore bibliographic and documentary rather than computational. In particular, the present revision draws on two local official documents used to strengthen the paper's motivation and deployment framing:
\begin{itemize}
    \item \textit{Research Note: Quantum Computing Applications in Financial Services} (Financial Conduct Authority, 2025).
    \item \textit{The UK's Modern Industrial Strategy} (Department for Business and Trade, 2025).
\end{itemize}

These documents remain in the project root and are cited in the main text as official background references. They are not reproduced inside the appendix package because the purpose here is to record provenance rather than duplicate source documents.

\clearpage
\section{Glossary of Specialised Terms and Abbreviations}
\label{sec:glossary}

\small
\begin{longtable}{p{0.18\linewidth}p{0.56\linewidth}p{0.16\linewidth}}
\caption{Core specialised terms and abbreviations used throughout the review.}
\label{tab:glossary}\\
\toprule
\textbf{Term} & \textbf{Meaning in this review} & \textbf{First substantive use} \\
\midrule
\endfirsthead
\toprule
\textbf{Term} & \textbf{Meaning in this review} & \textbf{First substantive use} \\
\midrule
\endhead
AES & Advanced Encryption Standard, used here as the benchmark symmetric cipher in post-quantum migration discussions. & \pagelink{term:aes} \\
CBDC & Central Bank Digital Currency. & \pagelink{term:cbdc} \\
CVA / DVA / FVA & Credit, Debit, and Funding Valuation Adjustments in derivative valuation and counterparty-risk workflows. & \pagelink{term:cva-dva-fva} \\
CVaR & Conditional Value-at-Risk, the expected loss conditional on being beyond the VaR threshold. & \pagelink{term:cvar} \\
DeFi & Decentralised Finance. & \pagelink{term:defi} \\
ECC & Elliptic-Curve Cryptography. & \pagelink{term:ecc} \\
GAS & Grover Adaptive Search, an oracle-based quantum search method for optimisation. & \pagelink{term:gas} \\
HNDL & Harvest-now, decrypt-later attack model for long-lived encrypted data. & \pagelink{term:hndl} \\
IQAE & Iterative Quantum Amplitude Estimation. & \pagelink{term:iqae} \\
Ising model & Spin-based Hamiltonian representation equivalent to many binary optimisation problems after affine transformation from QUBO variables. & \pagelink{term:ising} \\
MLQAE & Maximum-Likelihood Quantum Amplitude Estimation. & \pagelink{term:mlqae} \\
NISQ & Noisy intermediate-scale quantum, referring to near-term hardware with limited qubit counts and imperfect gates. & \pagelink{term:nisq} \\
PCA / QPCA & Principal Component Analysis and its quantum analogue, Quantum Principal Component Analysis. & \pagelink{term:pca-qpca} \\
PQC & Post-Quantum Cryptography. & \pagelink{term:pqc} \\
QAE & Quantum Amplitude Estimation, used here as the canonical quantum primitive for expectation estimation and Monte Carlo acceleration. & \pagelink{term:qae} \\
QAOA & Quantum Approximate Optimisation Algorithm. & \pagelink{term:qaoa} \\
QML & Quantum Machine Learning. & \pagelink{term:qml} \\
QNN & Quantum Neural Network, typically implemented through trainable parameterised quantum circuits. & \pagelink{term:qnn} \\
QROM / QROAM & Quantum Read-Only Memory and Quantum Random-Access Optimised Memory, used for structured loading of classical data into quantum circuits. & \pagelink{term:qrom-qroam} \\
QSVM / QSVC & Quantum Support Vector Machine / Quantum Support Vector Classifier. & \pagelink{term:qsvm-qsvc} \\
QUBO & Quadratic Unconstrained Binary Optimisation, the principal binary formulation used to map constrained financial optimisation problems to quantum solvers. & \pagelink{term:qubo} \\
QVC / VQC & Quantum or Variational Quantum Classifier. & \pagelink{term:qvc-vqc} \\
QVE / VQE & Variational Quantum Eigensolver. & \pagelink{term:qve-vqe} \\
VQS & Variational Quantum Simulation. & \pagelink{term:vqs} \\
VaR & Value-at-Risk, the loss threshold associated with a chosen confidence level. & \pagelink{term:var} \\
XVA & Collective shorthand for valuation adjustments such as CVA, DVA, and FVA. & \pagelink{term:xva} \\
\bottomrule
\end{longtable}
\normalsize

\phantomsection

\end{document}